\documentclass[journal,draftcls,onecolumn,12pt,romanappendices]{IEEEtran-v17}

\pdfoutput=1

\usepackage{xspace}
\usepackage{url}
\usepackage[cmex10]{amsmath}
\usepackage{bbm}
\usepackage{graphicx}
\usepackage{tikz}
\usepackage{pgfplots}

\usepackage{ifthen}

\newboolean{conf}
\setboolean{conf}{false} % boolvar=true or false
\usepackage{fancyref} % for \fref
\usepackage{amsmath}
\usepackage{amssymb}
\usepackage{footnote}
\usepackage{xcolor}
\usepackage[nosort]{cite}

\usepackage{color}
\definecolor{links}{rgb}{0.7,0,0}   % red
\definecolor{urls}{rgb}{0,0,0.8}    % blue
\definecolor{cites}{rgb}{0,0,0.8}   % blue

\usepackage[colorlinks,hyperindex,linkcolor=links,citecolor=cites,urlcolor=urls]{hyperref}

\usepackage{pgfplots}
\pgfplotsset{width=12cm, height=9cm,  compat=1.12}% <-- moves axis labels near ticklabels (respects tick label widths)
\pgfplotsset{major grid style={
color=black,
line width=0.1mm,
loosely dotted,
},
}
\pgfplotsset{
tick style={
color = black
},
}
 \pgfplotsset{every x tick label/.append style={font=\footnotesize}}
 \pgfplotsset{every y tick label/.append style={font=\footnotesize}}

\usepackage{bm}
\usepackage{upgreek}
\usepackage{amssymb}
\usepackage{amsfonts}
\usepackage{mathrsfs}
\usepackage{xspace}

\makeatletter
\def\amsbb{\use@mathgroup \M@U \symAMSb}
\makeatother

\newcommand{\lefto}{\mathopen{}\left}
\newcommand{\safemath}[2]{\newcommand{#1}{\ensuremath{#2}\xspace}}
\safemath{\opE}{\amsbb{E}}
\newcommand{\Ex}[2]{\ensuremath{\amsbb{E}_{#1}\mathopen{}\left[#2\right]}} 	% expectation
\safemath{\prob}{\amsbb{P}}
\safemath{\bigO}{\mathcal{O}}
\safemath{\littleo}{\mathit{o}}

\safemath{\extendreal}{\overline{\realset}}

\newtheorem{thm}{Theorem}
\newtheorem{lemma}[thm]{Lemma}

\newtheorem{rem}{Remark}

\newcommand{\indfun}[1]{\mathbbmss{1}\lefto\{#1\right\}}
\newtheorem{cor}[thm]{Corollary}

\safemath{\matA}{\mathsf{A}}
\safemath{\matB}{\mathsf{B}}
\safemath{\matC}{\mathsf{C}}
\safemath{\matD}{\mathsf{D}}
\safemath{\matE}{\mathsf{E}}
\safemath{\matF}{\mathsf{F}}
\safemath{\matG}{\mathsf{G}}
\safemath{\matH}{\mathsf{H}}
\safemath{\matI}{\mathsf{I}}
\safemath{\matJ}{\mathsf{J}}
\safemath{\matK}{\mathsf{K}}
\safemath{\matL}{\mathsf{L}}
\safemath{\matM}{\mathsf{M}}
\safemath{\matN}{\mathsf{N}}
\safemath{\matO}{\mathsf{O}}
\safemath{\matP}{\mathsf{P}}
\safemath{\matQ}{\mathsf{Q}}
\safemath{\matR}{\mathsf{R}}
\safemath{\matS}{\mathsf{S}}
\safemath{\matT}{\mathsf{T}}
\safemath{\matU}{\mathsf{U}}
\safemath{\matV}{\mathsf{V}}
\safemath{\matW}{\mathsf{W}}
\safemath{\matX}{\mathsf{X}}
\safemath{\matY}{\mathsf{Y}}
\safemath{\matZ}{\mathsf{Z}}
\safemath{\matSigma}{\mathsf{\Sigma}}
\safemath{\matPhi}{\mathsf{\Phi}}
\safemath{\matLambda}{\mathsf{\Lambda}}
\safemath{\matDelta}{\mathsf{\Delta}}

\safemath{\randveca}{\bm{A}}
\safemath{\randvecb}{\bm{B}}
\safemath{\randvecc}{\bm{C}}
\safemath{\randvecd}{\bm{D}}
\safemath{\randvece}{\bm{E}}
\safemath{\randvecf}{\bm{F}}
\safemath{\randvecg}{\bm{G}}
\safemath{\randvech}{\bm{H}}
\safemath{\randveci}{\bm{I}}
\safemath{\randvecj}{\bm{J}}
\safemath{\randveck}{\bm{K}}
\safemath{\randvecl}{\bm{L}}
\safemath{\randvecm}{\bm{M}}
\safemath{\randvecn}{\bm{N}}
\safemath{\randveco}{\bm{O}}
\safemath{\randvecp}{\bm{P}}
\safemath{\randvecq}{\bm{Q}}
\safemath{\randvecr}{\bm{R}}
\safemath{\randvecs}{\bm{S}}
\safemath{\randvect}{\bm{T}}
\safemath{\randvecu}{\bm{U}}
\safemath{\randvecv}{\bm{V}}
\safemath{\randvecw}{\bm{W}}
\safemath{\randvecx}{\bm{X}}
\safemath{\randvecy}{\bm{Y}}
\safemath{\randvecz}{\bm{Z}}

\safemath{\randvecLambda}{\bm{\Lambda}}

\safemath{\randmatA}{\amsbb{A}}
\safemath{\randmatB}{\amsbb{B}}
\safemath{\randmatC}{\amsbb{C}}
\safemath{\randmatD}{\amsbb{D}}
\safemath{\randmatE}{\amsbb{E}}
\safemath{\randmatF}{\amsbb{F}}
\safemath{\randmatG}{\amsbb{G}}
\safemath{\randmatH}{\amsbb{H}}
\safemath{\randmatI}{\amsbb{I}}
\safemath{\randmatJ}{\amsbb{J}}
\safemath{\randmatK}{\amsbb{K}}
\safemath{\randmatL}{\amsbb{L}}
\safemath{\randmatM}{\amsbb{M}}
\safemath{\randmatN}{\amsbb{N}}
\safemath{\randmatO}{\amsbb{O}}
\safemath{\randmatP}{\amsbb{P}}
\safemath{\randmatQ}{\amsbb{Q}}
\safemath{\randmatR}{\amsbb{R}}
\safemath{\randmatS}{\amsbb{S}}
\safemath{\randmatT}{\amsbb{T}}
\safemath{\randmatU}{\amsbb{U}}
\safemath{\randmatV}{\amsbb{V}}
\safemath{\randmatW}{\amsbb{W}}
\safemath{\randmatX}{\amsbb{X}}
\safemath{\randmatY}{\amsbb{Y}}
\safemath{\randmatZ}{\amsbb{Z}}
\safemath{\randmatSigma}{\mathbb{\Sigma}}
\safemath{\randmatPhi}{\mathbb{\Phi}}

\safemath{\pdff}{f}
\safemath{\pdfp}{p}
\safemath{\pdfq}{q}

\safemath{\cdfF}{F}
\safemath{\cdfP}{P}
\safemath{\cdfQ}{Q}

\safemath{\veca}{\bm{a}}
\safemath{\vecb}{\bm{b}}
\safemath{\vecc}{\bm{c}}
\safemath{\vecd}{\bm{d}}
\safemath{\vece}{\bm{e}}
\safemath{\vecf}{\bm{f}}
\safemath{\vecg}{\bm{g}}
\safemath{\vech}{\bm{h}}
\safemath{\veci}{\bm{i}}
\safemath{\vecj}{\bm{j}}
\safemath{\veck}{\bm{k}}
\safemath{\vecl}{\bm{l}}
\safemath{\vecm}{\bm{m}}
\safemath{\vecn}{\bm{n}}
\safemath{\veco}{\bm{o}}
\safemath{\vecp}{\bm{p}}
\safemath{\vecq}{\bm{q}}
\safemath{\vecr}{\bm{r}}
\safemath{\vecs}{\bm{s}}
\safemath{\vect}{\bm{t}}
\safemath{\vecu}{\bm{u}}
\safemath{\vecv}{\bm{v}}
\safemath{\vecw}{\bm{w}}
\safemath{\vecx}{\bm{x}}
\safemath{\vecy}{\bm{y}}
\safemath{\vecz}{\bm{z}}

\safemath{\veclambda}{\bm{\lambda}}
\safemath{\vecpi}{\bm{\pi}}
\safemath{\vecsigma}{\bm\sigma}              			% Singular value vector

\safemath{\setA}{\mathcal{A}}
\safemath{\setB}{\mathcal{B}}
\safemath{\setC}{\mathcal{C}}
\safemath{\setD}{\mathcal{D}}
\safemath{\setE}{\mathcal{E}}
\safemath{\setF}{\mathcal{F}}
\safemath{\setG}{\mathcal{G}}
\safemath{\setH}{\mathcal{H}}
\safemath{\setI}{\mathcal{I}}
\safemath{\setJ}{\mathcal{J}}
\safemath{\setK}{\mathcal{K}}
\safemath{\setL}{\mathcal{L}}
\safemath{\setM}{\mathcal{M}}
\safemath{\setN}{\mathcal{N}}
\safemath{\setO}{\mathcal{O}}
\safemath{\setP}{\mathcal{P}}
\safemath{\setQ}{\mathcal{Q}}
\safemath{\setR}{\mathcal{R}}
\safemath{\setS}{\mathcal{S}}
\safemath{\setT}{\mathcal{T}}
\safemath{\setU}{\mathcal{U}}
\safemath{\setV}{\mathcal{V}}
\safemath{\setW}{\mathcal{W}}
\safemath{\setX}{\mathcal{X}}
\safemath{\setY}{\mathcal{Y}}
\safemath{\setZ}{\mathcal{Z}}
\safemath{\emptySet}{\varnothing}

\safemath{\veczero}{\mathbf{0}} %vector font of 0,

\safemath{\diag}{\mathrm{diag}}

\safemath{\jpg}{\mathcal{CN}}			% jointly proper Gaussian
\safemath{\complexset}{\amsbb{C}}
\safemath{\realset}{\amsbb{R}}
\safemath{\natunum}{\mathbb{N}}
\safemath{\posrealset}{\realset_{+}}
\safemath{\integerset}{\amsbb{N}}

\newcommand{\given}{\,\vert\,}				% conditioning
\safemath{\define}{\triangleq}			% definition

\newcommand{\error}{\epsilon}
\newcommand{\secrecy}{\delta}
\newcommand{\tvar}{d}

\newcommand{\msg}{W}
\newcommand{\CS}{C_{\mathrm{S}}}

\newcommand{\unif}{\mathrm{unif}}

\newcommand{\VS}{V_{\mathrm{S}}}

\newcommand{\avg}{\mathrm{avg}}
\newcommand{\VBSC}{V_{\mathrm{BSC}}}
\newcommand{\Hb}{H_{\mathrm{b}}}

\newcommand{\der}{\mathrm{d}}

\begin{document}
\IEEEoverridecommandlockouts
 
\title{Wiretap Channels: Nonasymptotic \\Fundamental Limits }

\author{Wei Yang, \IEEEmembership{Member, IEEE}, Rafael F. Schaefer, \IEEEmembership{Senior Member, IEEE},\\
 and H. Vincent Poor, \IEEEmembership{Fellow, IEEE}
\thanks{The work of H. V. Poor and W. Yang was supported in part by the US  National Science Foundation under Grants CNS-1702808, ECCS-1343210
and ECCS-1647198.
 The work of R. F. Schaefer was supported by the German Research Foundation (DFG) under Grant WY 151/2-1. 
 The material of this paper was presented in part at the  IEEE International Symposium on Information Theory (ISIT), Barcelona, Spain, July 2016, and  the IEEE International Symposium on Information Theory (ISIT), Aachen, Germany, June 2017.
 }\\
\thanks{W. Yang and H. V. Poor are with the Department of Electrical Engineering, Princeton University, Princeton, NJ 08544 USA (email: $\{$weiy, poor$\}$@princeton.edu).}
\thanks{R. F. Schaefer is with the Information Theory and Applications Chair, Technische Universit\"{a}t Berlin, 10587 Berlin, Germany (email: rafael.schaefer@tu-berlin.de).}
}
\maketitle

\begin{abstract}
This paper investigates the maximal  secret communication rate over a wiretap channel subject to reliability and secrecy constraints at a given blocklength. 
New achievability and converse bounds are derived, which are uniformly tighter than existing bounds, and lead to the tightest bounds on the second-order coding rate for discrete memoryless and Gaussian wiretap channels. The exact second-order coding rate is established for semi-deterministic wiretap channels, which characterizes the optimal tradeoff between reliability and secrecy in the finite-blocklength regime. Underlying our achievability bounds are two new privacy amplification results, which not only refine the existing results, but also achieve stronger notions of secrecy.

\end{abstract}

\section{Introduction}

The pioneering work of Wyner~\cite{wyner1975-10a} and Csisz\'{a}r and K\"{o}rner~\cite{csiszar1978-05a} has characterized the maximal secret communication rate -- called the secrecy capacity -- at which information can be transmitted securely and reliably over a wiretap channel. 
In particular,  it was shown that both the error probability and the information leakage can be made as small as desired, as long as the transmission rate is below the secrecy capacity and as long as the data are mapped to sufficiently long codewords.  
Since then, there have been extensive studies on  the secrecy capacity of different wiretap models as well as schemes that achieve the secrecy capacity (see~\cite{Liang08} and~\cite{bloch11-b} for a comprehensive treatment on this subject).

Motivated by contemporary practical applications that often use short packets (e.g., smart metering and traffic safety), there is considerable interest in investigating the backoff from (secrecy) capacity incurred by coding at a finite blocklength~\cite{polyanskiy10-05,hayashi2009-11a,tan14}. 
In this paper, we study the maximal secret communication rate $R^*(n,\error,\secrecy)$ for a given blocklength $n$, error probability $\error$, and information leakage~$\delta$, over wiretap channels. 
An exact characterization of this fundamental quantity is computationally impossible for most wiretap channels of interest; instead, the aim of this paper is to develop tight upper and lower bounds as well as accurate approximations.  
Throughout this paper, the information leakage to the eavesdropper is measured by the total variation distance (between proper distributions). 
More specifically, we consider two secrecy metrics:  
\begin{itemize}
\item Average secrecy: the total variation distance between the joint distribution $P_{WZ}$ of the secret message $W$ and the eavesdropper's observation $Z$,  and an ideal distribution in which $W$ is uniformly distributed and is independent of $Z$.  
\item Maximum secrecy: the maximum total variation distance between $P_{Z|W=m}$ and $P_{Z}$, maximized over all messages $m$ in the message set. 
\end{itemize}
As we shall see in Section~\ref{sec:model},  the latter is equivalent to \emph{distinguishing security} and \emph{semantic security}~\cite{bellare12-00,goldwasser1984-04a}, which are two secrecy metrics widely accepted in the cryptography literature.

\emph{Prior work:} Hayashi~\cite{hayashi2006-04a} established general achievability bounds using the \emph{channel resolvability} technique~\cite{han93-03} (a.k.a. 
soft-covering~\cite{wyner1975-03a}) and studied the secrecy exponent (i.e., the exponential decreasing rate of the information leakage) for a fixed communication rate.  Later, he improved the secrecy exponent by leveraging the \emph{privacy amplification} technique~\cite{hayashi13-11a}. For the setting of fixed $\error$ and $\delta$ and $n\to\infty$, Yassaee \emph{et al}.~\cite{yassaee2013-07a} derived an achievability bound  on the second-order coding rate~\cite{hayashi2009-11a} (also known as dispersion~\cite{polyanskiy10-05}) by using the output statistics of random binning, which improves an earlier result by Tan~\cite{tan2012-11a}.
 Achievability bounds for wiretap channels with other secrecy metrics can be found in~\cite{hayashi2011-06a, yassaee2015-06a,tahmasbi2016-09a,parizi2017-01a}.
On the converse side,  Tyagi and Watanabe proposed a one-shot converse bound for the problem of secret key agreement~\cite{tyagi2015-05a}, which exploits binary hypothesis testing. Building upon the technique in~\cite{tyagi2015-05a}, Hayashi \emph{et al.}~\cite{hayashi2014-09a} established a converse bound for wiretap channels, which leads to the strong converse for the degraded case. This result strengthens a previous strong converse established by Tan and Bloch using the information spectrum method~\cite{tan2015-09a}.

\subsubsection*{Contributions}
In this paper, we develop new achievability and converse bounds on the maximal secrecy rate $R^*(n,\error,\delta)$ for general wiretap channels. 
Our achievability bounds are based on two new privacy amplification results, which yield secrecy codes that satisfy, respectively,  the   average  and maximum secrecy constraints.
These bounds are expressed in terms of  the $E_\gamma$ metric~\cite{liu2017-05a}, and establish a nice connection between two popular tools used in the achievability proofs of information-theoretic security---privacy amplification and channel resolvability. 
Our converse bound is motivated by the secret-key-agreement converse in~\cite{tyagi2015-05a} and the meta-converse in~\cite{polyanskiy10-05}. It relates the error probability and the  information leakage of a secrecy code  to  the error probability of a binary hypothesis testing, and relies on a test that is tailored to the wiretap channel. Both our achievability and converse bounds  are uniformly tighter than the best existing bounds (to the best of our knowledge).  

By analyzing the behavior of our bounds in the regime of fixed~$\error$ and $\delta$ and $n\to\infty$, we obtain upper and lower bounds on the second-order coding rate of discrete-memoryless wiretap channels (DM-WTCs) and Gaussian wiretap channels. More specifically, we show that 
\begin{IEEEeqnarray}{rCl}
\label{eq:review-bounds-2016}
&&\CS -\sqrt{\frac{V_1}{n}}Q^{-1}(\error) -\sqrt{\frac{V_2}{n}}Q^{-1}(\delta) \lessapprox 
R^*(n,\error,\delta) \lessapprox \CS -\sqrt{\frac{V_3}{n}}Q^{-1}(\error+\delta)  \IEEEeqnarraynumspace
\label{eq:bound-cs-v-introduction}
\end{IEEEeqnarray}
where $\CS$ is the secrecy capacity, and $V_1$, $V_2$, and $V_3$ are constants that depend on the stochastic variations of the legitimate and the eavesdropper's channel. 
This result holds under both the average secrecy and the  maximum secrecy constraint.  
Furthermore, the achievable second-order rate in~\eqref{eq:bound-cs-v-introduction} improves on the ones derived in~\cite{yassaee2013-07a} and \cite{tan2012-11a}.
Bridging the gap between the upper and lower bounds in~\eqref{eq:bound-cs-v-introduction} for general wiretap channels seems very difficult. 
This can be done, however,  for the class of semi-deterministic wiretap channels, in which the channel between the transmitter and the legitimate receiver is deterministic, while that between the transmitter and the eavesdropper is a discrete memoryless channel (DMC).
%
%
%
 %
%Since the legitimate channel is deterministic,  the transmitter can adjust its transmission to the legitimate receiver's output, and, thus, avoid the rate loss due to separation.
%
%
%
% 
%
Indeed, for semi-deterministic wiretap channels both the achievability and the converse bounds in~\eqref{eq:review-bounds-2016} can be tightened,  which yields the following asymptotic expansion  for $R^*(n,\error,\delta)$:
\begin{IEEEeqnarray}{rCl}
\label{eq:review-bounds-2017}
R^*(n,\error,\delta) = \CS -\sqrt{\frac{\VS}{n}}Q^{-1}\lefto(\frac{\delta}{1-\error}\right) + \bigO\lefto(\frac{\log n}{n}\right)  \IEEEeqnarraynumspace
\end{IEEEeqnarray}
for every~$\error$ and $\delta$ such that $\error +\delta <1$.
Here, $\VS$ is the conditional variance of an appropriate information density term.
The key idea behind the proof of~\eqref{eq:review-bounds-2017} is to understand the optimal tradeoff between reliability and secrecy. 
While it is possible to trade reliability for secrecy for a general wiretap channel, the deterministic nature of  the semi-deterministic wiretap channel makes it also possible to trade  secrecy for reliability (through additional privacy amplification).

\subsubsection*{Notation}
The cardinality of a finite set $\setA$ is denoted by~$|\setA|$, and the uniform distribution over $\setA$ is designated as $Q_{\setA}^{\unif}$.
For an input distribution $P_{X}$ and a random transformation $P_{Y|X}$, we let $P_{Y|X}\circ P_X$ denote the marginal distribution~of $P_{X}P_{Y|X}$ on $Y$.
If $P_X = Q_{\setC}^{\unif}$ for some finite set $\setC \subset \setX$, then we write $P_{Y|\setC} \define P_{Y|X}\circ Q_{\setC}^{\unif}$.
We shall consider the following ``metrics'' between two probability measures $P$ and $Q$ on a sigma-algebra~$\setF$ of subsets of the set~$\setA$:
\begin{itemize}
\ifthenelse{\boolean{conf}}{}{
\item  $\ell_p$ distance,\footnote{For notational convenience, we write summation over $\setA$. 
However, our bounds hold for arbitrary probability spaces, unless otherwise specified.} $p\in\{1,2\}$ 
\begin{equation}
\|P-Q\|_p \define \left(\sum\limits_{x\in\setA} |P(x)-Q(x)|^p\right)^{1/p}
\end{equation}}
\item  Total variation distance
\begin{IEEEeqnarray}{rCl} 
\tvar(P,Q) &\define& \sup_{\setE \in \setF} |P( \setE) - Q(\setE)| 
\label{eq:def-total-var}\\
 &=& \frac{1}{2} \|P-Q\|_1
\end{IEEEeqnarray}
\item $E_{\gamma}$ metric~\cite{liu2017-05a}
\begin{IEEEeqnarray}{rCl}
E_{\gamma}(P,Q) &\define& P\lefto[\frac{\der P}{\der Q} \geq \gamma\right] - \gamma Q\lefto[\frac{\der P}{ \der Q} \geq \gamma\right] \label{eq:def-e-gamma-1}\\
&=& \sup_{\setE \in \setF} \big\{ P[\setE] -\gamma Q[\setE] \big\}\label{eq:def-e-gamma-2}
\end{IEEEeqnarray}
with the understanding ${\der P}/{\der Q} =+\infty$  in the singular set if $P$ is not absolutely continuous with respect to $Q$
\item Neyman-Pearson $\beta$ function
\begin{IEEEeqnarray}{rCl}
\label{eq:def-beta}
\beta_\alpha(P,Q) \define \min\int P_{T\given X}(1\given x)  Q(d x)
\end{IEEEeqnarray}
where the minimum is over all random transformations $P_{T\given X}: \setA\mapsto\{0,1\}$ satisfying
\begin{IEEEeqnarray}{rCl}
 \int P_{T\given X} (1\given x) P(dx)\geq \alpha.
\end{IEEEeqnarray}
\end{itemize}

\section{Definitions and Preliminaries}
\label{sec:model}
We consider the wiretap channel (WTC) model $(\setX,  P_{YZ\given X}, \setY\times\setZ)$   with transmitter $X$, legitimate receiver $Y$ and eavesdropper $Z$ introduced by Wyner~\cite{wyner1975-10a}. 
The transmitter wishes to communicate  a message $W\in \setM\define \{1,...,M\}$ to the receiver while keeping it secret from the eavesdropper. 
We first introduce the secrecy metrics used in this paper. 

\subsection{Secrecy Metrics}

A secrecy metric measures the statistical independence between the message $W$ and the eavesdropper's observation $Z$.
In this paper, we consider two secrecy metrics:
\begin{itemize}
\item Average total variation secrecy:
\begin{equation}
S(W|Z) \define \tvar(P_{WZ}, Q^{\unif}_{\setM} P_Z) .
\label{eq:def-secrecy-index}
\end{equation}
Following~\cite[Eq.~7]{csiszar2004-12a} and~\cite[p.~400]{Csiszar11}, we shall also refer to $S(W|Z)$ as the \emph{security index} of $W$ against $Z$.
\item Maximum total variation secrecy: 
 \begin{IEEEeqnarray}{rCl}
S_{\max}(W|Z) \define  \max_{m\in \setM}  \tvar(P_{Z|\msg =m},Q_Z)  
 \label{eq:secrecy-max-tv}
\end{IEEEeqnarray}
where $Q_Z \define P_{Z|W} \circ Q_{\setM}^{\unif}$.
\end{itemize}

To the best of our knowledge, the maximum total variation secrecy has not been used in the literature. However, as we shall show in Theorem~\ref{thm:equivalance-btw-secrecy} below,  it is equivalent to \emph{distinguishing security} and to \emph{semantic security}~\cite{bellare12-00,goldwasser1984-04a}, which are secrecy metrics widely accepted in the cryptography literature.  
Formally, the distinguishing security metric is defined as 
\begin{IEEEeqnarray}{rCl}
\mathrm{DS}(W|Z) \define \max_{m_0,m_1\in\setM} \tvar(P_{Z|W=m_0}, P_{Z|W=m_1}) .
\end{IEEEeqnarray}
In words, $\mathrm{DS}(W|Z)$  measures the (maximum) distinguishability of two messages $m_0$ and $m_1$.
And the semantic security metric is defined as
\begin{IEEEeqnarray}{rCl}
\mathrm{SS}(W|Z) \define \sup_{P_W, f, g} \Big( \prob[f(W)=g(Z)] - \sup_{\setS}\prob[ \setS=f(W)] \Big).
\end{IEEEeqnarray}
Here, the outer supremum is over all probability distributions $P_W$ of the message $W$, and  (possibly randomized) functions $f:\setM\to \setA$ and $g:\setZ\to\setA$, where the cardinality of the common co-domain of $f$ and $g$ is unrestricted. The inner supremum is over all random simulators~$\setS$ (i.e., random variables) supported on $\setA$ that is independent of $f(W)$.
The first probability term measures the probability of success of the eavesdropper who tries to guess a function $f$ of the message $W$ upon observing $Z$. The second probability term is the probability of success of the eavesdropper before observing $Z$. 

 \begin{thm}
 \label{thm:equivalance-btw-secrecy}
 For every transition probability distribution $P_{Z|W}: \setM \to \setZ$, we have 
 \begin{IEEEeqnarray}{rCl}
 \mathrm{SS}(W|Z) \leq S_{\max}(W|Z)   \leq \mathrm{DS}(W|Z) \leq 2 \mathrm{SS}(W|Z) \leq 2 S_{\max}(W|Z) .
\label{eq:dis-secy-all}
\end{IEEEeqnarray}
\end{thm}
\begin{IEEEproof}
The following relation between distinguishing security and semantic security was established in~\cite{bellare12-00}:
\begin{IEEEeqnarray}{rCl}
 \mathrm{SS}(W|Z)  \leq \mathrm{DS}(W|Z) \leq 2 \mathrm{SS}(W|Z)  .
\end{IEEEeqnarray}
To conclude~\eqref{eq:dis-secy-all}, it remains to show that  
\begin{IEEEeqnarray}{rCl}
S_{\max}(W|Z)   \leq \mathrm{DS}(W|Z)
\label{eq:tv-bound-ds}
\end{IEEEeqnarray}
and that 
\begin{IEEEeqnarray}{rCl}
 \mathrm{SS}(W|Z) \leq S_{\max} (W|Z) .
 \label{eq:ss-bound-tv}
 \end{IEEEeqnarray}

The inequality~\eqref{eq:tv-bound-ds} can be established as follows:
\begin{IEEEeqnarray}{rCl}
\mathrm{DS}(W|Z) &=& \max_{m_0,m_1\in\setM} \tvar(P_{Z|W=m_0}, P_{Z|W=m_1})  \\
&\geq& \max_{m_0\in\setM} \frac{1}{M}\sum_{m\in\setM}  \tvar(P_{Z|W=m_0}, P_{Z|W=m}) \\
&\geq& \max_{m_0\in\setM}    \tvar(P_{Z|W=m_0}, P_{Z|W}\circ Q_{\setM}^{\unif})\label{eq:-convexity-lalala}\\
&=& S_{\max}(W|Z). 
\end{IEEEeqnarray}
Here,~\eqref{eq:-convexity-lalala} follows because from Jensen's inequality and because the total variation distance $  \tvar(P,Q)$ is convex in $Q$. 

To prove~\eqref{eq:ss-bound-tv}, we  observe that (recall that $Q_Z \define P_{Z|W} \circ Q_{\setM}^{\unif}$)
\begin{IEEEeqnarray}{rCl}
 \mathrm{SS}(W|Z) &\leq& \sup_{P_W, f, g} \Big( P_{WZ}[f(W)=g(Z)] -  (P_WQ_Z)[ g(Z)=f(W)] \Big) \label{eq:ss-t-step-1}\\
&\leq& \sup_{P_W} \tvar(P_{WZ}, P_W{Q}_Z)\label{eq:ss-t-step-4}\\
&=& \sup_{m\in \setM} \tvar(P_{Z|W=m}, {Q}_Z)\label{eq:ss-t-step-5}\\
&=& S_{\max}(W|Z). \label{eq:ss-t-step-6}
 \end{IEEEeqnarray}
Here,~\eqref{eq:ss-t-step-1} follows because under $P_WQ_Z$, the random variables $g(Z)$ and $f(W)$ are independent;~\eqref{eq:ss-t-step-4} follows from the definition~\eqref{eq:def-total-var}; and~\eqref{eq:ss-t-step-5} follows because the mapping $P_W\mapsto \tvar(P_{WZ},P_WQ_Z)$ is linear.
\end{IEEEproof}

\subsection{Secrecy Codes}

We next introduce the notion of secrecy codes for a wiretap channel.
An $(M,\error,\delta)_{\mathrm{avg}}$ secrecy code for the wiretap channel $P_{YZ|X}$ consists~of 
\begin{itemize}
\item a message $\msg$ which is distributed on the message set  $\setM = \{1,\ldots,M\}$,
\item a \emph{randomized} encoder that generates a codeword $X(m)$, $m\in\setM$, according to a conditional probability distribution $P_{X \given W =m}$, and
\item a decoder $g: \setY \to \setM $ that assigns an estimate $\hat{W}$ to each received signal $Y \in\setY$.
\end{itemize}
The encoder and decoder satisfy the average error probability constraint 
\begin{equation}
\prob[g(Y) \neq W ] \leq \error
\label{eq:error-constraint-wtc}
\end{equation}
and the average total variation secrecy constraint 
\begin{equation}
S(W|Z) \leq \secrecy.
\label{eq:def-secrecy-constraint-def1}
\end{equation}

\begin{rem}
Note that in the definition of average secrecy code above, we do not require the message $W$ to be uniformly distributed over $\setM$.  
However, the codes used in our achievability bounds satisfy this additional constraint.
Our converse bounds hold regardless of whether  the message is uniformly distributed or not. 
\end{rem}

An $(M,\error,\delta)_{\max}$ secrecy code is defined similarly except that \eqref{eq:error-constraint-wtc} is replaced by the maximum  error probability constraint
 \begin{IEEEeqnarray}{rCl}
\max_{m\in \setM} P_{Y|W=m}[g(Y) \neq m  ]  \leq \error 
 \label{eq:error-def-max}
\end{IEEEeqnarray}
 and that~\eqref{eq:def-secrecy-constraint-def1} is replaced by the maximum total variation secrecy constraint
 \begin{IEEEeqnarray}{rCl}
S_{\max}(W|Z) \leq \secrecy.
 \label{eq:secrecy-max}
\end{IEEEeqnarray}

An $(M,\error,\delta)_{\mathrm{avg}}$ secrecy code for the channel $ P_{Y^nZ^n \given X^n}$ will be called an $(n,M,\error,\secrecy)_{\mathrm{avg}}$ secrecy code.  Furthermore, the maximal secrecy rate (average probability of error and average secrecy) is defined as
\begin{equation}
R_{\avg}^*(n,\error,\secrecy) \define \max\lefto\{\frac{\log M}{n}\!\!:\exists\, (n, M,\error,\secrecy)_{\avg}\,\text{secrecy code} \right\}. \IEEEeqnarraynumspace
\end{equation}
The maximal secrecy rate (maximum probability of error and maximum secrecy) $R_{\max}^*(n,\error,\secrecy)$ can be defined analogously. 
By definition, we have 
\begin{IEEEeqnarray}{rCl}
R^*_{\max}(n,\error,\delta) \leq R^*_{\avg}(n,\error,\secrecy).
\end{IEEEeqnarray}

\section{Bounds on the Secrecy Rate}

\subsection{Privacy Amplification}
\label{sec:privacy-amplification}
Our achievability bounds rely on the privacy amplification technique. 
Privacy amplification is a method of distilling secret information from a random source that is only partially secret~\cite{bennett1995-11a}. 
More specifically, let Alice and Bob be given a random variable $X \in \setX$, such as a random key, which is correlated with an eavesdropper Eve's observation $Z$.
Alice and Bob wish to publicly choose a (hash) function $g: \setX \to \setK$ with $|\setK|< |\setX|$ such that $ g(X)$ is almost independent of Eve's observation  $Z$, and that the resulting random variable $g(X)$ is approximately uniformly distributed over $\setK$.
In this section, we establish two privacy amplification results that will be used to prove achievability bounds for wiretap channels.

Our first result refines the existing privacy amplification results in the literature.

\begin{lemma}
\label{lemma:privacy-amplification}
Let $\setX$, $\setK$ be finite sets, and $\setZ$ be an arbitrary set. 
Let $P_{XZ}$ be a probability distribution on  $\setX \times \setZ$. Then, 
% and let  $Q^{\unif}_{\setX}$ and $Q^{\unif}_{\setK}$ be the uniform distributions over $\setX$ and $\setK$, respectively.
%
%let $M$ be a positive integer, and let $P_0$ be the uniform distribution over $\{1,\ldots,M\}$. 
%
\begin{enumerate}
\item \label{item-lemma1-part1}
For every $\gamma>0$ and every probability distribution $Q_Z$ supported on $\setZ$, there exists a function $g: \setX \to \setK$ such that
\begin{IEEEeqnarray}{rCl}
S(g(X)|Z) &\leq&   E_{\gamma}(P_{XZ},  Q^{\unif}_{\setX} Q_Z)  +  \frac{1}{2}\sqrt{\frac{\gamma}{L} \Ex{}{\exp\lefto( - \big|\imath(X;Z)  - \log\gamma \big| \right)} } \IEEEeqnarraynumspace
\label{eq:privacy-amplification-lemma}
\end{IEEEeqnarray}
where $L\define |\setX|/|\setK|$, 
\begin{IEEEeqnarray}{rCl}
\imath(x;z) \define \log \frac{\der P_{XZ}}{\der (Q_{\setX}^{\unif}Q_Z)}(x,z)
\label{eq:info-den-qz}
\end{IEEEeqnarray}
and the expectation in~\eqref{eq:privacy-amplification-lemma} is taken with respect to $(X,Z)\sim P_{XZ}$. 
\item  \label{item-lemma1-part2} If in addition $L \in \mathbb{N}$ and $P_X = Q_{\setX}^{\unif}$, then there exists a function $g(\cdot)$ that satisfies~\eqref{eq:privacy-amplification-lemma} and that $g(X)$ is uniformly distributed over $\setK$.
\end{enumerate}
\end{lemma}
\begin{IEEEproof}
The proof is based on random hashing. Similar to a random coding argument, the idea is to  show that the average secrecy index of a properly chosen random hash function $G$ is upper-bounded by~\eqref{eq:privacy-amplification-lemma}. 
See Appendix~\ref{app:proof-of-lemma-privacy-amp} for the proof.
\end{IEEEproof}
 \begin{rem}
The bound~\cite[Cor.~5.6.1]{renner2005-09a}  can be obtained from~\eqref{eq:privacy-amplification-lemma} by upper-bounding the expectation term on the right-hand side (RHS) of~\eqref{eq:privacy-amplification-lemma} by $1$ and by using~\cite[Lem.~19]{hayashi2016-06a}. 
This implies that~\cite[Cor.~5.6.1]{renner2005-09a} is weaker than Lemma~\ref{lemma:privacy-amplification}. 
Similarly, the bounds in~\cite[Th.~1]{hayashi13-11a} and~\cite[Th.~6]{watanabe2013-07a} can also be obtained by weakening~\eqref{eq:privacy-amplification-lemma}, and are therefore weaker than Lemma~\ref{lemma:privacy-amplification}. 
\end{rem}

\begin{rem}
The bound~\eqref{eq:privacy-amplification-lemma} can be slightly improved if $P_{XZ}$ admits additional (symmetry) structures; see Theorem~\ref {thm:bsc-converse} for the case in which $P_{Z|X}$ is the channel law of a binary symmetric channel.
\end{rem}

Note that, the security index $S(g(X)|Z)$ measures the average total variation distance between $P_{Z|g^{-1}(K)}$ and $P_Z$ for the  random variable $K=g(X)$.
The next result shows that there exists a hash function~$g$ such that the maximum total variation secrecy  $S_{\max} ( g(X)|  P_Z)$ is very close to the average.
Its proof relies on Hoeffding's reduction argument for sample without replacement~\cite{hoeffding1963-03a} and on McDiarmid's inequality (also known as bounded difference inequality)~\cite{mcdiarmid-1989}.

\begin{lemma}
\label{prop:strong-pa}
Let $\setX$ and $\setK$ be   finite sets satisfying $L\define |\setX|/|\setK| \in \mathbb{N}$, and let $\setZ$ be an arbitrary set.
Let $P_X=Q_{\setX}^{\unif}$ be the uniform distribution over $\setX$, and let $P_Z=P_{Z|X} \circ P_X$. 
Let $\setG$ be the set of all functions $g :\setX\to \setK$ that satisfy $|g^{-1}(k)| = L$, $\forall k\in \setK$, and let $G$ be uniformly distributed over $\setG$. 
Then, we have 
\begin{IEEEeqnarray}{rCl}
\prob\lefto[  \max_{k\in\setK}  \tvar(P_{Z|G^{-1}(k) }, P_{Z}) \geq r + \mu\right] \leq  |\setK| e^{-2L r^2}  \IEEEeqnarraynumspace
\label{eq:strong-pa}
\end{IEEEeqnarray}
where 
\begin{IEEEeqnarray}{rCl}
\mu \define \Ex{}{\tvar( P_{Z|\setA},P_Z)}
\label{eq:def-mu-etvar-pza}
\end{IEEEeqnarray}
with $\setA=\{\bar{X}_1,\ldots,\bar{X}_L\}$ being a random codebook whose codewords are i.i.d. and $P_X$-distributed. 
Consequently, there exists a function $g\in\setG$ such that 
\begin{IEEEeqnarray}{rCl}
 \max_{k\in\setK}  \tvar(P_{Z|g^{-1}(k) }, P_{Z})  \leq \mu + \sqrt{ \frac{\log (|\setK| +1)}{2L}}.
 \label{eq:max-bound-tv}
\end{IEEEeqnarray}
\end{lemma}
 \begin{IEEEproof}
 See Appendix~\ref{app:proof-strong-pa}.
 \end{IEEEproof}
\begin{rem}
In the applications of Lemma~\ref{prop:strong-pa}, $|\setK|$ and $L$ are usually taken to be exponential in the dimension $n$. 
In such cases, $|\setK|e^{-2Lr^2}$ vanishes  double-exponentially fast in $n$. Such a double-exponential convergence was first observed by Ahlswede and Csisz\'{a}r in the analysis of random bits extractors~\cite{ahlswede1998-01a} (see also~\cite[p.~404]{Csiszar11}), and more recently in the setting of channel resolvability/soft-covering~\cite{cuff2015-09a,cuff2016-07a}. 
\end{rem}

Interestingly, Lemma~\ref{prop:strong-pa} connects privacy amplification to channel resolvability~\cite{han93-03}. 
Indeed, given a probability distribution $P_X$ on $\setX$ and a random transformation $P_{Z|X}:\setX\to\setZ$, the channel resolvability problem aims to find a set $\setA\subset \setX$ with  cardinality $|\setA|\leq L$ that minimizes the (total variation) distance between $P_{Z|\setA}$ and  the target $P_Z\define P_{Z|X}\circ P_X$.   
A good way to find $\setA$ is through random coding. Namely, we generate each element (i.e., codeword) in $\setA$ independently and identically according to $P_{X}$.
It turns out that the average total variation distance $\Ex{}{\tvar( P_{Z|\setA},P_Z)}$ for the random codebook $\setA$ is a tight upper bound on $\min_{\setA: |\setA| \leq L} d(P_{Z|\setA} ,P_Z)$. 

The next result shows that  $\Ex{}{\tvar(P_{Z\given \setA} ,P_Z)}$  admits an upper bound that is of the same form as~\eqref{eq:privacy-amplification-lemma},  which  indicate  that channel resolvability and privacy amplification are based on the same probabilistic principle.

\begin{lemma}
\label{lemma:channel-resolvability}
Let $\setX$  be a finite set, and let $\setZ$ be an arbitrary set.
Let~$P_{XZ}$ be a probability distribution supported on $\setX\times\setZ$ such that $P_X(x) >0$ for every $x\in\setX$. 
Let $\setA\define \{X_1,\ldots, X_{L}\}$ be a random codebook whose codewords are  independently and identically generated according to $P_X$.
For every $\gamma>0$, every probability distribution $Q_{Z}$ supported on $\setZ$, we have 
\begin{IEEEeqnarray}{rCl}
\Ex{}{\tvar (P_{Z|\setA} ,P_Z)} \leq E_{\gamma}(P_{XZ},Q_{\setX}^{\unif}Q_Z) + \frac{1}{2}\sqrt{ \frac{\gamma}{L} \Ex{}{\exp(-|\imath(\widetilde{X};Z)  -\log \gamma|)}} 
\label{eq:bound-channel-resolvability}
\end{IEEEeqnarray}
where $\imath(\cdot;\cdot)$ is defined in~\eqref{eq:info-den-qz}, and the expectation on the right-hand side of~\eqref{eq:bound-channel-resolvability} is taken with respect to $(\widetilde{X},Z) \sim Q_{\setX}^{\unif}P_{Z|X}$.
\end{lemma}
\begin{IEEEproof}
See Appendix~\ref{app:proof-lemma-resolvability}.
\end{IEEEproof}

In this paper, we will only use privacy amplification to prove achievability bounds for the wiretap channel. 
For completeness, though, we  present a converse result for privacy amplification below, which shows that the achievability bounds in Lemmas~\ref{lemma:privacy-amplification}--\ref{lemma:channel-resolvability} are tight up to some nuisance terms.

\begin{lemma}
\label{thm:converse-partition-egamma}
Let $\setX$, $\setZ$, $\setK$, and $P_{XZ}$ be as defined in Lemma~\ref{lemma:privacy-amplification}. Then, every function $g:\setX \to \setK$  satisfies 
\begin{IEEEeqnarray}{rCl}
S(g(X) |Z) &\geq&  E_{L}( P_{XZ} , Q_{\setX}^{\unif}Q_Z) .
\label{eq:thm-converse-pa}
\end{IEEEeqnarray}
\end{lemma}
\begin{IEEEproof}
See Appendix~\ref{app:proof-pa-converse-egamma}.
\end{IEEEproof}

As a corollary of Lemma~\ref{thm:converse-partition-egamma}, we obtain the following  converse bound for channel resolvability.

\begin{cor}
For every finite set $\setC=\{x_1,...,x_L\} \subset \setX$, every $P_{Z|X}:\setX \to \setZ$, and every $Q_{Z}$, we have 
\begin{equation}
\label{eq:converse-resolvability}
\tvar(P_{Z\given \setC}, Q_Z) \geq E_{L}( Q_{\setC}^{\unif} P_{Z\given X}, Q_{\setC}^{\unif} Q_{Z} ).
\end{equation}
\end{cor}
\begin{IEEEproof}
The proof follows by using Lemma~\ref{thm:converse-partition-egamma} with $\setX=\setC$, $\setK=\{1\}$, $P_X = Q_{\setC}^{\unif}$, and $g(\cdot)\equiv 1$.
\end{IEEEproof}

\subsection{Achievability Bound: Average Secrecy and Average Probability of Error}

The privacy amplification lemmas developed in Section~\ref{sec:privacy-amplification} allow us to convert an arbitrary (nonsecret) channel code for the legitimate channel $P_{Y|X}$ into a secrecy code with a guaranteed secrecy performance. 
%
%This step is commonly referred to as privacy amplification (see, e.g.,~\cite[p.~413]{Csiszar11}). 
By combining the random coding union (RCU) bound~\cite[Th.~16]{polyanskiy10-05} and the dependence testing (DT) bound~\cite[Th.~17, Eq.~(79)]{polyanskiy10-05} for channel coding with Lemma~\ref{lemma:privacy-amplification}, we obtain the following achievability bound for the wiretap channel. 

\begin{thm}
\label{thm:ach-wiretap}
Let $P_X$ be a probability distribution supported on $\setA \subset \setX$. For every $L \in \natunum$,  every $\gamma>0$,  and every $Q_Z$,  there exists an $(M,\error,\secrecy)_{\avg}$  code with a uniformly distributed message for the wiretap channel $(\setX, P_{YZ \given X},\setY\times\setZ)$ that satisfies
\begin{IEEEeqnarray}{rCl}
%\IEEEeqnarraymulticol{3}{l}{
\delta
%}\notag\\ \quad
  &\leq&  %\inf\limits_{\gamma>0 } \lefto\{ 
  \sup_{x\in \setA}  E_{\gamma}(P_{Z|X=x},Q_Z) 
 % \notag\\ && +\, 
  +\sup_{x\in \setA} \frac{1}{2}\sqrt{\frac{\gamma}{L} \Ex{P_{Z|X=x}}{ \exp\lefto( - \big|\imath(x;Z)  - \log\gamma \big| \right)}  } 
 % \right\}
 \IEEEeqnarraynumspace
\label{eq:thm-secrecy-bound}
\end{IEEEeqnarray}
and that
\begin{IEEEeqnarray}{rCl}
\error &\leq&   \min\mathopen{}\Big\{ \error_{\mathrm{RCU}}(ML) , \error_{\mathrm{DT}}\mathopen{}\Big(\frac{ML-1}{2}\Big)\Big\} \label{eq:min-rcu-beta-bound}.
\end{IEEEeqnarray}
Here, $\imath(\cdot;\cdot)$ is defined in~\eqref{eq:info-den-qz}, 
\begin{equation} 
%\IEEEeqnarraymulticol{3}{l}{
\error_{\mathrm{RCU}}(a)  \define \Ex{}{ \min\{1, (a-1)\prob[  i(\bar{X};Y) \geq i(X;Y) \given X, Y]\}} 
\label{eq:def-RCU} 
\end{equation}
where $i(x;y)\define \log \frac{\der P_{Y|X}(y|x)}{\der P_Y(y)}$,  $P_{XY\bar{X}}(x,y,\bar{x}) = P_{X}(x)P_{Y|X}(y|x)P_{X}(\bar{x})$, and 
\begin{IEEEeqnarray}{rCl}
\epsilon_{\mathrm{DT}} (a) \define 1- E_{a}(P_{XY},P_XP_Y).
\label{eq:def-DT-avg-1}
\end{IEEEeqnarray}
\end{thm}

\begin{IEEEproof}
By the RCU bound~\cite[Th.~16]{polyanskiy10-05} and the DT bound~\cite[Th.~17 and Eq.~(79)]{polyanskiy10-05},  there exists a code $\setC=\{x_1,\ldots,x_{LM}\}$ whose error probability over the legitimate channel $P_{Y|X}$ satisfies~\eqref{eq:min-rcu-beta-bound}. 
Next, we construct a secrecy code for the wiretap channel $P_{YZ|X}$ from this code  using Lemma~\ref{lemma:privacy-amplification}. 

Let $X\sim Q_{\setC}^{\unif}$.  Then,  by Lemma~\ref{lemma:privacy-amplification}, there exists a function $g: \setC \to \setM$ such that $g(X)$
is uniformly distributed over $\setM$, and that for every $\gamma>0$ and every $Q_Z$
\begin{IEEEeqnarray}{rCl}
%\IEEEeqnarraymulticol{3}{l}{
S(g(X)|Z)  &\leq&  E_{\gamma}( Q_{\setC}^{\unif}P_{Z|X}, Q_{\setC}^{\unif} Q_Z) \notag\\
 && +\, \frac{1}{2}\sqrt{\frac{\gamma}{L} \Ex{ Q_{\setC}^{\unif } P_{Z|X}   }{\exp\mathopen{}\big(\!-\!|\imath(X;Z) -\log \gamma |\big)}}  \\
&\leq&  \sup_{x\in\setA}  E_{\gamma}(P_{Z|X=x} ,Q_Z) \notag\\
&&+\, \sup_{x\in \setA} \frac{1}{2}\sqrt{\frac{\gamma}{L} \Ex{P_{Z|X=x}}{ \exp\lefto( - \big|\imath(x;Z)  - \log\gamma \big| \right)}  } . \IEEEeqnarraynumspace
\label{eq:privacy-amplification-lemma-application}
\end{IEEEeqnarray}
Here, the last step follows because the map $P_X\mapsto E_{\gamma}(P_{XZ},P_XQ_Z)$ is linear and because $\setC \subset \setA$.

Consider now a secrecy code with message $W=g(X)$ and with a random encoder $P_{X|W=m} = Q_{g^{-1}(m)}^{\unif}$.
In words, for each message $m\in\setM$, the encoder $P_{W|X}$  picks a codeword  from the set $g^{-1}(m)$ uniformly at random. 
The decoder first decodes the codeword~$\hat{X}$   and then outputs $\hat{W} = g(\hat{X})$. 
By construction, the message $W$ of the code is uniformly distributed, and the  average error probability  is upper-bounded by~\eqref{eq:min-rcu-beta-bound}, i.e.,
\begin{IEEEeqnarray}{rCl}
\prob[W\neq \hat{W}] \leq \prob[X\neq \hat{X}] \leq \min\mathopen{}\Big\{ \error_{\mathrm{RCU}}(ML) , \error_{\mathrm{DT}}\mathopen{}\Big(\frac{ML-1}{2}\Big)\Big\}.
\end{IEEEeqnarray}
Furthermore, we have 
\begin{IEEEeqnarray}{rCl}
S(W|Z) = S(g(X)|Z).
\label{eq:coincide-distribution}
\end{IEEEeqnarray}
From~\eqref{eq:privacy-amplification-lemma-application} and~\eqref{eq:coincide-distribution}, we conclude that the resulting code also satisfies the secrecy condition~\eqref{eq:thm-secrecy-bound}. Hence, it is  an $(M,\error,\delta)_{\avg}$ secrecy code. 
\end{IEEEproof}

\subsection{Achievability Bounds: Maximum Secrecy and Maximum Probability of Error}

By a standard  expurgation argument applied to the RCU bound~\cite[Th.~16]{polyanskiy10-05}, we observe that for the legitimate channel $P_{Y|X}$, there exists a code with $M L$ codewords whose maximum error probability is upper-bounded by (see~\cite[Eq.~(220)]{polyanskiy10-05})
\begin{IEEEeqnarray}{rCl}
\error_{\mathrm{RCU, max}}(ML) \define \inf_{\tau\in(0,1)} \tau^{-1} \frac{ \error_{\mathrm{RCU}}((1-\tau)ML)}{\tau} .
\label{eq:RCU-max-bound}
\end{IEEEeqnarray}
Combining Lemma~\ref{prop:strong-pa} and Lemma~\ref{lemma:channel-resolvability} with the achievability bounds in~\cite[Th.~21]{polyanskiy10-05} and~\eqref{eq:RCU-max-bound} for maximum error probability constrained channel codes, we obtain the following result.
\begin{thm}
\label{thm:ach-wiretap-max}
Let $P_X$ be a probability distribution supported on $\setA \subset \setX$. For every $L\in \natunum$,  every $\gamma>0$,  and every $Q_Z$,  there exists an $(M,\error,\secrecy)_{\max}$  secrecy code for the wiretap channel $(\setX, P_{YZ \given X},\setY\times\setZ)$ that satisfies
\begin{IEEEeqnarray}{rCl}
%\IEEEeqnarraymulticol{3}{l}{
\delta
%}\notag\\ \quad
  &\leq&  %\inf\limits_{\gamma>0 } \lefto\{ 
  \sup_{x\in \setA}  E_{\gamma}(P_{Z|X=x},Q_Z) + \sqrt{\frac{\log (M+1)}{2L}} \notag\\
  && +\, \sup_{x\in \setA} \frac{1}{2}\sqrt{\frac{\gamma}{L} \Ex{P_{Z|X=x}}{ \exp\lefto( - \big|\imath(x;Z)  - \log\gamma \big| \right)}  } 
 % \right\}
 \IEEEeqnarraynumspace
\label{eq:thm-secrecy-bound-max}
\end{IEEEeqnarray}
and that
\begin{IEEEeqnarray}{rCl}
\error &\leq& \min\{ \error_{\mathrm{DT,max}}\mathopen{}\big(ML-1\big) , \error_{\mathrm{RCU,max}}(ML)\}
\label{eq:min-rcu-beta-bound-max}.
\end{IEEEeqnarray}
Here, $\imath(\cdot;\cdot)$ is defined in~\eqref{eq:info-den-qz}, and
\begin{IEEEeqnarray}{rCl}
\epsilon_{\mathrm{DT,max}} (a) \define \inf_{\gamma \geq 0} \Big\{\prob[ i(X;Y) \leq \log \gamma] + a\sup_{x\in \setA} \prob[ i(x;Y) \geq \log \gamma] \Big\}
\label{eq:DT-max}
\end{IEEEeqnarray}
with $i(x;y)\define \log \frac{\der P_{Y|X}(y|x)}{\der P_Y(y)}$ and  $P_{XY}(x,y) = P_{X}(x)P_{Y|X}(y|x)$.

\end{thm}

In the proof of Theorem~\ref{thm:ach-wiretap-max}, we apply the privacy amplification lemma (Lemma~\ref{prop:strong-pa}) to a maximum error probability constrained code $\setC$ for the legitimate channel. Since each codeword of $\setC$ has a maximum error probability upper-bounded by $\error$ over $P_{Y|X}$, the maximum error probability of the resulting secrecy code is also upper-bounded by $\error$.
The remaining part of the proof is analogous to the proof of Theorem~\ref{thm:ach-wiretap} and is omitted.

An alternative bound   can be obtained by using the privacy amplification technique on an average error probability constrained code for the legitimate channel, and by using the  concentration of measure  technique employed in the proof of Lemma~\ref{prop:strong-pa}.  

\begin{thm}
\label{thm:ach-wiretap-max2}
Let $P_X$ be a probability distribution supported on $\setA \subset \setX$. For every $L\in \natunum$,  every $\gamma>0$,  and every $Q_Z$,  there exists an $(M,\error,\secrecy)_{\max}$  code for the wiretap channel $(\setX, P_{YZ \given X},\setY\times\setZ)$ that satisfies
\begin{IEEEeqnarray}{rCl}
%\IEEEeqnarraymulticol{3}{l}{
\delta
%}\notag\\ \quad
  &\leq&  %\inf\limits_{\gamma>0 } \lefto\{ 
  \sup_{x\in \setA}  E_{\gamma}(P_{Z|X=x},Q_Z) + \sqrt{\frac{\log (2M+1)}{2L}} \notag\\
  && +\, \sup_{x\in \setA} \frac{1}{2}\sqrt{\frac{\gamma}{L} \Ex{P_{Z|X=x}}{ \exp\lefto( - \big|\imath(x;Z)  - \log\gamma \big| \right)}  } 
 % \right\}
 \IEEEeqnarraynumspace
\label{eq:thm-secrecy-bound-max2}
\end{IEEEeqnarray}
and that
\begin{IEEEeqnarray}{rCl}
\error &\leq&\min\left\{\error_{\mathrm{RCU}}(ML) , \error_{\mathrm{DT}}\mathopen{}\Big(\frac{ML-1}{2}\Big) \right\} +  \sqrt{\frac{\log (2M+1)}{2L}} 
\label{eq:min-rcu-beta-bound-max2}.
\end{IEEEeqnarray}
Here, $\imath(\cdot;\cdot)$, $\error_{\mathrm{RCU}}(\cdot)$, and $\error_{\mathrm{DT}}$ are defined in~\eqref{eq:info-den-qz},~\eqref{eq:def-RCU}, and~\eqref{eq:def-DT-avg-1}, respectively.
\end{thm} 
 
 \begin{IEEEproof}
As shown in the proof of Theorem~\ref{thm:ach-wiretap}, there exists a code $\setC=\{x_1,\ldots,x_{LM}\}$ whose average error probability $\error_0$ on the channel $P_{Y|X}$ satisfies~\eqref{eq:min-rcu-beta-bound}. 
Next, we construct a secrecy code for the wiretap channel $P_{YZ|X}$ from this code $\setC$ using privacy amplification. 

Let $X\sim P_X=  Q_{\setC}^{\unif}$, and let $P_Z=P_{Z|X} \circ P_X$. 
Let $\setG$ be the set of all functions $g :\setC\to \setM$ that satisfy $|g^{-1}(m)| = L$ for every $ m\in \setM$, and let $G$ be uniformly distributed over $\setG$. 
Furthermore, let $P_e(x)$, $x\in\setC$, be the error probability of the codeword $x$ over the legitimate channel $P_{Y|X}$.
By definition, we have 
\begin{IEEEeqnarray}{rCl}
\Ex{X\sim Q_{\setC}^{\unif}}{P_e(X)} = \error_0.
\end{IEEEeqnarray}
The key step of the proof is to establish the following bound:
\begin{IEEEeqnarray}{rCl}
&&\prob\mathopen{}\Big[  \max_{m\in\setM}  \tvar(P_{Z|G^{-1}(m) }, P_{Z}) \geq r + \mu\, %\notag\\
%&&\qquad
 \cup \, \max_{m\in\setM} \Ex{ X\sim Q_{G^{-1}(m)}^{\unif} }{P_e(X)}  \geq r + \error_0\Big] \leq 2 M e^{-2L r^2}  \IEEEeqnarraynumspace
\label{eq:strong-pa-error}
\end{IEEEeqnarray}
where $\mu$ is defined in~\eqref{eq:def-mu-etvar-pza}, and $r$ is an arbitrary positive constant.
Setting $r = \sqrt{\frac{\log (2M+1)}{2L}}$ in~\eqref{eq:strong-pa-error}, we see that the RHS of~\eqref{eq:strong-pa-error} is strictly less than $1$. 
This in turn implies that there exists a hash function $g: \setC \to \setM$ such that 
\begin{IEEEeqnarray}{rCl}
\max_{m\in\setM} \tvar(P_{Z|g^{-1}(m)} ,P_Z ) \leq  \sqrt{\frac{\log (2M+1)}{2L}} + \mu
\label{eq:bound-max-error-mmmm}
\end{IEEEeqnarray}
and 
\begin{IEEEeqnarray}{rCl}
\max_{m\in\setM} \Ex{ X\sim Q_{g^{-1}(m)}^{\unif} }{P_e(X)}  &\leq& \sqrt{\frac{\log (2M+1)}{2L}} + \error_0 \\
 &\leq& \sqrt{\frac{\log (2M+1)}{2L}} + \min\left\{\error_{\mathrm{RCU}}(ML) , \error_{\mathrm{DT}}\mathopen{}\Big(\frac{ML-1}{2}\Big) \right\} .
 \label{eq:bound-max-secrecy-hahaha}
\end{IEEEeqnarray}
Furthermore, the RHS of~\eqref{eq:bound-max-error-mmmm} is upper-bounded by the expression on the RHS of~\eqref{eq:thm-secrecy-bound-max2}  according to Lemma~\ref{lemma:channel-resolvability}.

Consider now a secrecy code with message $W=g(X)$ and with a random encoder $P_{X|W=m} = Q_{g^{-1}(m)}^{\unif}$.
%
%
%
%In words, for each message $m\in\setM$, the encoder $P_{W|X}$  picks a codeword  from the set $g^{-1}(m)$ uniformly at random. 
%
%
The decoder first decodes the codeword~$\hat{X}$  and then set $\hat{W} = g(\hat{X})$. 
By construction, we have 
\begin{IEEEeqnarray}{rCl}
 d(P_{Z|W =m}, P_Z) =  \tvar(P_{Z|g^{-1}(m)} ,P_Z ), \quad\forall m\in\setM.
 \end{IEEEeqnarray}
Furthermore, the error probability of a message $m\in\setM$ is equal to 
\begin{IEEEeqnarray}{rCl}
P_e(m) = \Ex{ X\sim P_{X|W=m}}{P_e(X)} =\Ex{ X\sim Q_{g^{-1}(m)}^{\unif} }{P_e(X)}.
\end{IEEEeqnarray}
Therefore, the maximum error probability and maximum information leakage of the resulting secrecy code satisfy the bounds specified in~\eqref{eq:min-rcu-beta-bound-max2} and~\eqref{eq:thm-secrecy-bound-max2}.

To conclude the proof, it remains to show~\eqref{eq:strong-pa-error}.
By the union bound, we have 
\begin{IEEEeqnarray}{rCl}
&&\prob\mathopen{}\Big[  \max_{m\in\setM}  \tvar(P_{Z|G^{-1}(m) }, P_{Z}) \geq r + \mu\, %\notag\\
%&&\qquad
 \cup \, \max_{m\in\setM} \Ex{ X\sim Q_{G^{-1}(m)}^{\unif} }{P_e(X)}  \geq r + \error_0\Big] \notag\\
 &&\quad  \leq \prob\mathopen{}\Big[  \max_{m\in\setM}  \tvar(P_{Z|G^{-1}(m) }, P_{Z}) \geq r + \mu\Big]  + M \prob[  \Ex{ X\sim Q_{G^{-1}(1)}^{\unif} }{P_e(X)}  \geq r + \error_0 ]  \\
 && \quad  \leq M e^{-2L r^2}  + M  \prob[  \Ex{ X\sim Q_{G^{-1}(1)}^{\unif} }{P_e(X)}  \geq r + \error_0 ]  .
\label{eq:strong-pa-error-proof1}
\end{IEEEeqnarray}
Here,~\eqref{eq:strong-pa-error-proof1} follows from Lemma~\ref{prop:strong-pa}. 
Note that 
\begin{IEEEeqnarray}{rCl}
  \Ex{ X\sim Q_{G^{-1}(1)}^{\unif} }{P_e(X)} \stackrel{\mathrm{d}}{=}\frac{1}{L} \sum\limits_{i=1}^{L} P_e(X_i) 
\end{IEEEeqnarray}
where $X_1,...,X_L$ are  random samples \emph{without replacement} from $\setC$, and $ \stackrel{\mathrm{d}}{=}$ denotes equality in distribution. 
Since $0\leq P_e(X_i) \leq 1$ almost surely, and since $\Ex{}{P_e(X_i)} = \error_0$, 
we can invoke Hoeffding's inequality for samples without replacement~\cite{hoeffding1963-03a} and obtain 
\begin{IEEEeqnarray}{rCl}
 \prob\lefto[  \Ex{ X\sim Q_{G^{-1}(1)}^{\unif} }{P_e(X)}  \geq r + \error_0 \right]  \leq e^{-2L r^2} .
 \label{eq:hoeffding-without-replace-ment}
\end{IEEEeqnarray}
Substituting~\eqref{eq:hoeffding-without-replace-ment} into~\eqref{eq:strong-pa-error-proof1} we conclude~\eqref{eq:strong-pa-error}.
 \end{IEEEproof}

\subsection{Achievability: Trading Reliability for Secrecy}
\label{sec:trade-reli}
Our next result shows that one can trade reliability for secrecy for a wiretap channel.  
\begin{thm}
\label{thm:ach-relation}
Consider a wiretap channel $ P_{YZ|X}$. Every $(M,\error_0,\delta_0)_{\max}$ (resp. $(M,\epsilon_0,\delta_0)_{\avg}$) secrecy code for $P_{YZ|X}$ can be converted to an $(M,\error, \delta)_{\max}$ (resp. $(M,\error, \delta)_{\avg}$) secrecy code with 
$\error> \error_0$ and $\delta=\delta_0(1-\error)/(1-\error_0)$.
Consequently, 
\begin{IEEEeqnarray}{rCl}
R^*_{\max}(n,\error_0,\delta_0) &\leq & R^*_{\max}\lefto(n,\error,\frac{\delta_0 (1-\error)}{1-\error_0}\right) 
\label{eq:existence-code}\\
R^*_{\avg}(n,\error_0,\delta_0) &\leq& R^*_{\avg}\lefto(n,\error,\frac{\delta_0(1-\error)}{1-\error_0}\right) .
\label{eq:avg-ach-relation}
\end{IEEEeqnarray}
\end{thm}
 \begin{IEEEproof}
 Choose an arbitrary $(M,\error_0,\delta_0)_{\max}$ secrecy code~$\setC$.
Let $P_{X|W }^{\setC}$ be the stochastic encoder of this code, and let $P_{X}^{\setC}$ be the empirical output distribution of the encoder for a uniformly distributed message. 
We construct a new secrecy code $\widetilde{\setC}$ from~$\setC$ by using the following encoder
\begin{IEEEeqnarray}{rCl}
P_{X|W}^{\widetilde{\setC}} = (1- p) P_{X|W}^{\setC} + p P_{X}^{\setC}
\label{eq:code-tilde-encoder-def}
\end{IEEEeqnarray}
where 
\begin{IEEEeqnarray}{rCl}
p=\frac{\error-\error_0}{1-\error_0}.
\end{IEEEeqnarray}
The decoder of $\widetilde{\setC}$ is chosen to be the same  as that of~$\setC$.
We shall prove that the maximum error probability of the new code $\widetilde{\setC}$  is upper-bounded by $\error$, and its maximum information leakage is upper-bounded by $\delta$.
Consequently, $\widetilde{\setC}$ is an $(M,\error,\delta)_{\max}$ code, from which the claim of the theorem follows.

Let  $\setD_m$ be the decoding region corresponding to the message $m$.  
The error probability $P_{e,m}(\widetilde{\setC})$ of $\widetilde{\setC}$ for the message $m$ can be evaluated as follows:
\begin{IEEEeqnarray}{rCl}
P_{e,m}( \widetilde{\setC} ) &=& 1-  \big(P_{Y^n|X^n} \circ P_{X^n|W=m}^{\widetilde{\setC}}\big) [\setD_m] \\
&=&  1- (1-p)\error_0  -  \underbrace{ p \big( P_{Y^n|X^n} \circ P_{X^n}^{\setC} \big)[\setD_m]}_{\geq 0} \label{eq:error-codet-code1}\\
&\leq& \error\label{eq:error-codet-code2}
\end{IEEEeqnarray}
where~\eqref{eq:error-codet-code1} follows from~\eqref{eq:code-tilde-encoder-def} and because $\setC$ has error probability $\error_0$.

Let $P_{Z}^{\setC}$ and $P_{Z}^{\widetilde{\setC}}$ be the empirical output distributions of $\setC$ and $\widetilde{\setC}$ over the channel $P_{Z|X}$, respectively. 
For every message $m\in \setM$, we have 
\begin{IEEEeqnarray}{rCl}
\IEEEeqnarraymulticol{3}{l}{
\tvar\lefto(P_{Z|W=m }^{\widetilde{\setC}} ,  P_{Z}^{\widetilde{\setC}}\right) }\notag\\
\quad  &=& \tvar\lefto(P_{Z|W=m}^{\widetilde{\setC}} , P_{Z}^{ {\setC}}\right)\label{eq:TV-codet-code1} \\
&=& \tvar\lefto(   P_{Z|X} \circ \big((1-p) P_{X|W=m}^{\setC} + p P_{X}^{\setC} \big) ,  P_{Z}^{{\setC}} \right)\label{eq:TV-codet-code2}  \IEEEeqnarraynumspace \\
&=& (1-p) d(P_{Z|W=m}^{{\setC}} , P_{Z}^{\setC}) \\
&\leq & \frac{\delta_0(1-\error)}{1-\error_0} = \delta.\label{eq:TV-codet-code3}
\end{IEEEeqnarray}
Here,~\eqref{eq:TV-codet-code1} follows because $P_{Z}^{\setC} = P_{Z}^{\widetilde{\setC}}$,~\eqref{eq:TV-codet-code2}  follows from~\eqref{eq:code-tilde-encoder-def}, and~\eqref{eq:TV-codet-code3} follows because, by definition,  $d(P_{Z|W=m}^{{\setC}} ,  P_{Z}^{\setC}) \leq \delta_0$ for every $m\in \setM$.
The result for average probability of error and average secrecy constraint can be proved similarly. 
\end{IEEEproof}

\subsection{Converse Bounds}

We first present a general converse bound.
\begin{thm}
Every $(M,\error,\secrecy)_{\avg}$ secrecy code satisfies 
\begin{equation}
M \leq \sup\limits_{P_{X|W}} 
\inf_{0<\tau<1-\secrecy} \inf_{Q_Y}\frac{\beta_{\secrecy+\tau} (P_{WZ},Q_{\setM}^{\unif}P_Z )}{\tau \beta_{1-\error}(P_{WY} , Q_{\setM}^{\unif} Q_Y)} \IEEEeqnarraynumspace
\label{eq:thm-general-converse}
\end{equation}
where $P_W$ denotes the distribution of the message (not necessarily uniformly distributed), and $P_Y$ and $P_Z$ are the marginal distributions of the Markov chains $W\to X \to Y$ and $W \to X\to Z$, respectively.
\end{thm}
\begin{IEEEproof}
By the meta-converse bound for non-equiprobable messages~\cite{vazquezvilar2016-05a},  
every $(M,\error,\secrecy)_{\avg}$ secrecy code satisfies 
\begin{equation}
M \leq  \inf\limits_{Q_{Y}} \frac{1}{\beta_{1-\error}(P_{\msg Y} , Q_{\setX}^{\unif} Q_Y) }.
\label{eq:meta-converse-bound}
\end{equation}
Furthermore, by the secrecy constraint,  
\begin{IEEEeqnarray}{rCl}
\secrecy \geq d(P_{\msg Z}, Q_{\setM}^{\unif}P_Z) \geq \secrecy +  \tau - \beta_{\secrecy +\tau} (P_{\msg Z} , Q_{\setM}^{\unif} P_Z ) \IEEEeqnarraynumspace
\label{eq: secrecy-constraint-alpha-beta}
\end{IEEEeqnarray}
where the last step follows from~\eqref{eq:def-total-var}. 
Rearranging the terms in~\eqref{eq: secrecy-constraint-alpha-beta}, we conclude that 
\begin{equation}
\beta_{\secrecy +\tau } (P_{\msg Z}, Q_{\setM}^{\unif} P_Z ) \geq \tau.
\label{eq:beta-tau-geq-tau}
\end{equation}
Combining~\eqref{eq:beta-tau-geq-tau}  with~\eqref{eq:meta-converse-bound}, and optimizing the resulting bound over all stochastic encoders $P_{X|W}$,  we obtain~\eqref{eq:thm-general-converse}.
\end{IEEEproof}

The bound~\eqref{eq:thm-general-converse} is in general difficult to compute or analyze. 
Next, we prove a converse bound, which is motivated by the converse  in~\cite{tyagi2015-05a} for secret key generation. This bound is both numerically and analytically  tractable. 
\begin{thm}
\label{thm:converse-wei}
Let $Q_{Y|Z}: \setY\to\setZ$ be an arbitrary random transformation. Then, every $(M,\error,\secrecy)_{\avg}$ secrecy code for the wiretap channel  $(\setX, P_{YZ|X}, \setY\times\setZ)$ satisfies 
\begin{IEEEeqnarray}{rCl}
M\leq \inf_{\tau \in(0, 1-\error -\secrecy) } \frac{\tau +\secrecy}{\tau \beta_{1-\error-\secrecy -\tau} (P_{XYZ}, P_{XZ}Q_{Y|Z}) } \IEEEeqnarraynumspace
\label{eq:converse-bound-general}
\end{IEEEeqnarray} 
where $P_{XYZ}$ denotes the distribution induced by the code. 
\end{thm}

\begin{rem}
Using the result in~\cite{tyagi2015-05a}, Hayashi, Tyagi, and Watanabe recently derived the following converse bound~\cite{hayashi2014-09a} (for the case of uniformly distributed message): 
\begin{equation}
M \leq \inf_{\tau \in(0, 1-\error-\secrecy)} \frac{1}{\tau^2  \beta_{1-\error-\secrecy-\tau} (P_{XYZ}, P_{XZ}Q_{Y|Z}) } .
\label{eq:hayashi-bound-wtc}
\end{equation}
Our bound is stronger than~\eqref{eq:hayashi-bound-wtc} since $(\tau+\secrecy)/\tau < 1/\tau^2$.  %
%
%
%The improvement of our bound over~\eqref{eq:hayashi-bound-wtc} comes from the construction of the test, which is tailored to the wiretap channel. 
\end{rem}

\begin{IEEEproof}
\begin{figure}
%\begin{center}
%\includegraphics[scale=0.85]{./figure/markov.pdf}
%\end{center}
\centering
\begin{tikzpicture}

\draw (0, 0) node{\small  $P: W \rightarrow X\to (Y,\,Z) $};
\draw (1.6,-0.65) node{\small  $\hat{W}$};
\draw (1.2, -0.4) node {\small $\searrow$};
\draw(0.8,-1.5) node{\small $Q_{Y|Z}$};
	\draw(0, -2) node{\small $Q: W\to X \to Z\longrightarrow Y\to \hat{W}$};
%\caption{Markov}	
\end{tikzpicture}

\caption{A graphical illustration of the probability distributions $P_{WXYZ\hat{W}}$ and $Q_{WXYZ\hat{W}} $. \label{fig:markov}}
\end{figure}
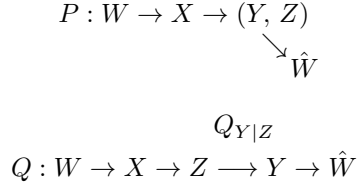
Fix an arbitrary $(M,\error,\secrecy)_{\avg}$ secrecy code  and let $P_{WXYZ\hat{W}} \define P_W P_{X|W} P_{YZ|X} P_{\hat{W} |Y}$ be the joint distribution induced by the code. And let $Q_{WXYZ\hat{W}} \define P_W P_{X|W}P_{Z|X} Q_{Y|Z} P_{\hat{W}|Y } $ (see Fig.~\ref{fig:markov} for a graphical illustration). 
As in the  meta-converse bound~\cite[Th.~26]{polyanskiy10-05} and in~\cite{tyagi2015-05a}, the idea is to relate the error probability and secrecy index of the secrecy code to the binary hypothesis testing between  $P_{W\hat{W}Z}$ and $Q_{W\hat{W}Z}$ through a suboptimal test. 
The following test $P_{T|W  \hat{W}Z}  : \setM^2 \times \setZ \mapsto\{0,1\}$ plays an prominent role in our proof:
\begin{IEEEeqnarray}{rCl}
T(m,\hat{m},z)=\indfun{ m = \hat{m} , P_{W|Z}(m|z) \leq 1/(\eta M) } \IEEEeqnarraynumspace
\end{IEEEeqnarray}
where $\eta\in(0,1)$.
One the one hand, we have
\begin{IEEEeqnarray}{rCl}
\IEEEeqnarraymulticol{3}{l}{
Q_{W\hat{W} Z} [T=1] } \notag\\
\quad &=& \sum_{m,\hat{m},z} Q_{Z\hat{W}}(z,\hat{m})P_{W|Z}(m|z) \notag\\
&& \quad \quad \cdot\, \indfun{m = \hat{m}, P_{W|Z}(m|z)\leq 1/(M\eta) } \\
&\leq & \frac{1}{ M \eta}\sum\limits_{m,z} Q_{Z\hat{W}}(z,m)  \indfun{ P_{W|Z}(m|z)\leq 1/(M\eta) } \IEEEeqnarraynumspace\\
&\leq & \frac{1}{M\eta}. \label{eq:ub-q-t-1}
\end{IEEEeqnarray}
One the other hand, by~\eqref{eq:error-constraint-wtc}, the probability $P_{W\hat{W}Z}[T=1]$ can be lower-bounded as
\begin{IEEEeqnarray}{rCl}
P_{W\hat{W}Z} [T=1] &\geq& 1-\error - P_{WZ}[P_{W|Z}(W|Z)\geq 1/(M\eta)] .\IEEEeqnarraynumspace
\label{eq:upper-bound-P-ac-1}
\end{IEEEeqnarray}
To further lower-bound the RHS of~\eqref{eq:upper-bound-P-ac-1}, we observe that
\begin{IEEEeqnarray}{rCl}
\delta &\geq& d(P_{WZ},Q_{\setM}^{\unif}P_Z)\\
 &\geq&  P_{WZ}[P_{W|Z}(W|Z)\geq 1/(M\eta)] \notag\\
&&- \, Q_{\setM}^{\unif}P_{Z}[P_{W|Z}(W|Z)\geq 1/(M\eta)] \IEEEeqnarraynumspace
\label{eq:bound-pwz-id-1}
\end{IEEEeqnarray}
and that
\begin{IEEEeqnarray}{rCl}
\IEEEeqnarraymulticol{3}{l}{
P_{WZ}[P_{W|Z}(W|Z)\geq 1/(M\eta)] }\notag\\
\quad & \geq & \frac{1}{\eta} Q_{\setM}^{\unif} P_{Z}[P_{W|Z}(W|Z)\geq 1/(M\eta)].
\label{eq:bound-pwz-id-2}
\end{IEEEeqnarray}
Here,~\eqref{eq:bound-pwz-id-2} follows from a standard change of measure argument.
Combining~\eqref{eq:bound-pwz-id-1} and~\eqref{eq:bound-pwz-id-2}, we obtain
\begin{IEEEeqnarray}{rCl}
P_{WZ}[P_{W|Z}(W|Z)\geq 1/(M\eta)] \leq \frac{\delta}{1-\eta}.
\label{eq:bound-pwz-id-3}
\end{IEEEeqnarray}
Substituting~\eqref{eq:bound-pwz-id-3} into~\eqref{eq:upper-bound-P-ac-1}, and using~\eqref{eq:ub-q-t-1}, we conclude that 
\begin{IEEEeqnarray}{rCl}
\beta_{1-\error-\secrecy/(1-\eta)} ( P_{W\hat{W}Z} , Q_{W\hat{W}Z}) \leq \frac{1}{M\eta}.
\label{eq:ub-beta-mmz}
\end{IEEEeqnarray}
The final bound~\eqref{eq:converse-bound-general} follows by rearranging the terms in~\eqref{eq:ub-beta-mmz}, by the change of variable $\tau =\delta/(1-\eta) -\delta$,  and by    observing that
\begin{equation}
\beta_{\alpha} ( P_{W\hat{W}Z} , Q_{W\hat{W}Z}) \geq \beta_{\alpha} ( P_{XYZ} , P_{XZ}Q_{Y|Z})
\end{equation}
which follows from the data-processing inequality for $\beta_\alpha$.
\end{IEEEproof}

\section{Bounds on the Second-Order Secrecy Rate}

\subsection{DM-WTC}

We shall use the following notation:
\begin{IEEEeqnarray}{rCl}
I(P_X,P_{Y|X}) &\define& I(X;Y)   \\
%V(P_X, P_{Y|X}) \define  \! \sum\limits_{x\in\setX}  \! P_X(x)\bigg(\! \sum\limits_{y\in \setY} P_{Y|X}(y|x)   \log^2\! \frac{P_{Y|X}(y|x)}{P_Y (y)}  \notag\\
% \qquad\qquad\qquad\quad - \,D(P_{Y|X=x}\| P_Y)^2 \bigg)\\
V(P_X, P_{Y|X}) &\define&  \! \sum\limits_{x\in\setX}  \! P_X(x)\bigg(\! \sum\limits_{y\in \setY} P_{Y|X}(y|x)   \log^2\! \frac{P_{Y|X}(y|x)}{P_Y (y)}  - \,D(P_{Y|X=x}\| P_Y)^2 \bigg) \IEEEeqnarraynumspace\\
\tilde{I}(P_X, P_{YZ|X}) &\define&  I(X;Y|Z) \\
%\end{IEEEeqnarray}
%and
%\begin{IEEEeqnarray}{rCl}
%\IEEEeqnarraymulticol{3}{l}{
\tilde{V}(P_X,P_{YZ|X})
%}\notag\\
  &\define&   \sum\limits_{x\in\setX}   P_X(x)\bigg(\! \sum\limits_{y,z }P_{ZY|X}(y,z|x)  \log^2\! \frac{P_{YZ|X}(y,z|x)}{P_{Z|X}(z|x)P_{Y|Z} (y|z)} \notag\\
 &&  \qquad\qquad\quad\quad  -\, D(P_{YZ|X=x}\| P_{Y|Z}P_{Z|X=x})^2\! \bigg).\IEEEeqnarraynumspace
\end{IEEEeqnarray}
The secrecy capacity of a  general DM-WTC is given by~\cite{csiszar1978-05a}
\begin{IEEEeqnarray}{rCl}
\CS = \max_{P_{VX}} \mathopen{}\Big( I(V;Y) - I(V;Z)\Big) 
\label{eq:secrecy-capacity-dmc}
\end{IEEEeqnarray}
where the maximization is over all probability distributions $P_{VX}$  for which $V\to X\to YZ $ form a Markov chain.  For simplicity, we shall assume that there exists a unique probability distribution $P_{VX}^* =P_{V}^* P_{X|V}^*$   that achieves the maximum in~\eqref{eq:secrecy-capacity-dmc}. Note that if the eavesdropper's channel $P_{Z|X}$  is less capable than the legitimate channel $P_{Y|X}$, then the secrecy capacity reduces to~\cite[Sec.~3.5.1]{bloch11-b}
\begin{IEEEeqnarray}{rCl}
\CS = \max_{P_{X}} \mathopen{}\Big( I(X;Y) - I(X;Z)\Big) .
\label{eq:secrecy-capacity-intro}
\end{IEEEeqnarray}

 The auxiliary random variable $V$ makes the evaluation of~\eqref{eq:secrecy-capacity-dmc} difficult. An upper bound on~\eqref{eq:secrecy-capacity-dmc} is given by~\cite{hayashi2014-09a}
\begin{equation}
\CS \leq \CS^{\mathrm{u}} \define  \max_{P_X} I(X;Y|Z)  .
\label{eq:ub-dmc-secrecy-capacity}
\end{equation}
 For simplicity, we shall also assume that there exists a unique probability distribution  $\tilde{P}_X^*$   that attains the maximum in~\eqref{eq:ub-dmc-secrecy-capacity}, and that $\tilde{V}(\tilde{P}_X^*, P_{YZ|X})>0$. 
Note that, the bound~\eqref{eq:ub-dmc-secrecy-capacity} is tight (i.e., $\CS = \CS^{\mathrm{u}}$) if the wiretap channel  is physically degraded~\cite[Def.~3.8]{bloch11-b}.

\begin{thm}
\label{thm:dm-wtc-asy}
Consider a DM-WTC $P_{YZ|X}$. If $\error +\delta <1$, then we have 
\begin{equation}
R^*_{\max}(n,\error,\secrecy) \geq \CS - \!\sqrt{\frac{V_1}{n}}Q^{-1}(\error) - \!\sqrt{\frac{V_2}{n}}Q^{-1}(\secrecy) + \bigO\lefto(\frac{\log n}{n}\right)
\label{eq:thm-ach-expansion}
\end{equation}
and 
\begin{equation}
R^*_{\avg} (n,\error,\secrecy)\leq \CS^{\mathrm{u}}- \sqrt{\frac{V_c}{n}}Q^{-1}(\error+\secrecy) +\bigO\lefto(\frac{\log n}{n}\right).
\label{eq:eq:thm-conv-expansion}
\end{equation}
Here, $Q^{-1}(\cdot)$ is the inverse of the Gaussian $Q$-function $Q(x) \define \int\nolimits_{x}^{\infty} \frac{1}{\sqrt{2\pi}} e^{-t^2/2} dt$ and
\begin{IEEEeqnarray}{rCl}
V_1 &\define& V(P_V^*, P_{Y|X}\circ P_{X|V}^*)\\
V_2 &\define& V(P_V^*, P_{Z|X}\circ P_{X|V}^*)\\
V_c &\define& \tilde{V}(\tilde{P}_X^*, P_{YZ|X}).
\end{IEEEeqnarray}
If $\error+\delta >1$, then 
\begin{IEEEeqnarray}{rCl}
R^*_{\max}(n,\error,\secrecy)  \geq C_{\mathrm{legit}} -\sqrt{\frac{V_{\mathrm{legit}}}{n}}Q^{-1}\lefto(\frac{\error+\delta -1}{\delta}\right) + \bigO\lefto(\frac{\log n }{n}\right)
\label{eq:strong-converse-does-not}
\end{IEEEeqnarray}
where $C_{\mathrm{legit}} $ and $V_{\mathrm{legit}}$ denote the capacity and channel dispersion of the legitimate channel.
\end{thm}

\begin{IEEEproof}
%Let $P_{U,X}$ be the joint distribution that achieves the secrecy capacity. 
See Appendix~\ref{app:proof-asy-dmc}.
\end{IEEEproof}

A few remarks are in order.
\begin{itemize}
\item   As implied by~\eqref{eq:strong-converse-does-not}, the strong converse does not hold if $\delta+\error > 1$.  

\item It is possible to slightly improve the achievability bound~\eqref{eq:thm-ach-expansion} by using Theorem~\ref{thm:ach-relation}; see Section~\ref{sec:2nd-sec-semideter}.

\item By Theorem~\ref{thm:equivalance-btw-secrecy}, every  achievability bound on $R^*_{\max}(n,\error,\secrecy)$ can be converted to achievability bounds on semantic security and distinguishing security. In particular, the bound~\eqref{eq:thm-ach-expansion} implies the semantic security capacity of wiretap channels recently studied in~\cite{hayashi2016-05b} and~\cite{goldfeld2016-07a}. 
\end{itemize}

The achievability bound~\eqref{eq:thm-ach-expansion} is tighter than the achievable second-order coding rate in~\cite{yassaee2013-07a} obtained by using output statistics of random binning, and is tighter than the one in~\cite{tan2012-11a} obtained via channel resolvability. 
The latter two approaches use a random coding argument  and bound the average error probability and average information leakage averaged over all random codebooks separately. They then invoke  Markov's inequality to show the existence of a code that satisfies \emph{simultaneously} the reliability and secrecy constraint. 
The use of Markov's inequality introduces a penalty to the second-order coding rate, which corresponds to the gap between~\eqref{eq:thm-ach-expansion} and~\cite[Eq.~(23)]{yassaee2013-07a}.
In contrast,  our result shows that \emph{every} code that satisfies the reliability constraint can be modified to satisfy the secrecy constraint, thereby avoiding the use of Markov's inequality. 
However, this does not mean that  channel resolvability based approaches have worse asymptotic performance than the privacy amplification based approach adopted in this paper. Indeed, the  same bound~\eqref{eq:thm-ach-expansion} can be proved by using the stronger channel resolvability result developed in~\cite{cuff2016-07a}, which shows that the random coding procedure in channel resolvability succeeds with a probability double-exponentially close to unity. 

\subsection{Gaussian wiretap channel}
Consider the Gaussian wiretap channel   
\begin{IEEEeqnarray}{rCl}
Y_i &=& X_i + U_{i},\quad  Z_i = X_i + \tilde{U}_i,\quad i=1,\ldots,n \label{eq:Gaussian-wiretap}
\end{IEEEeqnarray}
where $\{U_{i}\}$ are independent and identically distributed (i.i.d.)  $\mathcal{N}(0,N_1)$ random variables, and $\{\tilde{U}_i\}$ are i.i.d. $\mathcal{N}(0, N_2)$ random variables. 
Without loss of generality, we assume that $N_2 > N_1$ (otherwise the secrecy capacity is zero).
 Furthermore, we assume that each codeword~$x^n$ satisfies the power constraint
\begin{IEEEeqnarray}{rCl}
 \|x^n\|^2 \leq n P.  
 \end{IEEEeqnarray}
 %
%The next theorem studies the maximal secrecy rate $R^*(n,\error,\secrecy)$ for the Gaussian wiretap channel~\eqref{eq:Gaussian-wiretap}. 

\begin{thm}
\label{thm:gaussian-asy}
For the Gaussian wiretap channel~\eqref{eq:Gaussian-wiretap}, we have 
\begin{equation}
R^*_{\max}(n,\error,\secrecy) \geq \CS -\sqrt{\frac{V_1}{n}}Q^{-1}(\error) - \sqrt{\frac{V_2}{n}}Q^{-1}(\secrecy) + \bigO\lefto(\frac{\log n}{n}\right)
\label{eq:thm-ach-expansion-awgn}
\end{equation}
and 
\begin{equation}
R^*_{\avg}(n,\error,\secrecy) \leq  \CS -\sqrt{\frac{V_c}{n}}Q^{-1}(\error+\secrecy) +\bigO\lefto(\frac{\log n}{n}\right)
\label{eq:thm-conv-expansion-awgn}
\end{equation}
where 
\begin{IEEEeqnarray}{rCl}
\CS &=&\frac{1}{2}\log\lefto(1+\frac{P}{N_1}\right) - \frac{1}{2}\log\lefto(1+\frac{P}{N_2}\right)\\
V_i &=& \frac{\log^2 e}{2}\frac{P^2+2PN_i}{(P+N_i)^2}, \quad i\in\{1,2\}\\
V_c&=& V_1+V_2 - \frac{PN_1}{P+N_1} \left(\frac{1}{N_2} + \frac{1}{P+N_2}\right)\log^2 e . \label{eq:def-disp-gau-c} \IEEEeqnarraynumspace
\end{IEEEeqnarray}
\end{thm}
\ifthenelse{\boolean{conf}}{}{
\begin{IEEEproof}
See Appendix~\ref{app:proof-asy-gaussian}.
\end{IEEEproof}
}

\subsection{Numerical Results and Discussions}

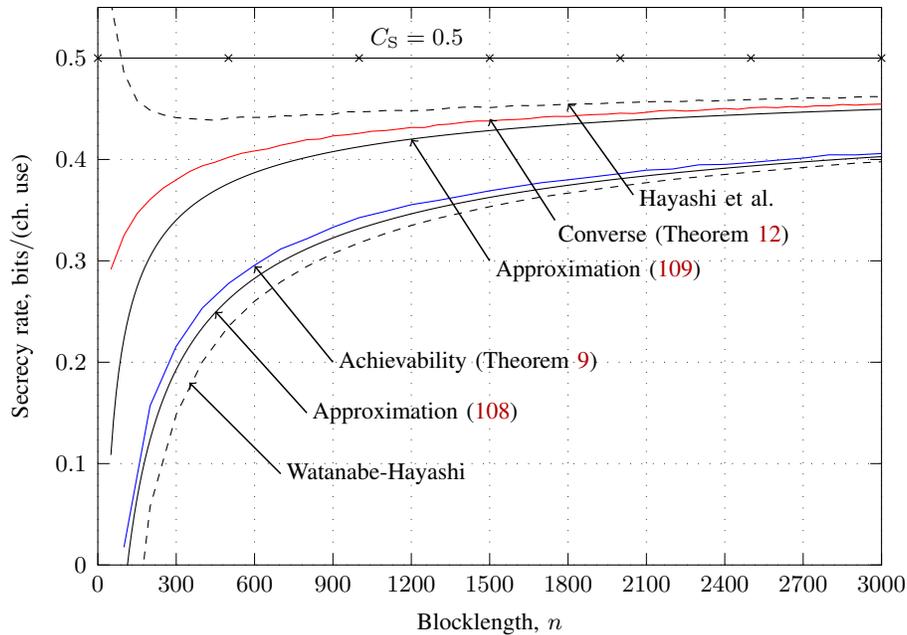
\begin{figure}
\centering

\begin{tikzpicture}
%\tikzstyle{every node}=[font=\small]
		\begin{axis}[
			xlabel={\footnotesize Blocklength, $n$},
			ylabel={\footnotesize Secrecy rate, bits$/$(ch. use)},
xmin=0, xmax=3000,
ymin=0, ymax=0.55,
grid=major,
xtick={0,300,...,3000},
xticklabels={%
$0$,
$300$,
$600$,
$900$,
$1200$,
$1500$,
$1800$,
$2100$,
$2400$,
$2700$,
$3000$},
%minor y tick num=1,
            %ylabel=y,
            ]
		\addplot[black] table {conv_na.dat};
		\addplot[black] table {ach_na.dat};
		%\addplot[black] table {nalt.dat};
\addplot[red] table {conv_wei.dat};
\addplot[black,dashed] table {conv_ha.dat};
\addplot[blue] table {ach_wei.dat};
\addplot[black,dashed ] table {ach_ha.dat};

\addplot[black,mark=x] coordinates {
(0, 0.5)
(500, 0.5)
(1000, 0.5)
(1500, 0.5)
(2000, 0.5)
(2500, 0.5)
(3000,0.5)
};

\draw (1000, 0.52) node[right]{\footnotesize $C_{\mathrm{S}}  = 0.5$};

\draw[line width = 0.2mm, <-] (1500, 0.44) -- (1750,0.34);
\draw (1720, 0.325) node[right]{\footnotesize Converse (Theorem~\ref{thm:converse-wei})};

\draw[line width = 0.2mm, <-] (1200, 0.42) -- (1500,0.3);
\draw (1480, 0.29) node[right]{\footnotesize Approximation~\eqref{eq:thm-conv-expansion-awgn}};

\draw[line width = 0.2mm, <-] (1800, 0.455) -- (2050,0.365);
\draw (2030, 0.36) node[right]{\footnotesize Hayashi et al.};

\draw[line width = 0.2mm, <-] (450,0.25) -- (800,0.15);
\draw (780, 0.15) node[right]{\footnotesize Approximation~\eqref{eq:thm-ach-expansion-awgn}};

\draw[line width = 0.2mm, <-] (350,0.18) -- (700,0.09);
\draw (680, 0.09) node[right]{\footnotesize Watanabe-Hayashi };

\draw[line width = 0.2mm, <-] (600,0.295) -- (900,0.2);
\draw (880, 0.2) node[right]{\footnotesize Achievability (Theorem~\ref{thm:ach-wiretap-max2})};

		\end{axis}
	\end{tikzpicture}

\caption{Secrecy rate for the Gaussian wiretap channel with $P/{N_1} = 3$\, dB, $P/{N_2} = -3$ dB, $\error=\delta=10^{-3}$. \label{fig:gaussian} }
 
\end{figure}

In this section, we compare the bounds proposed in this paper with existing bounds in~\cite{hayashi2014-09a} and \cite{watanabe2013-07a}, and with the approximations provided in Theorem~\ref{thm:gaussian-asy} for a Gaussian wiretap channel (with the $\bigO(\cdot)$ terms omitted). The results are shown in Fig.~\ref{fig:gaussian}. The bound labeled by ``Hayashi et al.'' is~\cite[Th.~6]{hayashi2014-09a} (see also~\eqref{eq:hayashi-bound-wtc}), and the one labeled by ``Watanabe-Hayashi'' is obtained by combining the privacy amplification bound~\cite[Cor.~2]{watanabe2013-07a} with Shannon's channel coding achievability bound~\cite{shannon59} (which is the tightest achievability bound for Gaussian channels~\cite[Sec.~III.J-4]{polyanskiy10-05}).\footnote{The numerical routine used to compute Shannon's channel coding achievability bound is available at https://github.com/yp-mit/spectre}

Several observations are in order. First of all, both our achievability and converse bounds are uniformly better than the ones in~\cite{hayashi2014-09a} and \cite{watanabe2013-07a}. Secondly, the expansions~\eqref{eq:thm-ach-expansion-awgn}  and~\eqref{eq:thm-conv-expansion-awgn} provide reasonable approximations for the  bounds in Theorems~\ref{thm:ach-wiretap-max2} and~\ref{thm:converse-wei}.  
Last but not least, there is a nontrivial gap between our achievability and converse bounds (which can also be inferred from the approximations~\eqref{eq:thm-ach-expansion-awgn}  and~\eqref{eq:thm-conv-expansion-awgn}). 

To understand the causes of the gap between the upper and lower bound in Theorems~\ref{thm:dm-wtc-asy} and~\ref{thm:gaussian-asy}, we note that our achievability schemes come with two limitations. 
\begin{itemize}
\item In our achievability schemes, the secrecy and reliability constraints are treated separately. In particular, the transmitter adds redundancy to the transmitted signals to  guarantee reliability, and uses (separated) randomness  to derive secrecy.
The amount of redundancy depends on the quality of the legitimate channel and the error probability constraint, whereas the amount of randomness depends on the quality of the eavesdropper's channel and the secrecy constraint. 
While separation-based schemes are secrecy-capacity-achieving, it is not clear whether they are second-order optimal.\footnote{A similar phenomenon occurs in the joint source-channel coding setting. Recently, it was shown that separate source-channel coding is first-order but not second-order optimal~\cite{kostina13-05}.} 

\item Assuming a separation-based structure, our achievability schemes further require the legitimate decoder to decode both the message and the randomness inserted by the transmitter. 
\end{itemize}
It would be interesting to construct  schemes that relax these requirements, and to see if such schemes achieve better second-order secrecy rate.

\section{Semi-Deterministic Wiretap Channel}

In this section, we focus on a special class of wiretap channels known as semi-deterministic DM-WTCs. 
More specifically, in a semi-deterministic DM-WTC $P_{YZ|X}:\setX\to\setY\times\setZ$, the  output of the legitimate channel is a deterministic function of the input, i.e., $P_{Y|X}(y|x)= \indfun{y=f(x)}$ for some deterministic function~$f$, and the eavesdropper's channel is a (noisy) DMC. 
Without loss of generality, we assume that the function $f$ is surjective. 
Note that, in general, a semi-deterministic wiretap channel is not (stochastically) degraded.

The secrecy capacity of a semi-deterministic wiretap channel is given by~\cite{grubb2007-07a}
\begin{equation}
\CS = \max_{P_X} H(Y|Z) . 
\label{eq:secrecy-capacity-semid}
\end{equation}
This can be achieved by setting $V=Y$ in the secrecy capacity formula~\eqref{eq:secrecy-capacity-dmc}.
An alternative expression for the secrecy capacity is given by
\begin{IEEEeqnarray}{rCl}
\CS =\max_{P_X} I(X;Y|Z).
\label{eq:secrecy-capacity-semid-2}
\end{IEEEeqnarray}
The equivalence between~\eqref{eq:secrecy-capacity-semid} and~\eqref{eq:secrecy-capacity-semid-2} follows because, for a semi-deterministic DM-WTC, 
\begin{IEEEeqnarray}{rCl}
I(X;Y|Z)- H(Y|Z) = H(Y|X,Z) = 0.
\end{IEEEeqnarray}

Throughout this section, we shall assume that there exists a unique input distribution $P_X^*$ that achieves the secrecy capacity~\eqref{eq:secrecy-capacity-semid}.
Let $P_{Y}^* $ and $P_{Z|Y}^*$ denote the marginal and conditional distributions of $ P_X^*P_{YZ|X}$. 
For semi-deterministic wiretap channels, the bounds in Theorem~\ref{thm:dm-wtc-asy} match for the special case $\error = 0$, which yields\footnote{Rigorously speaking, the bounds in Theorem~\ref{thm:dm-wtc-asy} do not directly imply~\eqref{eq:zero-error-dispersion}, since there it is required that $\error>0$ and that the dispersion of the legitimate channel is nonzero. 
Nevertheless, it is not difficult to prove~\eqref{eq:zero-error-dispersion} using the same arguments as in the proof of  Theorem~\ref{thm:dm-wtc-asy}. We omit these mechanical details here.} 
 \begin{IEEEeqnarray}{rCl}
 R^*_{i}(n,0,\delta) = \CS-\sqrt{\frac{\VS}{n}}Q^{-1}\lefto(\delta \right)  + \bigO\lefto(\frac{\log n}{n}\right) , \quad i \in \{\avg, \max\}\IEEEeqnarraynumspace 
 \label{eq:zero-error-dispersion}
\end{IEEEeqnarray}
where
\begin{IEEEeqnarray}{rCl}
\VS &\define& V(P_{Y}^*, P_{Z|Y}^*)\label{eq:dispersion-def-sdw}.
\end{IEEEeqnarray}
The main result of this section is  a complete characterization of the second-order secrecy rate of the semi-deterministic DM-WTC for arbitrary $\error$ and $\delta$. 

 \subsection{Nonasymptotic Converse Bound}
 
We first give a nonasymptotic converse bound for semi-deterministic wiretap channels.  
\begin{thm}
\label{thm:general-converse}
Every $(M,\error,\secrecy)_{\avg}$ secrecy code for the semi-deterministic DM-WTC $P_{YZ|X}$ satisfies 
\begin{IEEEeqnarray}{rCl} 
\min_{P_{X}\in\setP(\setX)} \!\min_{Q_{Z}\in\setP(\setZ)}\!\! E_{\frac{|\setY|}{M(1-\error)}} ( P_{Y Z|X} \circ P_{X} , Q_{\setY}^{\unif} Q_{Z}) \leq \frac{\delta}{1-\error} \IEEEeqnarraynumspace
\label{eq:egamma-delta-error}
\end{IEEEeqnarray}
where $Q_{\setY}^{\unif}$ is the uniform distribution over $\setY$.
\end{thm}

\begin{rem}
\label{rem:existence-optimizer}
By the triangular inequality for $E_{\gamma}(\cdot,\cdot)$~\cite[Eq.~(21)]{liu2017-05a}, the function $(P_X,Q_Z)\mapsto E_{\gamma}( P_{Y Z|X} \circ P_{X} , Q_{\setY}^{\unif}  Q_{Z})$ is continuous in the topology of total variation. Moreover, since the probability simplexes $\setP(\setX)$ and $\setP(\setZ)$ are compact, a minimizer for $E_{\gamma}( P_{Y Z|X} \circ P_{X} , Q_{\setY}^{\unif} Q_{Z})$ exists.
\end{rem}

\begin{IEEEproof} 
Choose an arbitrary $(M,\error,\delta)_{\avg}$ secrecy code. 
Without loss of generality, we can assume that the decoder at the legitimate receiver is deterministic, and that  the error probability constraint~\eqref{eq:error-constraint-wtc} holds with equality.
Denote by $\setD_m$, $m\in \setM$, the decoding region for message~$m$.
By definition, we have 
\begin{IEEEeqnarray}{rCl}
\prob[Y \in \setD_W ] &=& \sum\limits_{m=1}^{M}\sum_{y\in\setY} P_{WY} (m,y) \indfun{y \in \setD_m}  \IEEEeqnarraynumspace\\
&=& 1-\error.
\label{eq:error-speical-case-conv}
\end{IEEEeqnarray}
Next, we derive a lower bound on the security index $S(W|Z) = \tvar(P_{WZ}, Q_{\setW}^{\unif}P_{Z})$ by using a suboptimal set $\setE$ in~\eqref{eq:def-total-var}.
%
%
%With a slight abuse of notation, we use $f(x^n)$ to denote  the vector $[f(x_1),\ldots, f(x_n)] \in \setY^n$.
%%
%%
%Let $Q_{Y^{n}}$ be the uniform distribution over $\setY^n$, let 
%
Let
\begin{equation} 
P_{X}^{c}(x) \define  \frac{1}{1-\error} \sum\limits_{m=1}^{M} P_{WX} (m,x) \indfun{f(x) \in \setD_m} 
\label{eq:def-Xc-proof}
\end{equation}
and let $P_{YZ}^{c} \define P_{Y Z |X } \circ P_{X }^{c}$.
By~\eqref{eq:error-speical-case-conv},   both $P_{X }^{c}$ and $P_{YZ}^{c} $ are probability distributions.
Consider the following subset of $\setM \times \setZ$
\begin{IEEEeqnarray}{rCl}
\setE & \define& \Big\{(m,z) :  \max_{y  \in \setD_m}  \frac{P_{YZ }^{c}   }{ Q_{\setY }^{\unif} P_{Z }} (y ;z )\geq \gamma\Big\} 
\end{IEEEeqnarray}
where the threshold $\gamma$ will be specified later. 
On the one hand, we have
\begin{IEEEeqnarray}{rCl}
%\IEEEeqnarraymulticol{3}{l}{
P_{WZ} [ \setE ] 
%}\notag\\  
&=&\frac{1}{M}\sum\limits_{m=1}^{M}  \sum\limits_{y\in \setY } P_{WY} (m,y)  \notag\\
&&\cdot \, P_{Z |Y=y, W=m} \! \lefto[ \max_{\tilde{y}  \in \setD_m}  \frac{P_{Y  Z }^c }{  Q_{\setY }^{\unif}  \! P_{Z }} (\tilde{y} ;Z )\geq \gamma \right] \IEEEeqnarraynumspace\\ 
&\geq& \sum\limits_{m=1}^{M} \sum\limits_{y  \in\setD_m} P_{WY}(m,y)\notag\\
&& \cdot\, P_{Z |Y =y , W=m} \lefto[   \frac{P_{Y Z }^c  }{  Q_{\setY }^{\unif}  P_{Z }} (y ;Z )\geq \gamma \right] \\
&=& (1-\error) P_{Y Z }^{c} \lefto[\frac{P_{Y Z  }^c }{  Q_{\setY }^{\unif}  P_{Z } } (Y ;Z )\geq \gamma \right]. 
\label{eq:PMZ-T-1}
\end{IEEEeqnarray}
On the other hand, we have 
\begin{IEEEeqnarray}{rCl}
%\IEEEeqnarraymulticol{3}{l}{
(Q_{\setM}^{\unif}P_{Z})[ \setE ] 
%}\notag\\
&=& \frac{1}{M}\sum\limits_{m=1}^{M} \!P_{Z}\lefto[ \max_{y   \in D_m} \frac{P_{Y Z}^c  }{  Q_{\setY }^{\unif}  P_{Z}} (y;Z)\geq \gamma \right] \IEEEeqnarraynumspace\\
&\leq&  \frac{1}{M}\sum\limits_{m=1}^{M}\sum\limits_{y \in \setD_m} P_{Z}\lefto[ \frac{P_{YZ }^c  }{  Q_{\setY }^{\unif}  P_{Z}} (y;Z)\geq \gamma \right] \label{eq:PMPZ-T-1-union}\\
&=& \frac{|\setY|}{M}  \big( Q_{\setY }^{\unif}  P_{Z}\big)\lefto[  \frac{P_{YZ}^c  }{  Q_{\setY }^{\unif} P_{Z}} (Y;Z)\geq \gamma \right]
\label{eq:PMPZ-T-1}
\end{IEEEeqnarray}
where~\eqref{eq:PMPZ-T-1-union} follows from the union bound.
Setting $\gamma=|\setY|/(M(1-\error))$, and using~\eqref{eq:def-e-gamma-1},~\eqref{eq:PMZ-T-1}, and~\eqref{eq:PMPZ-T-1}, we finally obtain
\begin{IEEEeqnarray}{rCl}
\delta &\geq&  \tvar(P_{W Z}, P_W P_{Z})\\
 &\geq & P_{WZ} [ \setE ] - (Q_{\setM}^{\unif}P_{Z})[ \setE ] \\
&\geq&    (1-\error) P_{YZ}^{c} \lefto[   \log \frac{P_{YZ}^c   }{ Q_{\setY }^{\unif} P_{Z}} (Y;Z)\geq \gamma \right] \notag\\
&& -\, \frac{|\setY|}{M}  \big( Q_{\setY }^{\unif} P_{Z}\big)\lefto[  \frac{P_{Y Z}^c  }{  Q_{\setY }^{\unif} P_{Z}} (Y;Z)\geq \gamma \right]   \\
&=& (1-\error) E_{\frac{|\setY|}{M(1-\error)}} (P_{Y Z}^{c} ,  Q_{\setY }^{\unif}  P_{Z})
\label{eq:proof-thm-converse-semi}
\\ 
&\geq& (1-\error) \min_{P_{X}} \min_{Q_{Z}}  E_{\frac{|\setY|}{M(1-\error)}} ( P_{Y Z|X} \circ P_{X} , Q_{\setY }^{\unif}  Q_{Z}). \IEEEeqnarraynumspace
\end{IEEEeqnarray}
This concludes the proof. 
\end{IEEEproof}

 \subsection{Trading Secrecy for Reliability}
 In Section~\ref{sec:trade-reli}, we have shown that one can trade reliability for secrecy for an arbitrary  wiretap channel. 
 While the converse  is not true in general,  the following result shows that one can trade secrecy for reliability for a semi-deterministic wiretap channel.

  \begin{thm}
  \label{thm:conv-trade-sec-reli}
  Every $(M,\error,\delta)_{\avg}$ secrecy code for a semi-deterministic wiretap channel can be converted to an $(M',0, \delta')_{\avg}$ secrecy code for the same channel, where $M' \in \mathbb{N}$ and  
%  \begin{IEEEeqnarray}{rCl}
%  M'\define \left\lfloor\frac{4M(1-\error)}{n} \right\rfloor
%\end{IEEEeqnarray}
%and 
\begin{IEEEeqnarray}{rCl}
\delta' \define \frac{\delta}{1-\error} +\frac{1}{2}\sqrt{\frac{M'}{M(1-\error)}}.
\end{IEEEeqnarray}
  Consequently,  we have 
 \begin{IEEEeqnarray}{rCl}
 R^*_{\avg}(n,\error,\delta) \leq R^*_{\avg}\lefto(n, 0, \frac{\delta}{1-\error} + \tau_n\right) - \frac{1}{n}\log (4\tau_n^2(1-\error))
 \label{eq:trade-reli-sec-ee}
\end{IEEEeqnarray}
 where $\tau_n$ is an arbitrary positive constant that satisfies $4M(1-\error)\tau_n^2  \in \mathbb{N}$.
 \end{thm}
 
 \begin{IEEEproof}
 Choose an arbitrary $(M,\error,\delta)_{\avg}$ secrecy code. 
Note that, by~\eqref{eq:proof-thm-converse-semi} in the proof of Theorem~\ref{thm:general-converse}, this code must  satisfy 
\begin{IEEEeqnarray}{rCl} 
\min_{Q_Z}  E_{\frac{|\setY|}{M(1-\error)}} ( P_{Y Z|X} \circ P_{X}^c , Q_{\setY}^{\unif} Q_{Z}) \leq 
 \frac{\delta}{1-\error} \IEEEeqnarraynumspace
\label{eq:egamma-delta-error-n}
\end{IEEEeqnarray}
%
 %
%Let $(\hat{P}_{X^n},\hat{Q}_{Z^n})$ be a minimizer of the optimization problem on the LHS of~\eqref{eq:egamma-delta-error-n}.
%
%
%As argued in Remark~\ref{rem:existence-optimizer}, such a minimizer exists. 
%
where $P_{X}^{c}$ was defined in~\eqref{eq:def-Xc-proof}.
Let $P_{Y}^c$ and $P_{X|Y}^c$ be the marginal (conditional) distributions of $P^c_{X} P_{Y|X}$.
Then, using  $P_{X|Y}^c $  as a prefix to the channel $P_{Y  Z |X}$, we obtain an auxiliary semi-deterministic wiretap channel $P^c_{YZ |\hat{Y}} : \setY  \to \setY  \times \setZ $:
\begin{IEEEeqnarray}{rCl}
\hat{Y} \stackrel{P^c_{X|Y} }{\longrightarrow} X \stackrel{P_{Y Z|X}}{\longrightarrow} (Y,\,\,Z).
\end{IEEEeqnarray}
By construction, every secrecy code for the auxiliary channel $P^c_{YZ |\hat{Y}} $ is also a secrecy code for the original channel  $P_{Y  Z |X}$. Furthermore, $P_{Y|\hat{Y}}^c (y|\hat{y}) = \indfun{y = \hat{y}}$.

Next, we construct a secrecy code  for the auxiliary channel $P^c_{YZ |\hat{Y}} $ as follows.
Let $M' \in\mathbb{N}$ be a positive integer.  
Using Lemma~\ref{lemma:privacy-amplification} on the joint distribution $P_{YZ}$ with $\gamma = |\setY|/(M (1-\error))$, and upper-bounding the expectation term on the RHS of~\eqref{eq:privacy-amplification-lemma} by $1$, we conclude that  there exists a mapping $g_0 : \setY \to \{1,\ldots, M'\}$ satisfying   
\begin{IEEEeqnarray}{rCl}
 S(g_0(Y)|Z) & \leq& \min_{Q_Z} E_{ \frac{|\setY|}{ M (1-\error)}}     ( P_{YZ}^c, Q_{\setY}^{\unif } Q_Z )  + \frac{1}{2}\sqrt{\frac{M'}{M(1-\error)}}\\
&\leq& \frac{\delta}{1-\error} +\frac{1}{2}\sqrt{\frac{M'}{M(1-\error)}} =\delta'.
\end{IEEEeqnarray}
Let now the message $W$ be distributed according to $g(\hat{Y})$ where $\hat{Y} \sim P_Y^c$.
For each message $m\in\{1,\ldots,M'\}$, the stochastic encoder $P_{\hat{Y}|W =m}$ is chosen to be the conditional distribution of $\hat{Y} \sim P_{Y}^c$ given $\hat{Y} \in g^{-1}(m)$. 
Such a code has zero error over the legitimate channel since the sets $\{g^{-1}(m)\}$ are disjoint. 
Furthermore, $P^c_{\hat{Y}Z} (y,z) = P_{YZ}^c(y,z) $ for every $(y,z)\in\setY\times\setZ$.
Hence, we have 
\begin{IEEEeqnarray}{rCl}
S(W|Z) = S(g_0(\hat{Y})|Z) = S(g_0(Y) |Z) \leq \delta'.
\end{IEEEeqnarray}
This concludes the first part of the theorem.
The bound~\eqref{eq:trade-reli-sec-ee} follows by choosing $M =\exp(n R^*_{\avg}(n,\error,\delta))$ and by setting $M' = 4M(1-\error) \tau_n^2 $.
 \end{IEEEproof}
 
 \subsection{Second-Order Secrecy Rate}
 \label{sec:2nd-sec-semideter}
 
Using Theorems~\ref{thm:ach-relation} and~\ref{thm:conv-trade-sec-reli} in~\eqref{eq:zero-error-dispersion}, we obtain the following complete characterization of the second-order secrecy rate for semi-deterministic wiretap channels.
\begin{thm}
\label{th:semi-determin-special}
Consider a semi-deterministic DM-WTC $P_{YZ|X}$. Assume that
\begin{enumerate}
\item There exists a  unique input distribution $P_X^*$ that achieves the secrecy capacity~\eqref{eq:secrecy-capacity-semid}.
\item $V(P_Y^*, P_{Z|Y}^*)>0$.
\end{enumerate}
%
% Conditions~\ref{cond:1} and~\ref{cond:2}.
If $\delta+\error <1$, then 
\begin{IEEEeqnarray}{rCl}
R^*_{i}(n,\error,\delta) = \CS-\sqrt{\frac{\VS}{n}}Q^{-1}\lefto(\frac{\delta}{1-\error}\right)  + \bigO\lefto(\frac{\log n}{n}\right), \quad \forall i\in\{\avg, \,\max\} 
\label{eq:thm-semi-deter-dispersion}
\end{IEEEeqnarray}
where $\VS$ is given in~\eqref{eq:dispersion-def-sdw}.
If $\delta+\error \geq 1$, then
\begin{IEEEeqnarray}{rCl}
R_{\avg}^*(n,\error,\delta) \geq R^*_{\max}(n,\error,\delta) \geq  |\setY|.
\label{eq:rate-trivial-dmc}
\end{IEEEeqnarray}
\end{thm}
  
\begin{IEEEproof}
We first consider the case $\error+\delta <1$.
On the achievability side, we have 
\begin{IEEEeqnarray}{rCl}
%\IEEEeqnarraymulticol{3}{l}{
R^*_{\max}(n,\error,\delta) &\geq& R^*_{\max}\lefto(n, 0 , \frac{\delta}{1-\error}  \right)  \label{eq:ach-expansion-2nd-end-1}\\
 &=&\CS-\sqrt{\frac{\VS}{n}}Q^{-1}\lefto(\frac{\secrecy}{1-\error}\right)  + O\lefto(\frac{\log n}{n}\right)
 \label{eq:ach-expansion-2nd-end}
\end{IEEEeqnarray}
where~\eqref{eq:ach-expansion-2nd-end-1} follows Theorem~\ref{thm:ach-relation}, and~\eqref{eq:ach-expansion-2nd-end} follows from~\eqref{eq:zero-error-dispersion}.
On the converse side, we have 
\begin{IEEEeqnarray}{rCl}
R^*_{\avg}(n,\epsilon,\secrecy) &\leq& R^*\lefto(n,0\frac{\delta}{1-\error}  + \frac{1}{\sqrt{n}} \right) + \frac{1}{n}\log \frac{n}{4(1-\error)} \label{eq:conv-expansion-2nd-end-1}\\
&=& \CS-\sqrt{\frac{\VS}{n}}Q^{-1}\!\lefto(\frac{\delta}{1-\error}\right)  + \bigO\lefto(\frac{\log n}{n}\right)\label{eq:conv-expansion-2nd-end-2}
\end{IEEEeqnarray}  
where~\eqref{eq:conv-expansion-2nd-end-1} follows from  Theorem~\ref{thm:conv-trade-sec-reli}  with $\tau_n = 1/\sqrt{n}$, and~\eqref{eq:conv-expansion-2nd-end-2} follows from~\eqref{eq:zero-error-dispersion} and by Taylor-expanding $Q^{-1}(x)$ around $x= \delta/(1-\error)$. 
 Since $R^*_{\max}(n,\error,\secrecy) \leq R^*_{\avg}(n,\epsilon,\secrecy)$ by definition, together~\eqref{eq:ach-expansion-2nd-end} and~\eqref{eq:conv-expansion-2nd-end-2} imply the expansion~\eqref{eq:thm-semi-deter-dispersion} for the case $\error +\delta <1$.

By~\eqref{eq:existence-code}, if $\error + \delta \geq1$, then 
\begin{IEEEeqnarray}{rCl}
R^*_{\max}(n,\error,\delta) \geq R^*_{\max}(n,0,1).
\end{IEEEeqnarray}
Since the legitimate channel is noiseless and since the function~$f$ defining the channel $P_{Y|X}$ is surjective,  we can prefix an arbitrary random transformation $P_{\hat{Y}^n|X^n}$ to the channel $P_{Y^n|X^n}$, and transmit $|\setY|^n$ codewords in the channel $\hat{Y}^n \to X^n \to Y^n$ without making an error. 
Furthermore, since $d(P,Q) \leq 1$ for every pair of probability distributions $P$ and $Q$,  the secrecy constraint is trivially satisfied without any additional processing. This implies that 
$R^*_{\max}(n,0,1)  \geq |\setY|$.  
\end{IEEEproof}

A few remarks are in order.
\begin{enumerate}
\item An alternative expression for $\VS$ is
\begin{IEEEeqnarray}{rCl}
\VS= \mathrm{Var}[\log  P_{Y|Z}^*(f(X)|Z)  |X ] .
\end{IEEEeqnarray}

\item It follows from~\eqref{eq:thm-semi-deter-dispersion} that neither the achievability nor the converse bound in Theorem~\ref{thm:dm-wtc-asy} is tight for the semi-deterministic wiretap channels (unless $\error =0$). 

\item The expansion~\eqref{eq:thm-semi-deter-dispersion}  is still valid if one of the average  constraints  in the definition of $R^*_{\avg}(n,\error,\secrecy)$ is replaced with its maximum counterpart. 
\item The expansion~\eqref{eq:thm-semi-deter-dispersion}  is still valid  if the channel between the transmitter and the eavesdropper is discrete input and continuous output.
%
%\item The strong converse does not hold if $\delta+\error \geq 1$. This observation remains valid for general  wiretap channels. 
%%
%%
% 
\item  Interestingly, the second-order rate in Theorem~\ref{th:semi-determin-special} does not match  the second-order rate of secret key generation (although the first-order rates in the two settings do match each other). 
As shown in Hayashi \emph{et al.}~\cite{hayashi2016-07a}, the latter rate (for independent and identically distributed sources $X^n$, $Y^n$, and $Z^n$)  is
\begin{IEEEeqnarray}{rCl}
I(X;Y|Z) - \sqrt{\frac{V}{n}} Q^{-1}(\error +\delta) + \bigO\lefto(\frac{\log n}{n}\right) \IEEEeqnarraynumspace
\label{eq:second-order-rate}
\end{IEEEeqnarray}
where $\error$ denotes the probability that the keys generated  by the legitimate users do not coincide, $\delta$ represents a secrecy constraint similar to~\eqref{eq:def-secrecy-constraint-def1}, and $V$ is a constant depending on the stochastic variation of the sources. 
This comparison reveals a subtle difference in the tradeoffs between reliability and secrecy in the two settings.  
Another interesting observation is that, if the wiretap channel is augmented by an authenticated  public discussion channel, then the second-order secrecy rate is equal to the one in~\eqref{eq:second-order-rate}~\cite{tahmasbi2016-09a}.

\item In our conference paper~\cite{yang2017-06a}, we proved the achievability part of~\eqref{eq:thm-semi-deter-dispersion} using a stronger soft-covering lemma for constant-composition codes similar to the one developed in~\cite{cuff2016-07a}. Our proof here is based on the (strong) privacy amplification result in Lemma~\ref{prop:strong-pa}, and is much simpler.
\end{enumerate}

\subsection{The Binary Symmetric Wiretap Channel}

\label{sec:bswiretap}

\begin{figure}
\centering
	\begin{tikzpicture}
		\begin{axis}[
			xlabel={\footnotesize Blocklength, $n$},
			ylabel={\footnotesize Secrecy rate, bits$/$(ch. use)},
xmin=0, xmax=2000,
ymin=0, ymax=0.55,
grid=major,
xtick={0,200,...,2000},
ytick={0,0.1,...,0.5},
yticklabels={
$0$,
$0.1$,
$0.2$,
$0.3$,
$0.4$,
$0.5$,
},
xticklabels={%
$0$,
$200$,
$400$,
$600$,
$800$,
$1000$,
$1200$,
$1400$,
$1600$,
$1800$,
$2000$},
%minor y tick num=1,
            %ylabel=y,
            ]
		\addplot[red] table {rate_conv.dat};
		\addplot[blue] table {rate_ach.dat};
		%\addplot[black] table {nalt.dat};
\addplot[black,dashed ] table {rate_app.dat};

\addplot[red,dashed] table {rate_conv_old.dat};

\addplot[black,mark=x] coordinates {
(0, 0.5)
(400, 0.5)
(800, 0.5)
(1200, 0.5)
(1600, 0.5)
(2000,0.5)
};

\draw (1000, 0.52) node[]{\footnotesize $C_{\mathrm{S}}  = 0.5$};

\draw[line width = 0.2mm, <-] (220,0.3) -- (520,0.17);
\draw (500, 0.17) node[right]{\footnotesize Approximation~\eqref{eq:bsc-expansion-thm} };

\draw[line width = 0.2mm, <-] (100,0.25) -- (400,0.12);
\draw (380, 0.12) node[right]{\footnotesize Converse~\eqref{eq:converse-bsc-nonasy}};

\draw[line width = 0.2mm, <-] (550,0.4) -- (400,0.45);
\draw (400, 0.46) node[]{\footnotesize Converse~\eqref{eq:converse-bound-general}};

\draw[line width = 0.2mm, <-] (400,0.34) -- (700,0.21);
\draw (680, 0.21) node[right]{\footnotesize Achievability~\eqref{eq:achievability-bsc-nonasy}};

		\end{axis}
	\end{tikzpicture}
	
\caption{Maximum secrecy rate $R^*_{\avg}(n,\error,\delta)$ for a binary symmetric wiretap channel with crossover probability $p=0.11$; here, $\delta=\epsilon=10^{-3}$. 
\label{fig:bsc-plot}}

\end{figure}
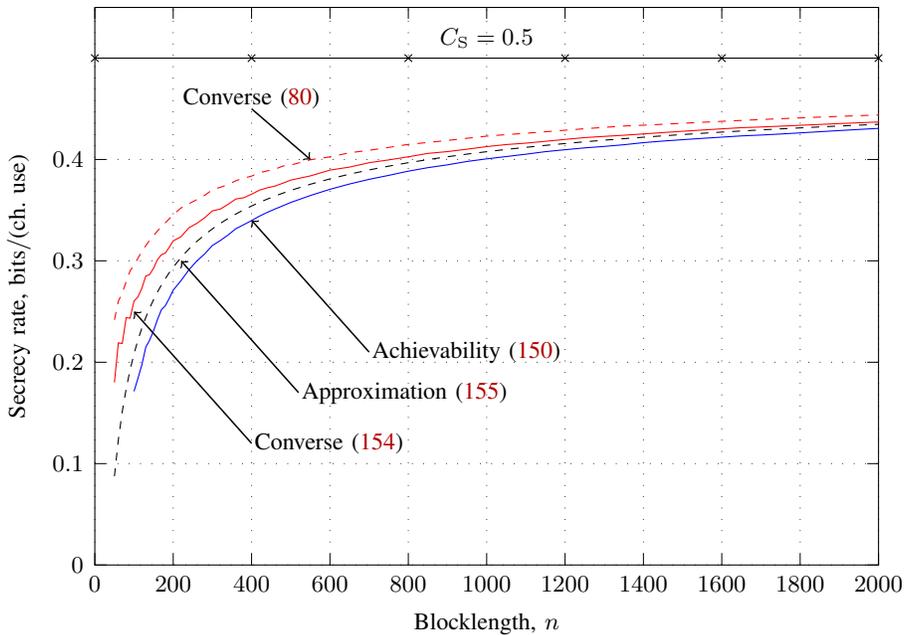

Consider a specific wiretap channel, where $\setX=\setY=\setZ = \mathbb{F}_2$, $P_{Y|X} = \indfun{Y=X}$, and $P_{Z|X}$ is a binary symmetric channel with crossover probability $p$. 
We shall refer to this channel as the binary symmetric wiretap channel (BS-WTC). 
Without loss of generality, we assume $p<1/2$.
%
%
%In this case, the converse bound in Theorem~\ref{thm:general-converse} and the achievability bound in~\cite[Th.~2]{yang2016-07a}   specialize to the following.
 \begin{thm} 
 \label{thm:bsc-converse}
For the BS-WTC with crossover probability $p$, there exists an $(n,2^k,\epsilon,\delta)_{\avg}$ secrecy code with uniformly distributed message such that
\begin{IEEEeqnarray}{rCl}
\frac{\delta}{1-\epsilon} \leq \frac{1}{2}\min_{\gamma} \left(g_n\lefto(\gamma\right) + \sqrt{g_n\lefto(\gamma\right) +  \frac{\gamma }{ 2^{n-k}} h_n(\gamma)}\right)
\label{eq:achievability-bsc-nonasy}
\end{IEEEeqnarray}
where $k\in \mathbb{N}$, 
\begin{IEEEeqnarray}{rcl}
%\IEEEeqnarraymulticol{3}{l}{ 
g_n(\gamma) \define
%}\notag\\ && 
1- \Ex{}{\exp\mathopen{}\Big\{\!\!-\!\Big[B\log\frac{p}{1-p}+  n \log(1-p) -\log\frac{ \gamma}{2^n} \Big]^+\!\Big\}}\IEEEeqnarraynumspace
\end{IEEEeqnarray}
with $[x]^{+} \define \max\{x,0\}$ and $B$ following the binomial distribution with parameters $n$ and $p$, and 
\begin{equation} 
h_n(\gamma) \define \Ex{}{\exp\mathopen{}\Big\{\!\!-\!\Big|B\log\frac{p}{1-p}+  n \log(1-p) -\log \frac{\gamma}{2^n} \Big|\Big\}} .  
\label{eq:def-hn-bsc-thm}
\end{equation}
Furthermore, there exists an $(n,2^k,\epsilon,\delta)_{\max}$ secrecy code such that
\begin{IEEEeqnarray}{rCl}
\frac{\delta}{1-\epsilon} \leq \frac{1}{2}\min_{\gamma} \left(g_n\lefto(\gamma\right) + \sqrt{g_n\lefto(\gamma\right) +  \frac{\gamma }{ 2^{n-k}} h_n(\gamma)}\right) + \sqrt{ \frac{\log (2^k +1)}{2^{n-k+1}}}.
\label{eq:achievability-bsc-nonasy-max}
\end{IEEEeqnarray}
Conversely, every $(n,M,\epsilon,\delta)_{\avg}$ secrecy code satisfies 
\begin{equation} 
g_n\lefto( \frac{2^n}{M(1-\error)} \right) \leq \frac{\delta}{1-\epsilon}.
\label{eq:converse-bsc-nonasy}
\end{equation}
  \end{thm}

\begin{IEEEproof}
The achievability bounds~\eqref{eq:achievability-bsc-nonasy} and~\eqref{eq:achievability-bsc-nonasy-max} follow from strengthened versions of Lemma~\ref{lemma:privacy-amplification} and Lemma~\ref{lemma:channel-resolvability}. 
The converse bound~\eqref{eq:converse-bsc-nonasy} follows by showing that the optimal  $P_X$ and $Q_Z$ in~\eqref{eq:egamma-delta-error} are both uniform distributions over $ \mathbb{F}_2^n$. 
Their proofs are given in Appendix~\ref{app:symmetry}.
\end{IEEEproof}

 For the BS-WTC, the reminder term in~\eqref{eq:thm-semi-deter-dispersion} can be improved to $\bigO(1 /n)$.

\begin{thm}
\label{thm:bsc-expansion}
For the BS-WTC with crossover probability $p$, and for every $\epsilon\geq 0$ and $\delta>0$ that satisfy $\error+\delta <1$, we have
\begin{IEEEeqnarray}{rCl}
%&& -\frac{1}{2}\frac{\log n}{n} +   \bigO\lefto(\frac{1}{n}\right) \leq 
R^*_{i}(n,\epsilon,\delta) =
 \Hb(p) - \sqrt{\frac{\VBSC}{n}}  Q^{-1}\!\lefto(\frac{\delta}{1-\error}\right) +  \bigO\lefto(\frac{1}{n}\right), \qquad % \IEEEeqnarraynumspace\notag\\
   i\in \{\max,\avg\}
   %\,\,\,\,\,\,\,\,\,\,\,\,\,
 \label{eq:bsc-expansion-thm}
\end{IEEEeqnarray}
where  $\Hb(\cdot)$ denotes the binary entropy function, and 
\begin{IEEEeqnarray}{rCl}
\VBSC \define p(1-p) \log^2((1-p)/p).
\end{IEEEeqnarray}

\end{thm}
\begin{IEEEproof}
See Appendix~\ref{app:proof-bsc-asy}.
\end{IEEEproof}
\begin{rem}
The strengthened achievability bounds~\eqref{eq:achievability-bsc-nonasy} and~\eqref{eq:achievability-bsc-nonasy-max} are crucial in establishing the correct  order $\bigO(1/\sqrt{n})$ in the achievability bound. 
For comparison, the bound~\eqref{eq:privacy-amplification-lemma} in Lemma~\ref{lemma:privacy-amplification} yields $-\frac{1}{2}\log n$ for the third-order term.
\end{rem}

In Fig.~\ref{fig:bsc-plot}, we plot the bounds~\eqref{eq:achievability-bsc-nonasy} and~\eqref{eq:converse-bsc-nonasy}   for $p=0.11$, and $\delta=\error=10^{-3}$.
For comparison, we  also plot the general converse bound in~\eqref{eq:converse-bound-general}  and the approximation~\eqref{eq:bsc-expansion-thm}.
From this figure, we see that the new converse bound in~\eqref{eq:converse-bsc-nonasy} is uniformly tighter than~\eqref{eq:converse-bound-general}.
Numerically, the maximum secrecy rate  at blocklength $400$ is between $0.34$ and  $0.36$ bits/ch. use.

 \section{Conclusion}
 
 In this paper, we have established nonasymptotic achievability and converse bounds on the maximum secret communication rate $R^*(n,\error,\delta)$ for a given blocklength $n$, error probability~$\error$, and information leakage~$\delta$ over wiretap channels. These bounds are shown to be tighter than the best previously known bounds.
The proofs of our achievability bounds are based on new privacy amplification results that may be useful for other security problems. We have also established connections between privacy amplification and channel resolvability, and have used this connection to derive tighter bounds for channel resolvability.

By analyzing the nonasymptotic bounds, we have further derived upper and lower bounds on the second-order secrecy rate for general discrete memoryless wiretap channels and for Gaussian wiretap channels. 
For the special case of semi-deterministic wiretap channels, we have established the \emph{exact}  second-order secrecy rate, which characterizes the optimal tradeoff between reliability and secrecy in the finite-blocklength regime. 
For general (non-deterministic) wiretap channels, the problem of establishing the second-order secrecy rate  is still open.  

\appendices

\section{Proof of Lemma~\ref{lemma:privacy-amplification}}

\label{app:proof-of-lemma-privacy-amp}

The proof is based on random hashing (also known as universal hashing) and the  left-over hash lemma~\cite{bennett1995-11a,renner2005-09a}. 
%which are standard techniques to prove privacy amplification results. 
%
%
%

\subsection{Proof of Part~\ref{item-lemma1-part1}}
Let $G$ be a random function  from $\setX$ to $\setK$ that satisfies the following collision probability bound 
\begin{IEEEeqnarray}{rCl}
\prob[G(x_1) = G(x_2)] \leq \frac{1}{|\setK|},\quad \forall x_1\neq x_2.
\label{eq:collision-prob}
\end{IEEEeqnarray}
In other words, $G$ is a \emph{universal}$_2$ hash  function~\cite{Carter79-02a,bennett1995-11a}.\footnote{One example of a \emph{universal}$_2$ hash function is the one that is uniformly distributed over the set of all functions from $\setX$ to $\setK$ (see~\cite[Sec. IV]{bennett1995-11a}).}
The average secrecy index of $G$ can be computed as follows:
\begin{IEEEeqnarray}{rCl}
\Ex{G}{S(G(X) |Z) } &=& \Ex{G}{d(P_{G(X)Z} ,Q_{\setK}^{\unif}P_Z)} \\
%&=& \frac{1}{2}\Ex{G}{\sum\limits_{k\in\setK, z\in\setZ} \bigg|\sum\limits_{x\in G^{-1}(k) } P_{XZ}(x,z) -\frac{1}{|\setK|} P_Z(z)\bigg|} \\
&=& \frac{1}{2} \sum\limits_{k\in\setK,z\in\setZ} \Ex{G}{\bigg|\sum\limits_{x\in G^{-1}(k) } P_{XZ}(x,z) -\frac{1}{|\setK|} P_Z(z)\bigg|}  \\
&=&  \frac{|\setK|}{2}\sum\limits_{z\in\setZ} \Ex{G}{\bigg|\sum\limits_{x\in G^{-1}(1) } P_{XZ}(x,z) -\frac{1}{|\setK|} P_Z(z)\bigg|} .
\label{eq:symmetry-step-11}
%&=&  \frac{|\setK|}{2} \sum\limits_{z} Q_Z(z) \Ex{G}{ \bigg|\sum\limits_{x\in G^{-1}(1) } \frac{P_{XZ}(x,z)}{Q_{Z}(z)} -\frac{1}{|\setK|} \frac{P_Z(z)}{Q_Z(z)}\bigg|} .  \label{eq:symmetry-step-12}
\end{IEEEeqnarray}
Here,~\eqref{eq:symmetry-step-11} follows from the symmetry of the set $\setG$.

Observe that the expectation on the RHS of~\eqref{eq:symmetry-step-11} is the first absolute  central moment of the random variable
\begin{IEEEeqnarray}{rCl}
A(z) \define \sum\limits_{x\in G^{-1}(1) }P_{XZ}(x,z) .
\end{IEEEeqnarray}
Indeed,  it is not difficult to verify that 
\begin{IEEEeqnarray}{rCl}
\Ex{}{A(z)} = \frac{P_Z(z)}{|\setK|}.
\end{IEEEeqnarray}
An exact computation of the first absolute moment of $A(z)$ is highly complex, and is not suitable for numerical evaluation or analytical analysis.   
%
%One natural way to upper-bound~\eqref{eq:symmetry-step-12} is to upper-bound the first absolute  central moment of $A(z)$ through the variance. 
%%
%%
%However, as observed in~\cite{bennett1995-11a}, the resulting bound is quite loose. 
%%
%
To get a simpler bound on~\eqref{eq:symmetry-step-11}, we shall split  $A(z)$ into two parts $A_1(z)$ and $A_2(z)$ that satisfy $\mathrm{Var}[A_1(z)] \gg \mathrm{Var}[A_2(z)]$, and bound  $\Ex{}{|A(z) -\Ex{}{A(z)}\!|}$  using the triangle inequality as in~\cite{renner2005-09a} and~\cite{hayashi13-11a}. 
Doing so, we obtain
\begin{IEEEeqnarray}{rCl}
\Ex{}{|A(z) -\Ex{}{A(z)}|} &=& \Ex{}{|A_1+ A_2 -\Ex{}{A_1(z)+A_2(z)}|} \\
&\leq& 2\Ex{}{|A_1(z)|} + \Ex{}{|A_2(z) -\Ex{}{A_2(z)} |}  \label{eq:bound-triangle-ineq}
\\ &\leq&2\Ex{}{|A_1(z)|} +  \sqrt{\mathrm{Var}[A_2(z)]}.\label{eq:bound-triangle-ineq-2} 
\end{IEEEeqnarray}
Without loss of generality, we can assume that $A_1(z)$ and $A_2(z)$ take the form 
\begin{IEEEeqnarray}{rCl}
A_2(z)  = \sum\limits_{x\in G^{-1}(z)} R_{XZ}(x,z) ,\quad A_1(z) = A(z)-A_2(z)
\label{eq:a1z-a2z-spec-form}
\end{IEEEeqnarray}
where $R_{XZ}$ is an arbitrary nonnegative measure (not necessary a probability measure).
Substituting~\eqref{eq:a1z-a2z-spec-form} into~\eqref{eq:bound-triangle-ineq-2}, then~\eqref{eq:bound-triangle-ineq-2} into~\eqref{eq:symmetry-step-11},  we obtain 
\begin{IEEEeqnarray}{rCl}
\Ex{}{S(G(X) |Z)} &\leq& |\setK| \sum_{k,z} \Ex{}{\Big|\sum\limits_{x\in G^{-1}(1)} (P_{XZ}  -R_{XZ})(x,z) \Big|}\label{eq:ub-sgz-hh} \notag\\
&& +\, \frac{1}{2}  |\setK| \sum_{z} \sqrt{\mathrm{Var}\mathopen{}\Big[\sum\limits_{x\in G^{-1}(1)} R_{XZ}(x,z) \Big]} \IEEEeqnarraynumspace \\ 
&\leq &\|P_{XZ} - R_{XZ}\|_1 + \frac{1}{2}\sum_{z} R_Z(z) \sqrt{|\setK|\Ex{}{\|R_{G(X) |Z=z} - Q_{\setK}^{\unif} \|_2^2}}
\label{eq:ub-sgz}
\end{IEEEeqnarray}
where $R_Z$ is the projection of $R_{XZ}$ on $\setZ$. 
Here, in~\eqref{eq:ub-sgz}, we have used the triangular inequality on the first term on the RHS of~\eqref{eq:ub-sgz-hh}  and applied the following identity
\begin{IEEEeqnarray}{rCl}
\Ex{}{\sum\limits_{x\in G^{-1}(z)} R(x,z)} = \frac{R_Z(z)}{|\setK|}.
\end{IEEEeqnarray}
The second term on the RHS of~\eqref{eq:ub-sgz} can be further bounded as
\begin{IEEEeqnarray}{rCl}
\IEEEeqnarraymulticol{3}{l}{
\sum_{z}R_Z(z) \sqrt{\Ex{}{\|R_{G(X) |Z=z} - Q_{\setK}^{\unif} \|_2^2} } }\notag \\
& =& \sum_{z}\sqrt{Q_Z(z)}\sqrt{\frac{R_Z^2(z)}{ Q_Z(z) }   \Ex{}{\|R_{G(X) |Z=z} - Q_{\setK}^{\unif} \|_2^2}} \\
&\leq& \sqrt{ \sum_{z} \frac{R_Z^2(z)}{ Q_Z(z) }  \Ex{}{\|R_{G(X) |Z=z} - Q_{\setK}^{\unif}   \|_2^2}}   \label{eq:rewrite-egamma-000} \\
&\leq&\sqrt{ \sum_{z} \frac{R_Z^2(z)}{ Q_Z(z)} \sum\limits_{x} R_{X|Z}^2(x|z)}  \label{eq:rewrite-egamma-001}  \\
&=&  \sqrt{ \exp( - H_{2}(R_{XZ}|Q_Z) ) } \label{eq:rewrite-egamma-0}.  
\end{IEEEeqnarray}
Here, $H_2(R_{XZ}|Q_Z) \define -\log \sum_{x,z} R_{XZ}(x,z)^2/Q_Z(z)$ denotes the conditional R\'{e}nyi entropy of order 2 relative to $Q_Z$,~\eqref{eq:rewrite-egamma-000} follows from the Cauchy-Schwarz inequality and because $\sum\nolimits_z Q_Z(z) =1$, and~\eqref{eq:rewrite-egamma-001}  follows from the left-over hash lemma by Bennett \emph{et al.}~\cite{bennett1995-11a}.
%
%
%Furthermore, by the left-over hash lemma~\cite[Lemma~5.4.3]{renner2005-09a} (see also~\cite[Sec. II.B]{watanabe2013-07a}), we obtain that for every probability distribution $Q_Z$ on~$\setZ$
%\begin{IEEEeqnarray}{rCl}
%a
%\end{IEEEeqnarray}
%Here, 
%%
%%
%%
%
Note that the steps in~\eqref{eq:symmetry-step-11}--\eqref{eq:rewrite-egamma-0} are essentially the ones employed in~\cite[Sec.~5.5]{renner2005-09a} and~\cite{hayashi13-11a}.
The reason that we present them here (in a different form) are two-fold. 
First, as will be shown in the proof of Lemma~\ref{lemma:channel-resolvability},  the same bounding techniques as above can be used to analyze channel resolvability. 
Second, as will be shown in Appendix~\ref{app:bsc-ach-nonasym}, if the distribution $P_{XZ}$ admits further structures (e.g., symmetry), then the bounds~\eqref{eq:bound-triangle-ineq} and \eqref{eq:bound-triangle-ineq-2} can be tightened.

We next minimize the RHS of~\eqref{eq:rewrite-egamma-0}   over $R_{XZ}$. This is unlike the existing privacy amplification results (for example, those discussed previously), in which one selects a convenient $R_{XZ}$ and then proves an upper bound on~\eqref{eq:rewrite-egamma-0}. 
%
%
%
%
%
%The difference between Lemma~\ref{lemma:privacy-amplification} and the privacy amplification results in~\cite{renner2005-09a,hayashi2011-06a} lies in the choice of $R_{XZ}$ that is used in the bound~\eqref{eq:ub-sgz}. 
%Instead of 
%
Consider the following optimization problem 
\begin{IEEEeqnarray}{rCl}
\min_{R_{XZ}: \| P_{XZ} -R_{XZ}  \|_1\leq \tilde{\epsilon} } \sum_{x,z} \frac{R_{XZ}(x,z)^2}{Q_{Z}(z)}
\label{eq:optimizing-h2-min}
\end{IEEEeqnarray}
where $\tilde{\epsilon}\in [0,2]$.
It is not difficult to see that the optimal  $R^*_{XZ}$ must satisfy $R_{XZ}^*(x,z) \leq P_{XZ}(x,z)$ for every pair $(x,z)$, i.e.,~\eqref{eq:optimizing-h2-min} is equivalent to 
\begin{IEEEeqnarray}{rCl}
\min_{R_{XZ}: \| P_{XZ} -R_{XZ}  \|_1\leq \tilde{\epsilon}, \, R_{XZ} \leq P_{XZ} } \sum_{x,z} \frac{R_{XZ}(x,z)^2}{Q_{Z}(z)}.
\label{eq:optimizing-h2-min-2}
\end{IEEEeqnarray}
 Indeed, suppose on the contrary that there exists $(x_0,z_0)$ which satisfy $R^*_{XZ}(x_0,z_0) > P_{XZ}(x_0, z_0)$. Since the objective function in~\eqref{eq:optimizing-h2-min} is monotonically increasing with $R_{XZ}(x_0,z_0)$,  we can further decrease it by setting $R^*_{XZ}(x_0,z_0) = P_{XZ}(x_0,z_0)$ without violating the constraint  in~\eqref{eq:optimizing-h2-min}. But this contradicts the assumption that $R^*_{X,Z}$ is the optimizer of~\eqref{eq:optimizing-h2-min}. Therefore, we must have $R^*_{X,Z} \leq P_{XZ}$. 
 Observe that the problem~\eqref{eq:optimizing-h2-min-2} is a convex optimization problem. Hence, by the Karush-Kuhn-Tucker  (KKT) condition~\cite[Sec.~5.5.3]{boyd04}, the optimizer of~\eqref{eq:optimizing-h2-min-2} must take the form
\begin{IEEEeqnarray}{rCl}
R^*_{XZ}(x,z) \define   \begin{cases}
P_{XZ}(x,z) & \imath(x;z)  \leq \log \gamma \\
\gamma Q_{\setX}^{\unif}(x)Q_Z(z)& \text{otherwise}\\
\end{cases}
\label{eq:def-rxz}
\end{IEEEeqnarray}  
where $\gamma$ is chosen such that $\|P_{XZ} -R_{XZ}^*\|_1 = \tilde{\epsilon}$.

Using~\eqref{eq:def-rxz} in the first term on the RHS of~\eqref{eq:ub-sgz}, we obtain
\begin{IEEEeqnarray}{rCl}
\|P_{XZ} -R^*_{XZ}\|_1 = E_{\gamma}(P_{XZ}, Q_{\setX}^{\unif}Q_Z).
\label{eq:evaluate-d1-egamma}
\end{IEEEeqnarray}
The bound~\eqref{eq:rewrite-egamma-0} can be evaluated as follows:
\begin{IEEEeqnarray}{rCl}
\IEEEeqnarraymulticol{3}{l}{
\exp(-H_2(R^*_{XZ}|Q_Z)) }\notag\\
 &=& \sum\limits_{x,z} \frac{R^*_{XZ}(x,z)^2}{Q_Z(z)}\\
&=& \frac{1}{|\setX|} \sum\limits_{x,z} P_{XZ}(x,z) \exp(\imath(x;z)) \indfun{\imath (x;z) \leq \log \gamma } \notag\\
&&+\,  \frac{\gamma^2}{|\setX|} Q_{\setX}^{\unif}Q_Z[\imath(X;Z) \geq \log \gamma]\\
&=& \frac{\gamma}{|\setX|} \Ex{P_{XZ}}{ \exp(-|\imath(X;Z) -\log \gamma|)} \notag\\
&&-\, \frac{\gamma}{|\setX|} \Ex{P_{XZ}}{ \exp(-|\imath(X;Z) -\log\gamma|) \indfun{\imath(X;Z) \geq \log\gamma}}\notag\\
&&+\,  \frac{\gamma^2}{|\setX|} Q_{\setX}^{\unif}Q_Z [\imath(X;Z) \geq \log \gamma]\\
&=&   \frac{\gamma}{|\setX|} \Ex{}{\exp\mathopen{}\big(\! - \! |\imath(X;Z)-\log \gamma |\big)}.\label{eq:evaluate-expe-term-pa}
\end{IEEEeqnarray}
Substituting~\eqref{eq:evaluate-expe-term-pa} into~\eqref{eq:rewrite-egamma-0}, and then~\eqref{eq:evaluate-d1-egamma}  and~\eqref{eq:rewrite-egamma-0} into~\eqref{eq:ub-sgz}, we conclude that for every $\gamma>0$,
\begin{IEEEeqnarray}{rCl}
 \Ex{G}{S(G(X) |Z)}  &\leq&   E_{\gamma}(P_{XZ} , Q_{\setX}^{\unif}Q_Z) + \frac{1}{2}\sqrt{ \frac{\gamma|\setK|}{|\setX|}  \Ex{}{\exp\mathopen{}\big(\! - \! |\imath(X;Z)-\log \gamma |\big)} } \label{eq:rewrite-egamma-3} .\IEEEeqnarraynumspace
\end{IEEEeqnarray}
Finally, the inequality~\eqref{eq:rewrite-egamma-3}  implies that there exists a $g\in\setG$ for which~\eqref{eq:privacy-amplification-lemma} holds.

\subsection{Proof of Part~\ref{item-lemma1-part2}}

Let $\setG$ be the set of all functions $g$ from $\setX$ to $\setK$ that satisfies 
$\big|g^{-1}(k)\big| =L$ for every $k \in \setK$. 
Let~$G$ be uniformly distributed over the set~$\setG$. 
It is not difficult to check that  such a random function satisfies the collision probability bound~\eqref{eq:collision-prob}, i.e., it is a \emph{universal}$_2$ function.
As such, it satisfies the bound~\eqref{eq:rewrite-egamma-3}. 
Furthermore,  for every $g\in \setG$ we have $g(X) \sim Q_{\setK}^{\unif}$, where $X\sim Q_{\setX}^{\unif}$.  
This concludes the proof of Part~\ref{item-lemma1-part2}.

\section{Proof of Lemma~\ref{prop:strong-pa}}
\label{app:proof-strong-pa}

By the union bound and the Chernoff bound, we have 
\begin{IEEEeqnarray}{rCl}
\prob\lefto[  \max_{k\in\setK}  \tvar(P_{Z|G^{-1}(k) }, P_{Z}) \geq r + \mu\right] &\leq&  |\setK|  \prob\lefto[    \tvar(P_{Z|G^{-1}(1) }, P_{Z}) \geq r + \mu\right]\\
&\leq &  |\setK|  e^{-t(r+\mu)}\Ex{}{ e^{t \tvar(P_{Z|G^{-1}(1) }, P_{Z})} }
\label{eq:Markov-inequality-sampling-wr}
\end{IEEEeqnarray}
where $t$ is an arbitrary  positive constant.
Observe that the random variable $\tvar (P_{Z|G^{-1}(1) }, P_{Z}) $ can be rewritten as
\begin{IEEEeqnarray}{rCl}
\tvar(P_{Z|G^{-1}(1) }, P_{Z}) &=&\frac{1}{2}\sum_{z\in\setZ} | L^{-1}\sum\limits_{x\in G^{-1}(1)} P_{Z|X}(z|x) -P_Z(z) | \\
&\stackrel{\der}{=}& \frac{1}{2}\sum_{z\in\setZ} | L^{-1} \sum\limits_{i=1,..., L} P_{Z|X}(z|X_i) -P_Z(z) | \\
&\define& h(X_1^L) \label{eq:def-hhhh}
\end{IEEEeqnarray}
where $\stackrel{\der}{=}$ denotes equality in distribution, and  $X_1,...,X_L$ are distributed according to the joint distribution
\begin{IEEEeqnarray}{rCl}
P_{X_1^L} (x_1^L)= Q_{\setX}^{\unif}(x_1) \prod_{i=2}^{L} Q_{\setX\setminus \{x_1^{L-1}\} }^{\unif}(x_i). 
\end{IEEEeqnarray}
In other words, $X_1^L$ are random samples \emph{without replacement} from $\setX$. 

We proceed to upper-bound $\Ex{}{e^{t h(X_1^L)}}$.
The key idea is to replace the dependent random variables $X_1^L$ with the i.i.d. samples $\bar{X}_1^L$ defined in the lemma. 
More specifically, we shall prove that 
\begin{IEEEeqnarray}{rCl}
\Ex{}{e^{t h(X_1^L)}} \leq \Ex{}{e^{t h(\bar{X}_1^L)}}.
\label{eq:proof-hoeffding-1}
\end{IEEEeqnarray}
This is done by using the elegant reduction argument of Hoeffding~\cite{hoeffding1963-03a}.  
Let 
\begin{IEEEeqnarray}{rCl}
f(x_1^L) \define e^{th(x_1^L)}.
\end{IEEEeqnarray}
Since $f(\cdot)$ is symmetric in its arguments,   there exists a symmetric function $\bar{f}: \setX^L \to  \realset$ that satisfies 
\begin{IEEEeqnarray}{rCl}
 \Ex{}{\bar{f}(X_1^L )} = \Ex{}{f(\bar{X}_1^L)}
\end{IEEEeqnarray}
 as observed in~\cite[Sec. 5]{hoeffding1963-03a}.
Such a function takes the form
\begin{IEEEeqnarray}{rCl}
\bar{f}(x_1^L) = \sum p_{k,r_1,...,r_k, i_1,...,i_k} f(\underbrace{x_{i_1},...,x_{i_1}}_{r_1},...., \underbrace{x_{i_k},...,x_{i_k}}_{r_k})
\end{IEEEeqnarray}
where the sum is taken over the integers $k,r_1,...,r_k, i_1,...,i_k$ such that $k\in\{1,...,L\}$, $r_1+\cdots+r_k=n$, $\{i_1,\ldots,i_k\} \subset\{1,...,L\} $ are distinct, and 
\begin{IEEEeqnarray}{rCl}
\sum p_{k,r_1,...,r_k, i_1,...,i_k} =1.
\end{IEEEeqnarray}

To prove~\eqref{eq:proof-hoeffding-1}, it suffices to show that 
\begin{IEEEeqnarray}{rCl}
\bar{f}(x_1^L) \geq f(x_1^L),\qquad \forall x_1^L \in \setX^L.
\label{eq:inter-mediate-hoeffding}
\end{IEEEeqnarray}
Consider the chain of inequalities
\begin{IEEEeqnarray}{rCl}
\bar{f}(x_1^L)  &\geq& \exp\mathopen{}\Big(  t\log (e) \sum p_{k,r_1,...,r_k, i_1,...,i_k} h(\underbrace{x_{i_1},...,x_{i_1}}_{r_1},...., \underbrace{x_{i_k},...,x_{i_k}}_{r_k}) \Big)
\label{eq:jensen-hodffding-1}
\\
&=& \exp\mathopen{}\Big(  \frac{t \log (e)}{2} \sum\limits_{z\in \setZ} \sum p_{k,r_1,...,r_k, i_1,...,i_k} \Big| g_z(\underbrace{x_{i_1},...,x_{i_1}}_{r_1},...., \underbrace{x_{i_k},...,x_{i_k}}_{r_k}) \Big|\Big) \label{eq:jensen-hodffding-2} \\
&\geq&\exp\mathopen{}\Big(  \frac{t \log (e)}{2} \sum\limits_{z\in \setZ}    \Big| \sum p_{k,r_1,...,r_k, i_1,...,i_k} g_z(\underbrace{x_{i_1},...,x_{i_1}}_{r_1},...., \underbrace{x_{i_k},...,x_{i_k}}_{r_k}) \Big|\Big) \IEEEeqnarraynumspace \label{eq:jensen-hodffding-3}\\
&=&f (x_1^L) \label{eq:jensen-hodffding-4}
\end{IEEEeqnarray}
where the functions $\{g_z(\cdot)\}$ are defined as 
\begin{IEEEeqnarray}{rCl}
g_z (x_1^L)\define   \frac{1}{L} \sum\limits_{i=1}^{L}P_{Z|X}(z|x_i) -P_Z(z).
\end{IEEEeqnarray}
Here,~\eqref{eq:jensen-hodffding-1} follows from  Jensen's inequality and the convexity of $x\mapsto e^{tx}$;~\eqref{eq:jensen-hodffding-2} follows from~\eqref{eq:def-hhhh};~\eqref{eq:jensen-hodffding-3} follows from the triangle inequality; and finally,~\eqref{eq:jensen-hodffding-4} follows because 
\begin{IEEEeqnarray}{rCl}
\sum p_{k,r_1,...,r_k, i_1,...,i_k} g_z(\underbrace{x_{i_1},...,x_{i_1}}_{r_1},...., \underbrace{x_{i_k},...,x_{i_k}}_{r_k})  = g_z(x_1^L).
\end{IEEEeqnarray}
This concludes~\eqref{eq:inter-mediate-hoeffding}, and, hence~\eqref{eq:proof-hoeffding-1}.

Now, observe that the partial discrete derivatives of $h(\cdot)$ are uniformly bounded, i.e.,  
\begin{IEEEeqnarray}{rCl}
\sup_{y,y' \in \setX}| h(x_1^{i-1} , y, x_{i+1}^L) -  h(x_1^{i-1} , y', x_{i+1}^L) | \leq \frac{1}{L} ,\quad  i=1,...,L. 
\end{IEEEeqnarray}
This allows us to conclude that (see the proof of McDiarmid's inequality~\cite{mcdiarmid-1989} and~\cite[Eq.~(2.2.25)]{raginsky13}, and recall that $\mu = \Ex{}{h(\bar{X}_1^L )}$)
\begin{IEEEeqnarray}{rCl}
\Ex{}{e^{t (h(\bar{X}_1^L) -\mu)}}  \leq e^{t^2/(8L)}.
\label{eq:bound-sample-wr}
\end{IEEEeqnarray}
Using~\eqref{eq:bound-sample-wr} and~\eqref{eq:proof-hoeffding-1} in~\eqref{eq:Markov-inequality-sampling-wr}, we conclude that 
\begin{IEEEeqnarray}{rCl}
\prob\lefto[  \max_{k\in\setK}  \tvar(P_{Z|G^{-1}(k) }, P_{Z}) \geq r + \mu\right] \leq |\setK| e^{-tr + t^2/(8L)}. 
\label{eq:McDir-lalala}
\end{IEEEeqnarray}
The choice $t = 4rL$ minimizes the RHS of~\eqref{eq:McDir-lalala} and yields~\eqref{eq:strong-pa}.

Finally, using $r = \sqrt{\log(1+|\setK|)/(2L)}$ on the RHS of~\eqref{eq:strong-pa}, we obtain 
\begin{IEEEeqnarray}{rCl}
\prob\lefto[  \max_{k\in\setK}  \tvar(P_{Z|G^{-1}(k) }, P_{Z}) \geq r + \mu\right]  \leq \frac{|\setK|}{|\setK|+1}<1
\end{IEEEeqnarray} 
which implies~\eqref{eq:max-bound-tv}.

 \section{Proof of Lemma~\ref{lemma:channel-resolvability}}
\label{app:proof-lemma-resolvability}

Consider 
\begin{IEEEeqnarray}{rCl}
\Ex{\setA}{d(P_{Z|\setA} , P_{Z})} =\frac{1}{2} \sum\limits_{z\in Z} \Ex{X_1,...,X_L}{\Big| \frac{1}{L} \sum\limits_{i=1,\ldots,L} P_{Z|X}(z|X_i) -P_Z(z)  \Big| }.
\end{IEEEeqnarray}
 As in the proof of Lemma~\ref{lemma:privacy-amplification}, we shall split the random variable 
\begin{IEEEeqnarray}{rCl}
B(z) \define \frac{1}{L} \sum\limits_{i=1,\ldots,L} P_{Z|X}(z|X_i) 
\end{IEEEeqnarray}
into two parts. 
Let $\{R_{Z|X=x}\}_{x\in\setX}$ be a sequence of arbitrary nonnegative measures indexed by~$x$. Let 
\begin{IEEEeqnarray}{rCl}
B_2(z) \define \frac{1}{L} \sum\limits_{i=1,\ldots,L} R_{Z|X} (z|X_i) ,\qquad B_1(z) \define B(z) - B_2(z). 
\end{IEEEeqnarray} 
It follows that 
\begin{IEEEeqnarray}{rCl}
\Ex{\setA}{d(P_{Z|\setA} , P_{Z})} &=& \frac{1}{2}\sum\limits_{z\in\setZ} \Ex{}{|B(z) - B_1(z)|} \\
&\leq& \sum\limits_{z\in\setZ} \Ex{}{|B_1(z)|}  + \frac{1}{2}\sum\limits_{z\in\setZ}\sqrt{\mathrm{Var}[B_2(z)]} \label{eq:channel-resolv-2}\\
&=& \|P_{XZ}- R_{Z|X}P_X\|_1 + \frac{1}{2}\sum\limits_{z\in\setZ}\sqrt{L^{-1}\mathrm{Var}[R_{Z|X}(z|X)]}.\label{eq:channel-resolv-3}
\end{IEEEeqnarray}
Here, in~\eqref{eq:channel-resolv-2} we have used the triangle inequality and Jensen's inequality.  
The second term on the RHS of~\eqref{eq:channel-resolv-3} can be evaluated as follows:
\begin{IEEEeqnarray}{rCl}
\IEEEeqnarraymulticol{3}{l}{
\sum\limits_{z\in\setZ}\sqrt{\mathrm{Var}[R_{Z|X}(z|X)]} }\notag \\
& =& \sum_{z} \sqrt{Q_Z(z)} \sqrt{\frac{1}{ Q_Z(z) }  \mathrm{Var}[R_{Z|X}(z|X)] } \\
&\leq& \sqrt{ \sum_{z} \frac{1}{ Q_Z(z) }  \mathrm{Var}[R_{Z|X}(z|X)] }   \label{eq:bound-var-z-cr-000} \\
&\leq&\sqrt{ \sum_{z} \frac{1}{ Q_Z(z)} \Ex{}{ R_{Z|X}^2(z|X)} } \label{eq:bound-var-z-cr-001}.\\
 &=&  \sqrt{ \sum\limits_{x,z} \frac{R_{Z|X}^2(z|x) P_X(z)}{Q_Z(z)} } \label{eq:bound-var-z-cr-0}.  
\end{IEEEeqnarray}

As in the proof of Lemma~\ref{lemma:privacy-amplification}, we shall minimize the RHS of~\eqref{eq:bound-var-z-cr-0} over all $R_{Z|X}$ that satisfy 
\begin{IEEEeqnarray}{rCl}
\|P_{XZ}- R_{Z|X}P_X\|_1 \leq \tilde{\epsilon}.
\end{IEEEeqnarray} 
The minimizer takes the form 
\begin{IEEEeqnarray}{rCl}
R^*_{Z|X}(z|x) \define   \begin{cases}
P_{Z|X}(z|x) & \imath(x;z)  \leq \log \gamma \\
\dfrac{\gamma Q_Z(z)}{|\setX| P_X(x)} & \text{otherwise}\\
\end{cases}
\label{eq:def-rxz-cr}
\end{IEEEeqnarray}  
where $\gamma$ is chosen such that $\|P_{XZ} -R_{Z|X}^* P_X\|_1 = \tilde{\epsilon}$.
Substituting~\eqref{eq:def-rxz-cr} into~\eqref{eq:bound-var-z-cr-0}, and then~\eqref{eq:def-rxz-cr} and~\eqref{eq:bound-var-z-cr-0} into~\eqref{eq:channel-resolv-3}, we conclude the proof of Lemma~\ref{lemma:channel-resolvability}.

\section{Proof of Lemma~\ref{thm:converse-partition-egamma}}
\label{app:proof-pa-converse-egamma}

Let $\imath(\cdot;\cdot)$ be  defined as in~\eqref{eq:info-den-qz}.
Let $K\define g(X)$. By definition, for every $m\in\setM$ and every $\gamma>0$, we have
\begin{IEEEeqnarray}{rCl} 
 S(g(X)|Z) &\geq& P_{KZ} \lefto[\max_{\bar{x}\in g^{-1}(K)} \imath (\bar{x};Z) \geq \log\gamma \right] \notag \\
 && -\, (Q_{\setK}^{\unif}P_{Z})\lefto[ \max_{\bar{x}\in g^{-1}(K)} \imath (\bar{x};Z) \geq \log\gamma \right].\IEEEeqnarraynumspace
\label{eq:lower-bound-vd-def}
\end{IEEEeqnarray}
The first term on the RHS of~\eqref{eq:lower-bound-vd-def} can be bounded as 
\begin{IEEEeqnarray}{rCl} 
\IEEEeqnarraymulticol{3}{l}{
P_{KZ}\lefto[\max_{\bar{x}\in g^{-1}(K)} \imath (\bar{x};Z) \geq \log \gamma \right] }\notag\\
&=&   \sum\limits_{k,z}\sum\limits_{x \in g^{-1}(k) } \! P_{XZ}(x,z) \mathbf{1}\mathopen{}\Big\{ \max_{ \bar{x}\in g^{-1}(k)}  \imath (\bar{x};z)\geq \log\gamma \Big\} \IEEEeqnarraynumspace\\
&\geq&   \sum\limits_{x ,z} P_{XZ}(x,z) \mathbf{1}\mathopen{}\Big\{  \tilde{\imath}(x;z)\geq\log \gamma \Big\} \\
&=& P_{XZ}\lefto[ \imath(X;Z) \geq \gamma \right].
\label{eq:lower-bound-vd-def-term1}
\end{IEEEeqnarray}
The second term on the RHS of~\eqref{eq:lower-bound-vd-def} can be bounded using the union bound as follows:
 \begin{IEEEeqnarray}{rCl} 
(Q_{\setK}^{\unif} P_{Z})\lefto[ \max_{\bar{x}\in g^{-1}(K)} \imath (\bar{x};Z) \geq \log\gamma \right]
 &\leq& \sum\limits_{k\in\setK}\frac{1}{|\setK|} \sum\limits_{x\in \pi^{-1}(k)}\!\! P_{Z}[\imath(x;Z)\geq \log\gamma] \\
 &=& \frac{|\setX|}{|\setK|} Q_{\setX}^{\unif} P_{Z}[\imath(X;Z)\geq \log\gamma].\IEEEeqnarraynumspace
\label{eq:lower-bound-vd-def-term2}
\end{IEEEeqnarray}
Substituting~\eqref{eq:lower-bound-vd-def-term1} and~\eqref{eq:lower-bound-vd-def-term2} into~\eqref{eq:lower-bound-vd-def},  setting $\gamma = L = |\setX|/|\setK|$,  we obtain 
\begin{IEEEeqnarray}{rCl}
 S(g(X)|Z)  &\geq& P_{XZ}\lefto[\imath(X;Z) \geq \log L \right] -  L \cdot (Q_{\setX}^{\unif}P_Z)[\imath(X;Z)\geq \log L] \\
 &=& E_{L}(P_{XZ}, Q_{\setX}^{\unif}P_Z). \IEEEeqnarraynumspace
 \label{eq:final-bound-dwz}
\end{IEEEeqnarray}

\section{Proof of Theorem~\ref{thm:dm-wtc-asy}}
\label{app:proof-asy-dmc}
\subsection{Achievability}
To prove  the achievability bound~\eqref{eq:thm-ach-expansion},  we shall use Theorem~\ref{thm:ach-wiretap-max} with a constant-composition code. 
The reason for using constant-composition codes instead of i.i.d. codes is two-fold. First, for a properly chosen $Q_{Z^n}$, all codewords $x^n$ of a constant-composition code have the same $E_{\gamma}(P_{Z^n|X^n=x^n }, Q_{Z^n})$. Secondly, constant-composition codes achieve the conditional variances $V_1$ and $V_2$, whereas i.i.d. codes achieve (the slightly bigger) unconditional variances.   
For simplicity of presentation, we only prove~\eqref{eq:thm-ach-expansion} for the special case in which 
\begin{IEEEeqnarray}{rCl}
\CS =\max_{P_X} I(X;Y) - I(X;Z).
\end{IEEEeqnarray}
Using channel prefixing, we can generalize the proof  in a straightforward manner to the case in which this condition does not hold.

Before presenting the proof, we first introduce some standard notation used in the method of types~\cite[Ch.~2]{Csiszar11}.   
We denote the simplex of probability distributions on a finite alphabet $\setA$ by $\setP(\setA)$.  The set of all $n$-types on $\setA$ is defined as 
\begin{IEEEeqnarray}{rCl}
\setP_n (\setA) \define \{P\in \setP: nP(a) \in \mathbb{Z}_{+} ,\,\, \forall a \in \setA \}.
\end{IEEEeqnarray}
The subset of sequences in $\setA^n$ with the same type $P$ is denoted by $\setT_{P}^n(\setA)$ or simply $\setT_{P}^n$ if the alphabet $\setA$ is clear from the context.
Let $P_n\in \setP_{n}(\setX)$ be the  type that is closest in total variation distance to the (unique) secrecy-capacity-achieving input distribution $P_X^*$. 
Furthermore, let $P_{X^n}=Q_{\setT_{P_n}(\setX)}^{\unif}$ denote the uniform distribution over the type class $\setT_{P_n}(\setX)$, and let $P_{Y^n} \define P_{Y^n|X^n} \circ P_{X^n}$. 
Using the permutation symmetry of the type class $\setT_{P_n}(\setX)$ and of the channel $P_{Z^n|X^n}$, it is not difficult to check that the probability term 
\begin{IEEEeqnarray}{rCl}
P_{Y^n}\lefto[ \log \frac{P_{Y^n|X^n}}{P_{Y^n}}(x^n,Y^n) \geq \log \gamma\right] 
\end{IEEEeqnarray}
takes the same value for every $x^n\in \setT_{P_n}(\setX)$. 
As a consequence, we have by~\cite[Th.~22]{polyanskiy10-05}
\begin{IEEEeqnarray}{rCl}
\error_{\mathrm{DT,max}}(a) = 1- E_{a}(P_{X^n}P_{Y^n|X^n},P_{X^n} P_{Y^n}).
\end{IEEEeqnarray}
Next, we evaluate $E_{a}(P_{X^n}P_{Y^n|X^n} , P_{X^n} P_{Y^n})$ for a given $a > 0$ as follows:
\begin{IEEEeqnarray}{rCl}
\IEEEeqnarraymulticol{3}{l}{1-  E_{a}(P_{X^n}P_{Y^n|X^n} , P_{X^n}P_{Y^n}) }\notag\\
&=& \inf_{\gamma_1 >0} \bigg\{ (P_{X^n}P_{Y^n|X^n})\lefto[  \frac{P_{Y^n | X^n} }{P_{Y^n}}(X^n,Y^n) \leq \gamma_1 \right] + a (P_{X^n}P_{Y^n})\lefto[ \frac{P_{Y^n | X^n} }{P_{Y^n}}(X^n,Y^n) >  \gamma_1 \right]\bigg\}  \label{eq:bound-E-gamma-xy-1} \IEEEeqnarraynumspace \\
&\leq & (P_{X^n}P_{Y^n|X^n})\lefto[ \frac{P_{Y^n | X^n} }{P_{Y^n}}(X^n,Y^n) \leq \sqrt{n} a \right]  + \frac{1}{\sqrt{n}}. \label{eq:bound-E-gamma-xy-2}
\end{IEEEeqnarray}
Here,~\eqref{eq:bound-E-gamma-xy-1} follows from~\eqref{eq:def-e-gamma-2}, and~\eqref{eq:bound-E-gamma-xy-2} follows by relaxing the infimum on the RHS of~\eqref{eq:bound-E-gamma-xy-1} with $\gamma_1 = \sqrt{n} a$ and by applying  the standard change of measure technique to the second term on the RHS of~\eqref{eq:bound-E-gamma-xy-1}. 
Suppose that $V_1 = V(P_X^* ,P_{Y|X})>0$.
Proceeding as in the proof of~\cite[Th.~4.2]{tan14}, we can further upper-bound the RHS of~\eqref{eq:bound-E-gamma-xy-2} by
\begin{IEEEeqnarray}{rCl}
Q\lefto( \frac{n I(P_X^*, P_{Y|X}) - \log (\sqrt{n} a)  }{\sqrt{nV_1}} \right) + \bigO\lefto(\frac{1}{\sqrt{n}}\right).
\label{eq: bd-feinstein-tan}
\end{IEEEeqnarray}
This implies that there exists an
\begin{IEEEeqnarray}{rCl}
 a_n = n^{-1/2}\exp\mathopen{}\Big( nI(X;Y) -\sqrt{nV_1}Q^{-1}\mathopen{}\big(\error  - \bigO(1/\sqrt{ n})\big) \Big)
 \label{eq:an_define-order-mang}
 \end{IEEEeqnarray}
 that satisfies 
\begin{equation}
\error \geq 1- E_{a_n}(P_{X^n} P_{Y^n|X^n} , P_{X^n}P_{Y^n}).
\label{eq: asy-error-eva}
\end{equation}
For the case  $V_1=0$, we have that 
\begin{equation}
V(P_n, P_{Y|X}) = \bigO(1/n)
\label{eq:bound-on-vp}
\end{equation}
which follows because $|P_n -P_X^*| =\bigO(1/n)$ and because  $P \mapsto V(P,P_{Y|X})$ are smooth maps on the interior of the probability simplex on $\setX$.  Proceeding step by step as in the proof of~\cite[Th.~4.2]{tan14},  and using Chebyshev's inequality and~\eqref{eq:bound-on-vp} in place of~\cite[Eq.~(4.53)]{tan14}, we conclude that~\eqref{eq: bd-feinstein-tan} and~\eqref{eq: asy-error-eva} remain to hold if $V_1=0$.

Let
\begin{IEEEeqnarray}{rCl}
\gamma_n(x_n,\delta) \define \inf\bigg\{\gamma: &\,&E_{\gamma}(P_{Z^n|X^n =x^n} ,Q_{Z^n})  \notag\\
&&+\, \frac{1}{2}\sqrt{\frac{1}{n} \Ex{}{\exp(-|\imath(x^n;Z^n) -\log \gamma|)}  }  + \sqrt{\frac{\log (1+a_n/(n\gamma))}{2n\gamma}} \leq \delta \bigg\} \IEEEeqnarraynumspace
\label{eq:def-gamma-n-xn-delta}
\end{IEEEeqnarray}
where $a_n$ is given in~\eqref{eq:an_define-order-mang} and $Q_{Z^n}$ is set to be the product distribution $(P_{Z|X} \circ P_n)^n$. 
Setting $\gamma = \sup_{x^n\in \setT_{P_n}}\gamma_n(x_n,\delta)$, $L=n\gamma$, $M_n=a_n/L$, and using~\eqref{eq:def-gamma-n-xn-delta},~\eqref{eq: asy-error-eva}, and Theorem~\ref{thm:ach-wiretap-max}, we conclude that there exists an $(n,M_n,\error,\delta)_{\max}$ secrecy code. Therefore,
\begin{IEEEeqnarray}{rCl}
R^*_{\max}(n,\error,\secrecy) &\geq &\frac{1}{n} \log M_n \\
&=& \frac{1}{n}\log a_n - \sup_{x^n\in \setT_{P_n}(\setX) } \log \gamma_n(x_n,\delta) -\frac{\log n }{n}.
\label{eq:max-r-dmc-lala}
\end{IEEEeqnarray}

It remains to evaluate $\gamma_n(x_n,\delta) $.  Using the permutation symmetry of $\setT_{P_n} $ and of $P_{Z^n|X^n}$, we see that $\gamma_n(x_n,\delta) $ takes the same value for every $x_n \in \setT_{P_n}$.
We next upper-bound  $E_{\gamma}(P_{Z^n|X^n =x^n} , Q_{Z^n})$.
In the analysis below, we shall assume that $V_2 = V(P^*_X, P_{Y|X})>0$. The case $V_2=0$ can be handled similarly as in~\eqref{eq:bound-on-vp}.
Consider the following chain of (in)equalities: 
\begin{IEEEeqnarray}{rCl}
E_{\gamma}(P_{Z^n|X^n =x^n} , Q_{Z^n}) &\leq& P_{Z^n |X^n =x^n}\lefto[\log \frac{P_{Z^n|X^n=x^n} }{ Q_{Z^n}} (Z^n) \geq \log \gamma \right] \\
&= &P_{Z^n |X^n =x^n}\lefto[ \sum\limits_{i=1}^{n} \log \frac{P_{Z|X=x_i}}{(P_{Z|X} \circ P_n)} (Z_i) \geq \log \gamma \right]\\
&\leq& Q\lefto(\frac{ \log \gamma -n I(P_n, P_{Z|X}) }{\sqrt{nV(P_n,P_{Z|X} )} }\right) + \bigO\lefto(\frac{1}{\sqrt{n}}\right)\label{eq:berry-esseen-pzx}\\
&=&  Q\lefto(\frac{ \log \gamma -n I(P_X^*, P_{Z|X}) }{\sqrt{nV_2} }\right) + \bigO\lefto(\frac{1}{\sqrt{n}}\right).\label{eq:berry-esseen-pzx-convergence}
\end{IEEEeqnarray}
Here,~\eqref{eq:berry-esseen-pzx} follows from the Berry-Esseen theorem~\cite[Sec. XVI.5]{feller70a} and~\cite[Lemma~46]{polyanskiy10-05}, and~\eqref{eq:berry-esseen-pzx-convergence} follows because  $|P_n - P_X^*| = \bigO(1/n)$ and because $P\mapsto I(P, P_{Z|X})$ and $P \mapsto V(P,P_{Z|X})$ are smooth maps on the interior of the probability simplex on $\setX$. 
Furthermore, we have 
\begin{IEEEeqnarray}{rCl}
\Ex{}{\exp(-|\imath(x^n;Z^n) -\log \gamma|)}  \leq 1
\label{eq:lalallala}
\end{IEEEeqnarray}
and 
\begin{IEEEeqnarray}{rCl}
\sqrt{\frac{\log (1+a_n/(n\gamma))}{2n\gamma}}  \leq \bigO\lefto(\frac{1}{n}\right)
\label{eq:lalallalalallal}
\end{IEEEeqnarray}
as long as $\gamma \geq n^2$.
Together~\eqref{eq:berry-esseen-pzx-convergence}--\eqref{eq:lalallalalallal} imply that 
\begin{IEEEeqnarray}{rCl}
\log \gamma_n(x_n,\delta) \leq  nI(P_X^*, P_{Z|X}) + \sqrt{nV_2}Q^{-1}\mathopen{}\big(\delta \big) + \bigO\lefto(\frac{1}{n}\right).
\label{eq:estimate-gamma-n-lala}
\end{IEEEeqnarray}
Substituting~\eqref{eq:estimate-gamma-n-lala} into~\eqref{eq:max-r-dmc-lala} we conclude~\eqref{eq:thm-ach-expansion}.

\subsection{Converse}

We next prove the converse bound~\eqref{eq:eq:thm-conv-expansion} using~\eqref{eq:converse-bound-general}. 
%We shall assume that $Y^n$ and $Z^n$ are conditionally independent of $X^n$. This comes without loss of generality since the maximal rate  $R^*(n,\error,\delta)$ depends on the channel law $P_{Y^nZ^n\given X^n}$ only through the marginal transition probabilities $P_{Y^n\given X^n}$ and $P_{Z^n\given X^n}$~\cite[Lemma~3.4]{bloch11-b}. 
%
In order to apply~\eqref{eq:converse-bound-general}, we need to select a $Q_{Y^n \given Z^n}$. 
Before doing so, we remark that in the point-to-point channel coding setting, a converse bound is usually proved by reducing a code to a constant-composition subcode (see, e.g.,~\cite{polyanskiy10-05} and~\cite{shannon67}). 
The rationale behind this reduction is that removing all codewords except those of a dominant type  reduces the coding rate by at most $\bigO((\log n)/n)$, and at the same time it  does not increase the error probability. 
A converse bound can be then proved by using the meta-converse bound~\cite[Th.~27]{polyanskiy10-05} on the dominant type, with the auxiliary output distribution  $Q_{Y^n}$ chosen to be the output distribution induced by this type. 
This reduction argument, however, does not work for the wiretap channel, because it is not clear how removing codewords will affect the secrecy index  $S(W|Z)$.
Instead, we shall choose $Q_{Y^n|Z^n}$ to be a mixture of conditional distributions $P_{Y^n\given Z^n}$ induced by all types in $\setP_n(\setX)$. We now proceed with the proof.  

For each type $P_X^{(t)}\in \setP_n$, $t=1,\ldots, |\setP_n|$, let $P_{YZ}^{(t)} \define P_{YZ|X} \circ P_X^{(t)}$ and let $P_{Y|Z}^{(t)}$ be the induced conditional distribution.  Furthermore, let 
\begin{IEEEeqnarray}{rCl}
Q_{Y^n|Z^n} (y^n \given z^n)= \frac{1}{|\setP_n|}\sum\limits_{t=1}^{|\setP_n|} \prod\limits_{i=1}^{n} P_{Y|Z}^{(t)}(y_i | z_i).
\label{eq:dmc-q-channel}
 \end{IEEEeqnarray}
Using this conditional distribution in the bound~\eqref{eq:converse-bound-general}, we obtain that, for every $\tau \in(0,1-\error-\delta)$ and every $\gamma\in\realset$,
\begin{IEEEeqnarray}{rCl}
nR^*_{\avg}(n,\error,\delta) &\leq&   - \inf_{P_{X^n}}\log \beta_{1-\error -\delta-\tau}  (P_{X^nY^nZ^n},  P_{X^nZ^n}Q_{Y^n\given Z^n}) + \log \frac{\tau+\delta}{\tau} \\
&\leq& \gamma -  \log\lefto( 1-\error-\secrecy-\tau - \sup_{P_{X^n}}
P_{X^nY^nZ^n}\lefto[ \log  \frac{P_{X^nY^nZ^n}}{P_{X^nZ^n} Q_{Y^n\given Z^n}} \geq \gamma \right] \right) + \log \frac{\tau+\delta}{\tau}  \notag\\
&&\label{eq:bound-max-rate-conv-1}
\end{IEEEeqnarray}
where the second step follows from~\cite[Eq.~(102)]{polyanskiy10-05}.
The probability term on the RHS of~\eqref{eq:bound-max-rate-conv-1} can be evaluated as follows:
\begin{IEEEeqnarray}{rCl}
\IEEEeqnarraymulticol{3}{l}{
P_{X^nY^nZ^n}\lefto[ \frac{P_{X^nY^nZ^n}}{P_{X^nZ^n} Q_{Y^n\given Z^n}} \geq \gamma \right] }\notag\\
%\,\,&=&  P_{X^nY^nZ^n}\lefto[ \log \frac{P_{Y^nZ^n|X^n}}{ P_{Z^n\given X^n} Q_{Y^n\given Z^n}} \geq \gamma \right]   \label{eq:bound-idcdf-conv-1} \\
\quad &\leq& \sup\limits_{x^n \in \setX^n}  P_{Y^nZ^n\given X^n = x^n}\lefto[ \log \frac{P_{Y^nZ^n|X^n=x^n}}{P_{Z^n\given X^n=x^n}Q_{Y^n\given Z^n}}\geq  \gamma \right] . \label{eq:bound-idcdf-conv-2}
\end{IEEEeqnarray}
%Here,~\eqref{eq:bound-idcdf-conv-1} follows because by assumption $P_{Y^n\given X^nZ^n} =P_{Y^n\given X^n }$.
 For an arbitrary $x^n\in \setX^n$,  let $t \in \{1,\ldots, |\setP_n|\}$ denote the index of the type of  $x^n$. 
Using~\eqref{eq:dmc-q-channel} and using that $\log |\setP_n| \leq |\setX|\log (n+1)$, we obtain
\begin{IEEEeqnarray}{rCl}
\IEEEeqnarraymulticol{3}{l}{
P_{Y^nZ^n\given X^n = x^n}\lefto[ \log \frac{P_{Y^nZ^n|X^n=x^n}}{P_{Z^n\given X^n =x^n}Q_{Y^n\given Z^n}}\geq  \gamma \right]  }\notag\\
&\leq&  P_{Y^nZ^n\given X^n = x^n}\lefto[ \log \frac{P_{Y^nZ^n|X^n=x^n}}{ P_{Z^n\given X^n = x^n}(P_{Y|Z}^{(t)})^n }\geq \gamma -|\setX|\log (n+1)  \right] \\
&=& P_{Y^nZ^n\given X^n = x^n}\lefto[ \sum\limits_{i=1}^{n} \log \frac{P_{YZ|X} (Y_i,Z_i |x_i)}{P_{Z|X}(Z_i|x_i) P_{Y|Z}^{(t)}(Y_i|Z_i) } \geq  \gamma - |\setX|\log (n+1) \right].
\label{eq:bound-PYZ}
\end{IEEEeqnarray}

Let now
\begin{IEEEeqnarray}{rCl}
\tilde{\gamma}(\alpha) \define   \inf   \left\{\xi:   \sup_{t \{1,\ldots, |\setP_n|\},\, x^n\in \setT_{P_{X}^{(t)} }} P_{Y^nZ^n\given X^n = x^n}\lefto[ \sum\limits_{i=1}^{n} \log \frac{P_{ZY|X} (Y_i,Z_i |x_i)}{P_{Z|X}(Z_i|x_i)P_{Y|Z}^{(t)}(Y_i|Z_i) } \geq  \xi   \right] \geq \alpha\right\}. \notag\\
\end{IEEEeqnarray}
Following similar steps as in the proof of~\cite[Th.~48]{polyanskiy10-05}, and using that $\tilde{P}_X^*$ is the unique maximizer of~\eqref{eq:ub-dmc-secrecy-capacity}, that $\tilde{V}(\tilde{P}_X^*,P_{YZ|X})>0$, and that $P_X\mapsto I(X;Y|Z)$ is concave,  we obtain  
\begin{IEEEeqnarray}{rCl}
\tilde{\gamma} (\alpha) = n \CS^{\mathrm{u}} -\sqrt{nV_c}Q^{-1}(1-\alpha) + \bigO(\log n).  
\label{eq:expand-gamma'-conv-dmc}
\end{IEEEeqnarray}
Finally, setting~$\gamma = |\setX|\log(n+1) +\tilde{\gamma}(1-\error-\secrecy-\tau)$ and $\tau =1/\sqrt{n}$, and using~\eqref{eq:expand-gamma'-conv-dmc},~\eqref{eq:bound-PYZ}, and~\eqref{eq:bound-idcdf-conv-2} in~\eqref{eq:bound-max-rate-conv-1}, we conclude the proof of~\eqref{eq:eq:thm-conv-expansion}.

\subsection{The $\error+\delta >1$ Case}

By~\eqref{eq:existence-code}, if $\error + \delta >1 $, then 
\begin{IEEEeqnarray}{rCl}
R^*_{\max}(n,\error,\delta) \geq R^*_{\max}\lefto(n, \frac{\error+\delta -1}{\delta}, 1\right).
\end{IEEEeqnarray}
By~\cite[Sec. IV.A]{polyanskiy10-05}, the maximum communication rate over the legitimate channel  for a given blocklength $n$ and maximum error probability 
$(\error+\delta -1)/\delta$ is lower-bounded by 
\begin{IEEEeqnarray}{rCl}
C_{\mathrm{legit}} -\sqrt{\frac{V_{\mathrm{legit}}}{n}}Q^{-1}\lefto( \frac{\error+\delta -1}{\delta}\right) + \bigO\lefto(\frac{\log n }{n}\right).
\end{IEEEeqnarray}
Furthermore, since $d(P,Q) \leq 1$ for every pair of probability distributions $P$ and $Q$,  the secrecy constraint is trivially satisfied without any additional processing. This implies~\eqref{eq:strong-converse-does-not}.

\section{Proof of Theorem~\ref{thm:gaussian-asy}}
\label{app:proof-asy-gaussian}

To prove~\eqref{eq:thm-ach-expansion-awgn}, we shall use Theorem~\ref{thm:ach-wiretap-max} with $P_{X^n}$ being the uniform distribution over the power sphere $\setS_n \define \{x^n\in \realset^n:  \|x^n\|^2 = nP\}$ and with $Q_{Z^n} \sim \mathcal{N}(\mathbf{0}, (P+N_2)\matI_n) $. 
For such a $P_{X^n}$ it was shown in~\cite{tan15-05-a} that there exists an
\begin{equation}
\tilde{a}_n = \exp\mathopen{}\Big(\frac{n}{2}\log(1+P/N_1) -\sqrt{n V_1}Q^{-1}(\error-\error/\sqrt{n}) + \bigO(\log n)\Big)
\label{eq:def-at_n}
\end{equation}
that satisfies 
\begin{IEEEeqnarray}{rCl}
\error_{\mathrm{RCU}} (\tilde{a}_n)\leq (1-1/\sqrt{n})\error-\error/\sqrt{n}. 
\end{IEEEeqnarray}
Setting $a_n=\tilde{a}_n /\sqrt{n}$, $\tau=1- 1/\sqrt{n}$, and using~\eqref{eq:RCU-max-bound}, we conclude that 
\begin{IEEEeqnarray}{rCl}
\error_{\mathrm{RCU,max}}(a_n) \leq \tau^{-1} \error_{\mathrm{RCU}}(\tilde{a}_n) \leq \error.
\end{IEEEeqnarray}
Furthermore, from~\eqref{eq:def-at_n}, we see that 
\begin{IEEEeqnarray}{rCl}
\log a_n = \frac{n}{2}\log(1+P/N_1) -\sqrt{n V_1}Q^{-1}(\error) + \bigO(\log n).
\end{IEEEeqnarray}

We next evaluate~\eqref{eq:thm-secrecy-bound-max}. Due to the spherical symmetry of $\setS_n$ and $Q_{Z^n}$, we have that for every $x^n \in \setS_n$
\begin{IEEEeqnarray}{rCl}
E_{\gamma}(P_{Z^n\given X^n =x^n}, Q_{Z^n}) = E_{\gamma}(P_{Z^n\given X^n =\bar{x}^n} , Q_{Z^n})
\label{eq:eval-egamma-gaussian}
\end{IEEEeqnarray}
where
\begin{equation}
\bar{x}^n\define[\sqrt{P},\ldots, \sqrt{P}].
\label{eq:special-cwd-gaussian}
\end{equation}
 The RHS of~\eqref{eq:eval-egamma-gaussian} can be evaluated 
 \begin{IEEEeqnarray}{rCl}
\IEEEeqnarraymulticol{3}{l}{
E_{\gamma}(P_{Z^n\given X^n =\bar{x}^n} , Q_{Z^n})}\notag\\
 \quad &\leq & P_{Z^n\given X^n= \bar{x}^n}\lefto[ \log \frac{dP_{Z^n\given X^n= \bar{x}^n}}{dQ_{Z^n}} (Z^n) \geq \log \gamma \right] \\
 &= & Q\lefto(\frac{\log \gamma -n \log(1+P/N_2)}{\sqrt{nV_2}}\right) + \bigO\lefto( \frac{1}{\sqrt{n}}\right) \IEEEeqnarraynumspace
 \label{eq:bound-e-gamma-gaussian}
\end{IEEEeqnarray}
where the last step follows from~\cite[p.~2357]{polyanskiy10-05}. The proof of~\eqref{eq:thm-ach-expansion-awgn} follows by repeating the steps~\eqref{eq:def-gamma-n-xn-delta}--\eqref{eq:estimate-gamma-n-lala} with~\eqref{eq:bound-e-gamma-gaussian} used in place of~\eqref{eq:berry-esseen-pzx-convergence}.

To prove the converse bound~\eqref{eq:thm-conv-expansion-awgn}, we assume  that the channel $P_{Y^nZ^n|X^n}$ is physically degraded. 
This assumption comes without loss of generality since the maximal coding rate  $R^*(n,\error,\delta)$ depends on the channel law $P_{Y^nZ^n\given X^n}$ only through the marginal transition probabilities $P_{Y^n\given X^n}$ and $P_{Z^n\given X^n}$~\cite[Lemma~3.4]{bloch11-b}. 
We shall use Theorem~\ref{thm:converse-wei} with 
\begin{IEEEeqnarray}{rCl}
Q_{Y^n\given Z^n} = \mathcal{N} \lefto(\frac{P+N_1}{P+N_2} z^n , \frac{(P+N_1)(N_2 -N_1)}{P+N_2} \matI_n\right)
\end{IEEEeqnarray}
which coincides with the marginal conditional distribution $P_{Y^n|Z^n}$ of $P_{X^nY^nZ^n}$  for the case $X^n \sim \mathcal{N}(0, P\matI_n)$.  
Observe that, by the data-processing inequality for $\beta_{\alpha}(\cdot,\cdot)$, we have 
\begin{IEEEeqnarray}{rCl}
 \beta_{\alpha} (P_{X^nY^nZ^n}, P_{X^nZ^n}Q_{Y^n|Z^n}) \geq \beta_{\alpha} (P_{X^{n+1}Y^{n+1}Z^{n+1}}, P_{X^{n+1}Z^{n+1}}Q_{Y^{n+1}|Z^{n+1}}). 
 \label{eq:data-pro-beta-conv}
\end{IEEEeqnarray}
By~\eqref{eq:data-pro-beta-conv} and~\cite[Lemma~39]{polyanskiy10-05}, it suffices to consider the case where $P_{X^n}$ is supported on the power sphere $\setS_n$. Furthermore, by the spherical symmetry of $P_{Y^nZ^n|X^n}$, $Q_{Y^n|X^n}$, and~$\setS_n$, and by using~\cite[Lemma~29]{polyanskiy10-05}, we obtain that, for every $P_{X^n}$ supported on $\setS_n$,
\begin{IEEEeqnarray}{rCl}
 \beta_{\alpha} (P_{X^nY^nZ^n}, P_{X^nZ^n}Q_{Y^n|Z^n})  =  \beta_{\alpha} (P_{Y^nZ^n|X^n=\bar{x}^n}, P_{Z^n|X^n=\bar{x}^n}Q_{Y^n|Z^n}) 
 \label{eq:beta-n-n1}
\end{IEEEeqnarray} 
where~$\bar{x}^n$ is defined in~\eqref{eq:special-cwd-gaussian}.

To evaluate the asymptotic behavior of the RHS of~\eqref{eq:beta-n-n1} we observe that, under $P_{Y^nZ^n\given X^n=\bar{x}^n }$, the random variable $\log \frac{\der P_{Y^nZ^n\given X^n=\bar{x}^n}}{\der (P_{Z^n\given X^n = \bar{x}^n} Q_{Y^n\given Z^n})}(Y^n,Z^n)$ has the same distribution as 
\begin{IEEEeqnarray}{rCl}
n\CS + \frac{\log e}{2} \sum\limits_{i=1}^{n}\left( \frac{(U_i + \bar{U}_i)^2}{N_2} -   \frac{U_i^2}{N_1} + \frac{(\sqrt{P} + U_i)^2}{P+N_1} -\frac{(\sqrt{P} + U_i + \bar{U}_i)^2}{P+N_2}\right)
\label{eq:infod-gaussian}
\end{IEEEeqnarray}
where~$\{\bar{U}_i\}$ are i.i.d. $\mathcal{N}(0, N_2-N_1)$-distributed, and $\{U_i\}$ are i.i.d. $\mathcal{N}(0,N_1) $ distributed. 
 As in~\cite[Sec.~IV.B]{polyanskiy10-05},  a central limit theorem analysis of~\eqref{eq:infod-gaussian} shows that
 \begin{IEEEeqnarray}{rCl}
-\log \beta_{\alpha} (P_{Y^nZ^n|X^n=\bar{x}^n}, P_{Z^n|X^n=\bar{x}^n}Q_{Y^n|Z^n}) =  n\CS  - \sqrt{n V_c}Q^{-1}(1-\alpha) + \bigO(\log n) \IEEEeqnarraynumspace
\label{eq:beta-gau-c-exp}
\end{IEEEeqnarray}
where $V_c$ is given in~\eqref{eq:def-disp-gau-c}. Setting $\alpha=1-\error-\delta-\tau$ and $\tau =1/\sqrt{n}$, and substituting~\eqref{eq:beta-gau-c-exp} and~\eqref{eq:beta-n-n1} into~\eqref{eq:converse-bound-general}, we conclude~\eqref{eq:thm-conv-expansion-awgn}.

\section{Proof of Theorem~\ref{thm:bsc-converse}}
\label{app:symmetry}

 \subsection{Achievability}
 \label{app:bsc-ach-nonasym}
To prove the achievability bound~\eqref{eq:achievability-bsc-nonasy}, we shall use a strengthened version of the privacy amplification lemma (Lemma~\ref{lemma:privacy-amplification}).
 Fix a positive integer $k$ and let $M\define 2^k$.
Let  $ \setX= \setZ =  \mathbb{F}_2^n $,  let $\setK = \setM = \{1,...,M\}$, and let $P_X$ and $Q_Z$ be the uniform distributions over $\setX$ and $\setZ$, respectively.
Let $P_{Z|X}$ denote the channel law of $n$ uses of a BSC. It follows that $P_{Z|X}\circ P_X = Q_Z$.
Furthermore, let $G$ be uniformly distributed over the set of functions $g$ from $\setX \to \setK$ satisfying $|g^{-1}(\ell)|= 2^{n-k}$, for every $\ell\in\setK$.
As  in the proof of Lemma~\ref{lemma:privacy-amplification}, we can rewrite  the average security index as 
\begin{IEEEeqnarray}{rCl}
S(G(X)|Z) = \frac{|\setK|}{2}\sum\limits_{z}\Ex{}{|A_1(z) + A_2(z) -\Ex{}{A_1(z)  + A_2(z)}\!|}
\label{eq:rewrite-SG-bsc1}
\end{IEEEeqnarray}
where $A_1(z)$ and $A_2(z)$ are defined in~\eqref{eq:a1z-a2z-spec-form} with $R_{XZ}$ taking the form~\eqref{eq:def-rxz}.
Different from~\eqref{eq:bound-triangle-ineq-2}, we now upper-bound the absolute value in~\eqref{eq:rewrite-SG-bsc1} as follows:
\begin{IEEEeqnarray}{rCl}
\IEEEeqnarraymulticol{3}{l}{
\Ex{}{|A_1(z) + A_2(z) -\Ex{}{A_1(z)  + A_2(z)}|} }\notag\\
 &\leq& \Ex{}{|A_1(z)|} + \Ex{}{|A_2(z) -\Ex{}{A_1(z)  + A_2(z)}|} \\
&\leq& \Ex{}{|A_1(z)|} + \sqrt{\Ex{}{|A_2(z) -\Ex{}{A_1(z)  + A_2(z)}|^2}} \\
&\leq &\Ex{}{|A_1(z)|} +  \sqrt{\Ex{}{|A_1(z)|}^2 +\mathrm{Var}[A_2(z)]}.
\label{eq:bound-a1z-a2z}
\end{IEEEeqnarray}
Now, by the symmetry of $P_{Z|X}$, $Q_Z$, $P_X$, and $R_{XZ}$,  it can be shown that  $\Ex{}{|A_1(z)|}$  is independent of $z$. 
This guarantees that 
\begin{IEEEeqnarray}{rCl}
\sum\limits_{z\in\setZ} \frac{1}{Q_Z(z)}\Ex{}{|A_1(z)|}^2 = \left(\sum\limits_{z\in\setZ} \Ex{}{|A_1(z)|}\right)^2.
\label{eq:using-setz-symmetry}
\end{IEEEeqnarray}
Substituting~\eqref{eq:bound-a1z-a2z} into~\eqref{eq:rewrite-SG-bsc1}, using~\eqref{eq:using-setz-symmetry}, and following straightforward computations, we obtain
\begin{IEEEeqnarray}{rCl}
S(G(X)|Z) &\leq& \frac{1}{2}\min_{\gamma} \left(g_n\lefto(\gamma\right) + \frac{1}{2}\sqrt{g_n\lefto(\gamma\right) +  \frac{\gamma }{ 2^{n-k}} h_n(\gamma)}\right).
\label{eq:achievability-bsc-nonasy-proo}
\end{IEEEeqnarray}
This implies that there exists a function $g:\setX\to\setM$ such that $g(X)\sim Q_{\setM}^{\unif}$ and that $S(g(X)|Z)$ is upper-bounded by the RHS of~\eqref{eq:achievability-bsc-nonasy-proo}. 
Setting $W = g(X)$, $P_{X|W} = Q_{g^{-1}(W)}^{\unif}$, and using that the legitimate channel is noiseless, we conclude that there exists an $(n,2^{k},0,\tilde{\delta})_{\avg}$ secrecy code with $\tilde{\delta}$ upper-bounded by the RHS of~\eqref{eq:achievability-bsc-nonasy-proo}. The claimed achievability bound~\eqref{eq:achievability-bsc-nonasy} follows then from~\eqref{eq:avg-ach-relation}.

Similarly, Lemma~\ref{lemma:channel-resolvability} can be strengthened to show that for a random codebook~$\setA$ with $2^{n-k} $ i.i.d. $P_X$-distributed codewords, if $P_{Z|X}$ is a BSC, then 
\begin{IEEEeqnarray}{rCl}
\Ex{\setA}{\tvar(P_{Z|\setA} ,P_Z)} \leq  \frac{1}{2}\min_{\gamma} \left(g_n\lefto(\gamma\right) + \frac{1}{2}\sqrt{g_n\lefto(\gamma\right) +  \frac{\gamma }{ 2^{n-k}} h_n(\gamma)}\right).
\end{IEEEeqnarray}
Using this result and Lemma~\ref{prop:strong-pa} on the BS-WTC, we conclude the bound~\eqref{eq:achievability-bsc-nonasy-max}.

 \subsection{Converse}
The proof of the converse bound~\eqref{eq:converse-bsc-nonasy} follows by showing that the optimal $P_X$ and $Q_Z$ of~\eqref{eq:egamma-delta-error} are both uniform distributions over $\mathbb{F}_2^n$.
For the BS-WTC with $n$ channel uses,  the left-hand side (LHS) of~\eqref{eq:egamma-delta-error} is equivalent to 
\begin{IEEEeqnarray}{rCl}
\min_{P_{X^n}\in\setP(\mathbb{F}_2^n)} \min_{Q_{Z^n}\in\setP(\mathbb{F}_2^n)}  E_{\frac{|\setX|^n}{M(1-\error)}} ( P_{X^nZ^n}, Q_{\mathbb{F}_2^n}^{\unif} Q_{Z^n}).
\label{eq:egamma-delta-error-bsc}
\end{IEEEeqnarray}
Our proof relies on the following properties of the function $E_{\gamma} ( P_{X^nZ^n}, Q_{\mathbb{F}_2^n}^{\unif}   Q_{Z^n})$.
\begin{itemize}
\item Translation invariance: for every $P_{X^n}$ and every $Q_{Z^n}$ supported on $ \mathbb{F}_2^n$,  the value of the function $E_{\gamma} ( P_{X^nZ^n}, Q_{\mathbb{F}_2^n}^{\unif} Q_{Z^n})$ does not change if we  translate  $P_{X^n} $ and $Q_{Z^n} $ by a fixed vector $v^n\in \{0,1\}^n$ simultaneously, i.e.,
\begin{IEEEeqnarray}{rCl}
X^n \to X^n \oplus v^n\\
Z^n \to Z^n \oplus v^n.
\end{IEEEeqnarray}
\item Convexity: the map $(P_{X^n}, Q_{Z^n}) \mapsto E_{\gamma} ( P_{X^nZ^n} , Q_{\mathbb{F}_2^n}^{\unif}  Q_{Z^n})$ is convex for every~$\gamma>0$. This follows because  the $E_{\gamma}$ divergence is an $f$-divergence, and every $f$-divergence is jointly convex in its arguments~\cite{csiszar2004-04a}.
\end{itemize}
%
%
%
%

%
%
%
%
%
%
%
%
%Since the finite-dimensional simplexes of probability measures on $\setX^n$ and  $\setZ^n$ are compact, we conclude that there exist $P_{X^n}^*$ and $Q_{Z^n}^*$ that attains the infimum in~\eqref{eq:egamma-delta-error}. 
%
For a given vector $v^n\in \{0,1\}^n $ and a given probability distribution $X^n \sim P_{X^n}$, we define  $P_{X^n}^{v^n}$ as the probability distribution of $X^n \oplus v^n$, and $\bar{P}_{X^n}$ as the distribution of $X^n \oplus V^n$ with $V^n$ uniformly distributed over $\mathbb{F}_2^n$. 
It is not difficult to see that $\bar{P}_{X^n}$ is also a uniform distribution. 
The probability distributions $ Q_{Z^n}^{v^n}$ and $ \bar{Q}_{Z^n}$ are defined similarly.
Now, by Jensen's inequality and the translation invariance and convexity of $E_\gamma(\cdot,\cdot)$, we have
\begin{IEEEeqnarray}{rCl}
 \IEEEeqnarraymulticol{3}{l}{
 E_{\gamma} (P_{X^n} P_{Z^n|X^n} , Q_{\mathbb{F}_2^n}^{\unif}  Q_{Z^n} ) }\notag\\
\quad &=& \frac{1}{2^n} \sum_{v^n\in \{0,1\}^n} E_{\gamma} (  P_{X^n}^{v^n} P_{Z^n|X^n}, Q_{\mathbb{F}_2^n}^{\unif}   Q_{Z^n}^{v^n}) \IEEEeqnarraynumspace\\
 &\geq& E_{\gamma} (  \bar{ P}_{X^n}P_{Z^n|X^n}   , Q_{\mathbb{F}_2^n}^{\unif}   \bar{Q}_{Z^n}) .
\end{IEEEeqnarray}
Therefore,~\eqref{eq:egamma-delta-error-bsc} is indeed minimized by uniform distributions. 
Note that, the function $g_n(\gamma)$ is equal to
\begin{IEEEeqnarray}{rCl}
g_n(\gamma) = E_\gamma(Q_{\mathbb{F}_2^n}^{\unif}  P_{Z^n|X^n}, Q_{\mathbb{F}_2^n}^{\unif}   Q_{\mathbb{F}_2^n}^{\unif} ).
\label{eq:gn-gamma-equal-egamma}
\end{IEEEeqnarray}
 This concludes the converse bound~\eqref{eq:converse-bsc-nonasy}.

\section{Proof of Theorem~\ref{thm:bsc-expansion}}
\label{app:proof-bsc-asy}

The converse part of~\eqref{eq:bsc-expansion-thm} follows from~\eqref{eq:converse-bsc-nonasy} and the following estimate 
\begin{equation} 
\log g^{-1}_{n}(t) = n\log 2-n\Hb(p)+\sqrt{n \VBSC }  Q^{-1}\!\lefto(t \right) +  \bigO\lefto(1\right)
\label{eq:expansion-egamma}
\end{equation}
which can be distilled from~\cite[Sec.~IV.A.1]{polyanskiy10-05}. Here, $g^{-1}_{n}(t): [0,1] \to [0,\infty)$ denotes the pseudo-inverse of the non-increasing function $g_n(\cdot)$
\begin{IEEEeqnarray}{rCl}
g^{-1}_{n}(t)  \define \inf \{x \in [0,\infty): g_n(x) \leq t\}. 
\end{IEEEeqnarray}

To prove the achievability part of~\eqref{eq:bsc-expansion-thm},  we shall evaluate the upper bound in~\eqref{eq:achievability-bsc-nonasy-max} with the parameters  $k =\lfloor n- \log_2(g_n^{-1}(\tilde{\delta}))\rfloor$ and $\gamma =2^{n-k}$ for some arbitrary $\tilde{\delta}\in(0,1)$.
Since $\gamma \mapsto g_n(\gamma)$ is non-increasing (because $g_n(\gamma)$ corresponds to the $E_{\gamma}$ ``distance'' between two probability distributions, which is non-increasing in $\gamma$; see~\eqref{eq:gn-gamma-equal-egamma}), and since $\gamma \geq g_n^{-1}(\tilde{\delta})$,  it follows that 
\begin{IEEEeqnarray}{rCl}
g_n(\gamma) \leq g_n(g_n^{-1}(\tilde{\delta})) \leq \tilde{\delta}. 
\label{eq:bound-on-g-n-1}
\end{IEEEeqnarray}
The term $h_n(\gamma)$ in~\eqref{eq:achievability-bsc-nonasy-max} can be rewritten as
\begin{IEEEeqnarray}{rCl}
h_n(\gamma) &=& \Ex{}{ \exp\mathopen{}\bigg\{ - \sum\limits_{i=1}^{n} \!Z_i  +  n \log(1-p)  +  \log 2^k \bigg\} \indfun{\sum\limits_{i=1}^{n} Z_i  >  - n \log(1-p)  - \log 2^k } }\notag\\ 
&& + \, \Ex{}{\exp\mathopen{}\bigg\{ \sum\limits_{i=1}^{n} \!Z_i  -  n \log(1-p)  -  \log 2^k \bigg\} \indfun{\sum\limits_{i=1}^{n} Z_i  \leq-   n \log(1-p) -\log 2^k} } \IEEEeqnarraynumspace
\label{eq:rewrite-Hn}
 \end{IEEEeqnarray}
where $Z_i \define B_i \log (p/(1-p))$ with $\{B_i\}$ i.i.d. $\mathrm{Bern}(p)$ distributed.
Using~\cite[Lemma~47]{polyanskiy10-05} on both terms on the RHS of~\eqref{eq:rewrite-Hn}, we obtain 
\begin{IEEEeqnarray}{rCl}
h_n(\gamma) \leq \frac{c_1(p)}{ \sqrt{n}}
\label{eq:bound-on-hn}
\end{IEEEeqnarray} 
 where
\begin{equation} 
c_1(p)   \define \frac{4}{\sqrt{\VBSC}}\left( \frac{\log 2}{\sqrt{2\pi \VBSC}} +  \frac{12 p|\Hb(p) -\log p|^3}{\VBSC} + \frac{12(1-p)|\Hb(p)+\log(1-p)|^3}{\VBSC} \right)
 \label{eq:c2p-b2}
\end{equation}
is a constant independent of $n$.
Using~\eqref{eq:bound-on-g-n-1} and~\eqref{eq:bound-on-hn} in~\eqref{eq:achievability-bsc-nonasy-max}, we conclude that there exists an $(n,2^k, 0, \hat{\delta})_{\max}$  secrecy code that satisfies  
\begin{IEEEeqnarray}{rCl}
k &=&\lfloor n- \log_2(g_n^{-1}(\tilde{\delta}))\rfloor 
\end{IEEEeqnarray}
and 
\begin{IEEEeqnarray}{rCl}
\hat{\delta} &=& \frac{1}{2} \tilde{\delta} + \frac{1}{2}\sqrt{\tilde{\delta}^2 + \frac{c_1(p)}{\sqrt{n}}} + \sqrt{ \frac{\log (2^k +1)}{2^{n-k+1}}}  \\
&=& \tilde{\delta} + \bigO\lefto(\frac{1}{\sqrt{n}}\right).
\end{IEEEeqnarray}
This implies that 
\begin{IEEEeqnarray}{rCl}
R^*_{\max}\lefto(n, 0,\hat{\delta} \right) & \geq& \frac{1}{n}\log(2)  \left( n- (\log_2) g_n^{-1}\mathopen{}\Big(\hat{\delta}- \bigO\mathopen{}\Big(\frac{1}{\sqrt{n}}\Big)  \Big) - 1 \right)\\
&= &  \Hb(p) - \sqrt{\frac{\VBSC}{n}}  Q^{-1}\!\lefto(\hat{\delta}\right) +  \bigO\lefto(\frac{1}{n}\right)  .
\label{eq:finally-final-step}
\end{IEEEeqnarray}
Here, the second step follows from~\eqref{eq:expansion-egamma} and by Taylor-expanding $Q^{-1}(x)$ around $x = \hat{\delta}$. 
Combining~\eqref{eq:existence-code} and~\eqref{eq:finally-final-step} we conclude the proof.



\begin{thebibliography}{10}
\providecommand{\url}[1]{#1}
\csname url@samestyle\endcsname
\providecommand{\newblock}{\relax}
\providecommand{\bibinfo}[2]{#2}
\providecommand{\BIBentrySTDinterwordspacing}{\spaceskip=0pt\relax}
\providecommand{\BIBentryALTinterwordstretchfactor}{4}
\providecommand{\BIBentryALTinterwordspacing}{\spaceskip=\fontdimen2\font plus
\BIBentryALTinterwordstretchfactor\fontdimen3\font minus
  \fontdimen4\font\relax}
\providecommand{\BIBforeignlanguage}[2]{{%
\expandafter\ifx\csname l@#1\endcsname\relax
\typeout{** WARNING: IEEEtran.bst: No hyphenation pattern has been}%
\typeout{** loaded for the language `#1'. Using the pattern for}%
\typeout{** the default language instead.}%
\else
\language=\csname l@#1\endcsname
\fi
#2}}
\providecommand{\BIBdecl}{\relax}
\BIBdecl

\bibitem{wyner1975-10a}
A.~D. Wyner, ``The wire-tap channel,'' \emph{Bell Syst. Tech. J.}, vol.~54,
  no.~8, pp. 1355--1367, Oct. 1975.

\bibitem{csiszar1978-05a}
I.~Csisz\'{a}r and J.~K\"{o}rner, ``Broadcast channels with confidential
  messages,'' \emph{{IEEE} Trans. Inf. Theory}, vol.~24, no.~3, pp. 339--348,
  May 1978.

\bibitem{Liang08}
Y.~Liang, H.~V. Poor, and S.~{Shamai (Shitz)}, ``Information theoretic
  security,'' in \emph{Foundations and Trends in Communications and Information
  Theory}.\hskip 1em plus 0.5em minus 0.4em\relax NOW, 2008, vol.~5, no. 4--5,
  pp. 355--580.

\bibitem{bloch11-b}
M.~Bloch and J.~Barros, \emph{Physical-Layer Security: from Information Theorey
  to Security Engineering}.\hskip 1em plus 0.5em minus 0.4em\relax Cambridge,
  UK: Cambridge University Press, 2011.

\bibitem{polyanskiy10-05}
Y.~Polyanskiy, H.~V. Poor, and S.~Verd{\'u}, ``Channel coding rate in the
  finite blocklength regime,'' \emph{{IEEE} Trans. Inf. Theory}, vol.~56,
  no.~5, pp. 2307--2359, May 2010.

\bibitem{hayashi2009-11a}
M.~Hayashi, ``Information spectrum approach to second-order coding rate in
  channel coding,'' \emph{{IEEE} Trans. Inf. Theory}, vol.~55, no.~11, pp.
  4947--4966, Nov. 2009.

\bibitem{tan14}
V.~Y.~F. Tan, ``Asymptotic estimates in information theory with non-vanishing
  error probabilities,'' in \emph{Foundations and Trends in Communications and
  Information Theory}.\hskip 1em plus 0.5em minus 0.4em\relax Delft, The
  Netherlands: now Publishers, 2014, vol.~11, no. 1--2, pp. 1--184.

\bibitem{bellare12-00}
M.~Bellare, S.~Tessaro, and A.~Vardy, ``Semantic security for the wiretap
  channel,'' in \emph{Advances in Cryptology (Lecture Notes in Computer
  Science)}.\hskip 1em plus 0.5em minus 0.4em\relax Berlin, Germany: Springer,
  2012, vol. 7417, pp. 2940--311.

\bibitem{goldwasser1984-04a}
S.~Goldwasser and S.~Micali, ``Probabilistic encryption,'' \emph{J. Comput.
  Syst. Sci.}, vol.~28, no.~2, pp. 270--299, Apr. 1984.

\bibitem{hayashi2006-04a}
M.~Hayashi, ``General nonasymptotic and asymptotic formulas in channel
  resolvability and identification capacity and their application to the
  wiretap channel,'' \emph{{IEEE} Trans. Inf. Theory}, vol.~52, no.~4, pp.
  1562--1575, Apr. 2006.

\bibitem{han93-03}
T.~S. Han and S.~Verd{\'u}, ``Approximation theory of output statistics,''
  \emph{{IEEE} Trans. Inf. Theory}, vol.~39, no.~3, pp. 752--772, May 1993.

\bibitem{wyner1975-03a}
A.~Wyner, ``The common information of two dependent random variables,''
  \emph{{IEEE} Trans. Inf. Theory}, vol.~21, no.~2, pp. 163--179, Mar. 1975.

\bibitem{hayashi13-11a}
M.~Hayashi, ``Tight exponential analysis of universally composable privacy
  amplification and its applications,'' \emph{{IEEE} Trans. Inf. Theory},
  vol.~59, no.~11, Nov. 2013.

\bibitem{yassaee2013-07a}
M.~H. Yassaee, M.~R. Aref, and A.~Gohari, ``Non-asymptotic output statistics of
  random binning and its applications,'' in \emph{Proc. IEEE Int. Symp. Inf.
  Theory (ISIT)}, Istanbul, Turkey, Jul. 2013, pp. 1849--1853.

\bibitem{tan2012-11a}
V.~Y.~F. Tan, ``Achievable second-order coding rates for the wiretap channel,''
  in \emph{Proc. IEEE Int. Conf. Comm. Syst.}, Singapore, Nov. 2012, pp.
  65--69.

\bibitem{hayashi2011-06a}
M.~Hayashi, ``Exponential decreasing rate of leaked information in universal
  random privacy amplification,'' \emph{{IEEE} Trans. Inf. Theory}, vol.~57,
  no.~6, pp. 3989--4001, Jun. 2011.

\bibitem{yassaee2015-06a}
M.~H. Yassaee, ``One-shot achievability via fidelity,'' in \emph{Proc. IEEE
  Int. Symp. Inf. Theory (ISIT)}, Hong Kong, China, Jun. 2015.

\bibitem{tahmasbi2016-09a}
M.~Tahmasbi and M.~R. Bloch, ``Second order asymptotics for degraded wiretap
  channels: How good are existing codes?'' in \emph{Proc. 54th Allerton Conf.
  Commun., Contr., Comp.}, UIUC, Illinois, USA, Sep. 2016, pp. 830--837.

\bibitem{parizi2017-01a}
M.~B. Parizi, E.~Telatar, and N.~Merhav, ``Exact random coding secrecy
  exponents for the wiretap channel,'' \emph{{IEEE} Trans. Inf. Theory},
  vol.~63, no.~1, pp. 509--531, Jan. 2017.

\bibitem{tyagi2015-05a}
H.~Tyagi and S.~Watanabe, ``Converses for secret key agreement and secure
  computing,'' \emph{{IEEE} Trans. Inf. Theory}, vol.~61, no.~9, pp.
  4809--4827, Sep. 2015.

\bibitem{hayashi2014-09a}
M.~Hayashi, H.~Tyagi, and S.~Watanabe, ``Strong converse for a degraded wiretap
  channel via active hypothesis testing,'' in \emph{Proc. Allerton Conf.
  Commun., Contr., Comput.}, Monticello, IL, Sep. 2014, pp. 148--151.

\bibitem{tan2015-09a}
V.~Y.~F. Tan and M.~Bloch, ``Information spectrum approach to strong converse
  theorems for degraded wiretap channels,'' \emph{{IEEE} Trans. Inf. Forensics
  Security}, vol.~10, no.~9, pp. 1891--1904, Sep. 2015.

\bibitem{liu2017-05a}
J.~Liu, P.~Cuff, and S.~Verd\'{u}, ``${E}_{\gamma}$-resolvability,''
  \emph{{IEEE} Trans. Inf. Theory}, vol.~63, no.~5, pp. 2629--2658, May 2017.

\bibitem{csiszar2004-12a}
I.~Csisz\'{a}r and P.~Narayan, ``Secrecy capacities for multiple terminals,''
  \emph{{IEEE} Trans. Inf. Theory}, vol.~50, no.~12, pp. 3047--3061, Dec. 2004.

\bibitem{Csiszar11}
I.~Csisz\'{a}r and J.~K\"{o}rner, \emph{Information Theory: Coding Theorems for
  Discrete Memoryless Systems}, 2nd~ed.\hskip 1em plus 0.5em minus 0.4em\relax
  Cambridge, U.K.: Cambridge Univ. Press, 2011.

\bibitem{bennett1995-11a}
C.~H. Bennett, G.~Brassard, C.~Cr\'{e}peau, and U.~M. Maurer, ``Generalized
  privacy amplification,'' \emph{{IEEE} Trans. Inf. Theory}, vol.~41, no.~6,
  pp. 1915--1923, Nov. 1995.

\bibitem{renner2005-09a}
R.~Renner, ``Security of quantum key distribution,'' Ph.D. dissertation, ETH,
  Zurich, Switzerland, Sep. 2005.

\bibitem{hayashi2016-06a}
M.~Hayashi, ``Security analysis of $\varepsilon$-almost dual universal$_2$ hash
  functions: Smoothing of min entropy vs. smoothing of {R}\'{e}nyi entropy of
  order 2,'' \emph{{IEEE} Trans. Inf. Theory}, vol.~62, no.~6, pp. 3451--3476,
  Jun. 2016.

\bibitem{watanabe2013-07a}
S.~Watanabe and M.~Hayashi, ``Non-asymptotic analysis of privacy amplification
  via {R}\'{e}nyi entropy and inf-spectral entropy,'' in \emph{Proc. IEEE Int.
  Symp. Inf. Theory (ISIT)}, Istanbul, Turkey, Jul. 2013, pp. 2715--2719.

\bibitem{hoeffding1963-03a}
W.~Hoeffding, ``Probability inequalities for sums of bounded random
  variables,'' \emph{J. Amer. Statist. Assoc}, vol.~58, no. 301, pp. 13--30,
  Mar. 1963.

\bibitem{mcdiarmid-1989}
C.~McDiarmid, ``On the method of bounded differences,'' in \emph{Surveys in
  Combinatorics}.\hskip 1em plus 0.5em minus 0.4em\relax Cambridge, UK:
  Cambridge University Press, 1989, vol. 141, pp. 148--188.

\bibitem{ahlswede1998-01a}
R.~Ahlswede and I.~Csisz\'{a}r, ``Common randomness in information theory and
  cryptography---part ii. cr capacity,'' \emph{{IEEE} Trans. Inf. Theory},
  vol.~44, no.~1, pp. 225--240, Jan. 1998.

\bibitem{cuff2015-09a}
P.~Cuff, ``A stronger soft-covering lemma and applications,'' in \emph{{IEEE
  Commun. Netw. Security (CNS)}}, Florence, Italy, Sep. 2015, pp. 40--43.

\bibitem{cuff2016-07a}
------, ``Soft covering with high probability,'' in \emph{Proc. IEEE Int. Symp.
  Inf. Theory (ISIT)}, Barcelona, Spain, Jul. 2016, pp. 2963--2967.

\bibitem{vazquezvilar2016-05a}
G.~{Vazquez-Vilar}, A.~T. Campo, A.~{Guill\'{e}n i F\`{a}bregas}, and
  A.~Martinez, ``Bayesian $m$-ary hypothesis testing: The meta-converse and
  {Verd\'{u}-Han} bounds are tight,'' \emph{{IEEE} Trans. Inf. Theory},
  vol.~62, no.~5, pp. 2324--2333, May 2016.

\bibitem{hayashi2016-05b}
M.~Hayashi and R.~Matsumoto, ``Secure multiplex coding with dependent and
  non-uniform multiple messages,'' \emph{{IEEE} Trans. Inf. Theory}, vol.~62,
  no.~5, pp. 2355--2409, May 2016.

\bibitem{goldfeld2016-07a}
Z.~Goldfeld, P.~Cuff, and H.~H. Permuter, ``Semantic-security capacity for
  wiretap channels of type {II},'' \emph{{IEEE} Trans. Inf. Theory}, vol.~62,
  no.~7, pp. 3863--3879, Jul. 2016.

\bibitem{shannon59}
C.~E. Shannon, ``Probability of error for optimal codes in a {G}aussian
  channel,'' \emph{Bell Syst. Tech.~J.}, vol.~38, no.~3, pp. 611--656, May
  1959.

\bibitem{kostina13-05}
V.~Kostina and S.~Verd\'{u}, ``Lossy joint source-channel coding in the finite
  blocklength regime,'' \emph{{IEEE} Trans. Inf. Theory}, vol.~59, no.~5, pp.
  2545--2575, May 2013.

\bibitem{grubb2007-07a}
J.~Grubb, S.~Vishwanath, Y.~Liang, and H.~V. Poor, ``Secrecy capacity of
  semi-deterministic wire-tap channels,'' in \emph{Proc. IEEE Inf. Theory
  Workshop Inf. Theory Wireless Netw.}, Solstrand, Norway, Jul. 2007.

\bibitem{hayashi2016-07a}
M.~Hayashi, H.~Tyagi, and S.~Watanabe, ``Secret key agreement: General capacity
  and second-order asymptotics,'' \emph{{IEEE} Trans. Inf. Theory}, vol.~62,
  no.~7, pp. 3796--3810, Jul. 2016.

\bibitem{yang2017-06a}
W.~Yang, R.~F. Schaefer, and H.~V. Poor, ``Secrecy-reliability tradeoff for
  semi-deterministic wiretap channels at finite blocklength,'' in \emph{Proc.
  IEEE Int. Symp. Inf. Theory (ISIT)}, Aachen, Germany, Jun. 2017.

\bibitem{Carter79-02a}
L.~Carter and M.~Wegman, ``Universal classes of hash functions,'' \emph{J.
  Comput. Syst. Sci.}, vol.~18, no.~2, pp. 143--154, 1979.

\bibitem{boyd04}
S.~Boyd and L.~Vandenberghe, \emph{Convex Optimization}.\hskip 1em plus 0.5em
  minus 0.4em\relax Cambridge, U.K.: Cambridge Univ. Press, 2004.

\bibitem{raginsky13}
M.~Raginsky and I.~Sason, ``Concentration of measure inequalities in
  information theory, communications and coding,'' in \emph{Foundations and
  Trends in Communications and Information Theory}.\hskip 1em plus 0.5em minus
  0.4em\relax now Publishers, 2013, vol.~10, no. 1--2, pp. 1--246.

\bibitem{feller70a}
W.~Feller, \emph{An Introduction to Probability Theory and Its
  Applications}.\hskip 1em plus 0.5em minus 0.4em\relax New York, NY, USA: John
  Wiley \& Sons, 1970, vol.~1.

\bibitem{shannon67}
C.~E. Shannon, R.~G. Gallager, and E.~R. Berlekamp, ``Lower bounds to error
  probability for coding on discrete memoryless channels~{I},'' \emph{{Inf.
  Contr.}}, vol.~10, pp. 65--103, 1967.

\bibitem{tan15-05-a}
V.~Y.~F. Tan and M.~Tomamichel, ``The third-order term in the normal
  approximation for the {AWGN} channel,'' \emph{{IEEE} Trans. Inf. Theory},
  vol.~61, no.~5, pp. 2430--2438, May 2015.

\bibitem{csiszar2004-04a}
I.~Csisz\'{a}r and P.~C. Shields, ``Information theory and statistics: A
  tutorial,'' in \emph{Foundations and Trends in Communications and Information
  Theory}.\hskip 1em plus 0.5em minus 0.4em\relax now Publishers, 2004, vol.~1,
  no.~4, pp. 417--528.

\end{thebibliography}
\end{document}